\documentstyle[12pt,epsf]{article}
\def\fo{\hbox{{1}\kern-.25em\hbox{l}}}
\def\fnote#1#2{\begingroup\def\thefootnote{#1}\footnote{#2}\addtocounter
{footnote}{-1}\endgroup}

\renewcommand{\thefootnote}{\fnsymbol{footnote}}

\def\drawbox#1#2{\hrule height#2pt 
        \hbox{\vrule width#2pt height#1pt \kern#1pt 
               \vrule width#2pt}
               \hrule height#2pt}
\def\Dbox{\vbox{\drawbox{6.0}{0.4}}}

\def\beq{\begin{equation}}
\def\eeq{\end{equation}}
\def\eq{\end{equation}}
\def\to{\rightarrow}

\def\mA{m_{A^0}}
\def\mH{m_{H^0}}
\def\mh{m_{h^0}}
\def\mW{m_{W^{\pm}}}
\def\mZ{m_{Z^0}}
\def\mHu{m_{H_u}}
\def\mHd{m_{H_d}}

\def\bino{\tilde{B}}
\def\mbino{m_{\bino}}

\def\EmissT{\not \! \!  E_{T}}

\def\Emiss{\not  \! \! E}

\def\Rslash{\not \! \! R}

\def\ai{\alpha}
\def\bi{\beta}
\def\aid{\dot{\alpha}}
\def\bid{\dot{\beta}}

\newcommand{\newc}{\newcommand}

\newc{\mpl}{M_{PL}}
 
\newc{\eegg}{e^+e^-\gamma\gamma}
\newc{\mmgg}{\mu \mu \gamma\gamma}
\newc{\ttgg}{\tau \tau \gamma\gamma}

\newc{\leplep}{l^+l^-}
\newc{\llgg}{l^+l^- \gamma \gamma}
\newc{\lllgg}{l^+l^-l^{\prime \pm} \gamma \gamma}
\newc{\ljjgg}{l^{\pm}jj \gamma \gamma}
\newc{\lljjgg}{l^+l^-jj \gamma \gamma}

\newc{\eeggE}{ee \gamma \gamma + \EmissT}
\newc{\llggE}{l^+l^- \gamma \gamma + \EmissT}

\newc{\gag}{\gamma\gamma}
\newc{\lgg}{l^{\pm} \gamma \gamma}

\newc{\jjgg}{jj \gamma \gamma}
\newc{\jjjjgg}{4j \gamma \gamma}

\newc{\gsim}{\lower.7ex\hbox{$\;\stackrel{\textstyle>}{\sim}\;$}}
\newc{\lsim}{\lower.7ex\hbox{$\;\stackrel{\textstyle<}{\sim}\;$}}
\newc{\ie}{{\it i.e.}}		
\newc{\etal}{{\it et al.}}
\newc{\eg}{{\it e.g.}}		
\newc{\kev}{\hbox{\rm\,keV}}		
\newc{\mev}{\hbox{\rm\,MeV}}		
\newc{\gev}{\hbox{\rm\,GeV}}		
\newc{\tev}{\hbox{\rm\,TeV}}
\newc{\xpb}{\hbox{\rm\, pb}}
\newc{\xfb}{\hbox{\rm\, fb}}

\newc{\mtop}{m_t}
\newc{\mbot}{m_b}
\newc{\mz}{m_Z}
\newc{\mw}{m_W}
\newc{\alphasmz}{\alpha_s(m_Z^2)}
\newc{\swsq}{\sin^2\theta_W}
\newc{\tw}{\tan\theta_W}
\newc{\cw}{\cos\theta_W}
\newc{\sw}{\sin\theta_W}
\newc{\BR}{\hbox{\rm BR}}
\newc{\zbb}{Z\to b\bar}
\newc{\Gb}{\Gamma (Z\to b\bar b)}
\newc{\Gh}{\Gamma (Z\to \hbox{\rm hadrons})}
\newc{\rbsm}{R_b^\hbox{\rm sm}}
\newc{\rbsusy}{R_b^\hbox{\rm susy}}
\newc{\drb}{\delta R_b}

\newc{\tbeta}{\tan\beta}
\newc{\uL}{{\tilde u_L}}
\newc{\uR}{{\tilde u_R}}
\newc{\cL}{{\tilde c_L}}
\newc{\cR}{{\tilde c_R}}
\newc{\tL}{{\tilde t_L}}
\newc{\tR}{{\tilde t_R}}
\newc{\dL}{{\tilde d_L}}
\newc{\dR}{{\tilde d_R}}
\newc{\sL}{{\tilde s_L}}
\newc{\sR}{{\tilde s_R}}
\newc{\bL}{{\tilde b_L}}
\newc{\bR}{{\tilde b_R}}
\newc{\eL}{{\tilde e_L}}
\newc{\eR}{{\tilde e_R}}
\newc{\mhp}{m_{H^\pm}}
\newc{\mhalf}{m_{1/2}}

\newc{\ta}{\tilde t_1}
\newc{\tb}{\tilde t_2}
\newc{\mta}{m_{\tilde t_1}}
\newc{\mtb}{m_{\tilde t_2}}

\newc{\stauR}{\tilde \tau_R}

\newc{\lR}{\tilde{l}_R}
\newc{\lL}{\tilde{l}_L}
\newc{\nL}{\tilde{\nu}_L}
\newc{\na}{\chi^0_1}
\newc{\nb}{\chi^0_2}
\newc{\nc}{\chi^0_3}
\newc{\nd}{\chi^0_4}
\newc{\ca}{\chi^{\pm}_1}
\newc{\cb}{\chi^{\pm}_2}
\newc{\capos}{\chi^{+}_1}
\newc{\caneg}{\chi^{-}_1}

\newc{\mna}{m_{\na}}

\newc{\mlR}{m_{\lR}}
\newc{\mlL}{m_{\lL}}
\newc{\meR}{m_{\eR}}
\newc{\mstauR}{m_{\stauR}}
\newc{\stau}{\tilde{\tau}}

\def\NPB#1#2#3{Nucl. Phys. B {\bf #1} (19#2) #3}
\def\PLB#1#2#3{Phys. Lett. B {\bf #1} (19#2) #3}

\def\PRD#1#2#3{Phys. Rev. D {\bf #1} (19#2) #3}
\def\PRL#1#2#3{Phys. Rev. Lett. {\bf#1} (19#2) #3}

\def\beq{\begin{equation}}
\def\eeq{\end{equation}}
\def\bea{\begin{eqnarray}}
\def\eea{\end{eqnarray}}
\def\slashchar#1{\setbox0=\hbox{$#1$}           
   \dimen0=\wd0                                 
   \setbox1=\hbox{/} \dimen1=\wd1               
   \ifdim\dimen0>\dimen1                        
      \rlap{\hbox to \dimen0{\hfil/\hfil}}      
      #1                                        
   \else                                        
      \rlap{\hbox to \dimen1{\hfil$#1$\hfil}}   
      /                                         
   \fi}                                         %
\catcode`@=11
\long\def\@caption#1[#2]#3{\par\addcontentsline{\csname
  ext@#1\endcsname}{#1}{\protect\numberline{\csname
  the#1\endcsname}{\ignorespaces #2}}\begingroup
    \small
    \@parboxrestore
    \@makecaption{\csname fnum@#1\endcsname}{\ignorespaces #3}\par
  \endgroup}
\catcode`@=12

\def\jfig#1#2#3{
 \begin{figure}
 \centering
 \epsfysize=2.75in
 \hspace*{0in}
 \epsffile{#2}
 \caption{#3}
 \label{#1}
 \end{figure}}
\def\sfig#1#2#3{
 \begin{figure}
 \centering
 \epsfysize=2.0in
 \hspace*{0in}
 \epsffile{#2}
 \caption{#3}
 \label{#1}
 \end{figure}}

\begin{document}

\begin{titlepage}

\begin{flushright}
SLAC-PUB-7237 \\
CERN-TH/96-173 \\
SU-ITP 96-29\\
hep-ph/9609434 
\end{flushright}


\huge
\begin{center}
  Sparticle Spectroscopy and\\
  Electroweak Symmetry Breaking \\
  with Gauge-Mediated \\
  Supersymmetry Breaking
\end{center}

\large

\vspace{.1in}
\begin{center}

Savas Dimopoulos$^{a,b}$, Scott Thomas$^c$\fnote{\dagger}{Work 
supported by the Department of Energy
under contract DE-AC03-76SF00515.}, James D. Wells$^{c \dagger}$ \\
\vspace{.15in}
\normalsize
$^a${\it Theoretical Physics Division,
CERN\\
CH-1211 Geneva 23, Switzerland}
\\
\vspace{.1in}
$^b${\it Physics Department, 
Stanford University, 
Stanford, CA  94305}\\
\normalsize
\vspace{.1in}
$^c${\it Stanford Linear Accelerator Center,
Stanford, CA 94309\\}

\end{center}
 
 
\vspace{0.1in}

\normalsize

\begin{abstract}

\baselineskip=16pt

The phenomenology associated with gauge-mediated supersymmetry
breaking is presented. 
A renormalization group analysis of the minimal model
is performed in which the constraints of radiative electroweak
symmetry breaking are imposed. 
The resulting superpartner and Higgs boson spectra are highly 
correlated and depend on only a few parameters. 
Superpartner mass ratios and sum rules are identified
which can be tested at future colliders. 
Some of these relations are logarithmically sensitive
to the messenger scale, 
while others allow gauge-mediation to be distinguished from other
schemes for transmitting supersymmetry breaking.
Deviations from the minimal model, such as larger messenger 
representations and additional contributions to Higgs
sector masses, can in some circumstances dramatically
modify the low energy spectrum. 
These modifications include a slepton or Higgsino as the lightest
standard model superpartner, or exotic mass relations among
the scalars and gauginos. 
The contribution to $b \to s \gamma$ and resulting
bound on superpartner masses are also presented for the minimal model. 
Finally, the unique collider signatures of 
heavy charged particle production, or
decay to the Goldstino within a detector are discussed. 

\end{abstract}

\end{titlepage}

\baselineskip=18pt

\section{Introduction}

Supersymmetry provides an elegant framework 
in which physics at the electroweak scale
can be decoupled from Planck scale physics. 
The electroweak scale arises dynamically as the effective scale
of supersymmetry breaking in the visible sector. 
The breaking of supersymmetry must be transmitted
from a breaking sector to the visible sector through
a messenger sector. 
Most phenomenological studies of low energy 
supersymmetry implicitly assume that messenger
sector interactions are of gravitational strength. 
The intrinsic scale of supersymmetry breaking is 
then necessarily of order $\sqrt{F} \sim 10^{11}$ GeV,
giving an electroweak scale of ${G_F^{-1/2}} \sim F / M_p$. 
While gravitational strength interactions represent a
lower limit, it is certainly possible that the messenger
scale, $M$, is anywhere between the Planck and just above
the electroweak scale,
with supersymmetry broken at 
an intermediate scale, ${G_F^{-1/2}} \sim F / M$.

If the messenger scale is well below the Planck scale, 
it is likely that the usual gauge interactions of the
standard model play some role in the messenger sector. 
This is because standard model gauginos couple
at the renormalizable level only through gauge interactions. 
If the Higgs bosons received masses 
predominantly from non-gauge interactions in the messenger sector, 
with only a small contribution from
gauge interactions,
the standard model gauginos would be unacceptably
lighter than the electroweak scale.\footnote{ 
The argument for light gauginos in the absence of 
standard model gauge interactions within a messenger
sector well below the Planck scale
only applies if the gauginos
are elementary degrees of freedom. 
If the standard model gauge group is magnetic at or below
the messenger scale the gauginos could in principle
receive a large mass from operators suppressed by the 
confining magnetic scale.} 
It is therefore interesting to consider theories in which 
the standard model gauge interactions act as messengers of supersymmetry
breaking \cite{lsgauge,hsgauge,dnmodels}.
This mechanism occurs if supersymmetry is realized non-linearly
in some sector which transforms under the standard model 
gauge group.
Supersymmetry breaking in the visible sector spectrum 
then arises as a radiative correction.

In this paper we consider the superpartner spectroscopy and 
important phenomenological signatures that result
from gauge-mediated supersymmetry breaking.
Since within this anzatz
the gauge interactions transmit supersymmetry breaking,
the standard model soft masses arise in proportion to 
gauge charges squared. 
This leads to a sizeable hierarchy among the superpartner masses
according to gauge quantum numbers. 
In addition, for a large class of models, there are a number
of relations and sum rules among the superpartner masses.
Electroweak symmetry breaking is driven by negative radiative
corrections to the up-type Higgs mass squared from the large top quark
Yukawa coupling and large stop masses
\cite{dnmodels}.
With the constraint of electroweak symmetry breaking 
the minimal model of gauge-mediation is highly constrained
and very predictive,
with the superpartner spectrum depending primarily
on two parameters -- the overall scale and $\tan \beta$. 
In addition, there is a logarithmic dependence 
of various mass relations on the messenger scale.


The precise form of the low lying superpartner spectrum 
determines the signatures that can be observed at a high
energy collider. 
With gauge-mediated supersymmetry breaking, either 
a neutralino or slepton is the lightest standard model
superpartner. 
The signature for supersymmetry is then either the traditional
missing energy or heavy charged particle production. 
In a large class of models the general form of the cascade decays
to the lightest standard model superpartner is largely fixed
by the anzatz of gauge-mediation. 
In addition, 
for a low enough supersymmetry breaking scale, the lightest
standard model superpartner can decay to its partner plus
the Goldstino component of the gravitino within the 
detector.

In the next subsection the natural lack of flavor
changing neutral currents with gauge-mediated supersymmetry
breaking is discussed. 
The minimal model of gauge-mediated supersymmetry breaking
(MGM) and its variations are presented in section 2.
A renormalization group analysis of the minimal model
is performed in section 3,
with the constraint of proper radiative electroweak symmetry
breaking enforced. 
Details of the resulting superpartner and Higgs boson spectra
are discussed.
Mass relations and sum rules are identified which can distinguish
gauge mediation from other theories for the soft terms.
Some mass relations allow a logarithmically sensitive probe of  
the messenger scale.  
In section 4 variations of the minimal model are studied.
With larger messenger sector representations
the lightest standard model superpartner is naturally a slepton. 
Alternately, additional sources for Higgs sector masses, can lead
in some instances to a Higgsino as the lightest standard model
superpartner. 
The phenomenological consequences of gauge-mediated supersymmetry
breaking are given in section 5. 
The supersymmetric contribution to 
${\rm Br}(b \to s \gamma)$ in the minimal model, and resulting
bound on the overall scale for the superpartners, are quantified. 
The collider signatures for superpartner production in both
the minimal model, and models with larger messenger sector
representations, are also detailed. 
In the latter case, the striking signature of heavy charged
particles exiting the detector can result, rather than 
the traditional missing energy. 
The signatures resulting from decay of the lightest standard 
model superpartner to its partner plus the Goldstino are 
also reviewed. 
In section 6 we conclude with a few summary remarks
and a comment about tuning.

The general expression for scalar and gaugino masses in a large
class of models is given in appendix A. 
A non-minimal model is presented in appendix B which 
demonstrates an approximate $U(1)_R$ symmetry, and has
exotic scalar and gaugino mass relations, even though
it may be embedded in a GUT theory. 
Finally, in appendix C the couplings of the Goldstino
component of the gravitino 
are reviewed. 
In addition to the general expressions for the decay 
rate of the lightest standard model superpartner 
to its partner plus the Goldstino,
the severe suppression of the branching ratio to Higgs boson final 
states in the minimal model is quantified. 




\subsection{Ultraviolet Insensitivity}

\label{UVinsensitive}

Low energy supersymmetry removes power law sensitivity
to ultraviolet physics. 
In four dimensions with ${\cal N}=1$ supersymmetry the parameters
of the low energy theory are renormalized however. 
Infrared physics can therefore be logarithmically sensitive
to effects in the ultraviolet.
The best example of this is the value of the weak
mixing angle at the electroweak scale in supersymmetric grand unified 
theories \cite{dimgeo}.
Soft supersymmetry breaking terms in the low energy theory
also evolve logarithmically with scale. 
The soft terms therefore remain ``hard'' up to the 
messenger scale at which they are generated. 
If the messenger sector interactions are of gravitational strength,
the soft terms are sensitive to ultraviolet physics all the 
way to the Planck or compactification scale. 
In this case patterns within the soft terms might give an 
indirect window to the Planck scale \cite{peskin}.
However, the soft terms are then also sensitive
to flavor violation at all scales. 
Flavor violation at any scale can then in principle 
lead to unacceptable flavor violation at the electroweak 
scale \cite{hcr}.
This is usually avoided by postulating precise relations
among squark masses at the high scale, such as 
universality \cite{dimgeo} or proportionality. 
Such relations, however, do not follow from any unbroken
symmetry, and are 
violated by Yukawa interactions. 
As a result the relations only hold at a single scale.
They are ``detuned'' under renormalization
group evolution, and can be badly violated in extensions
of the minimal supersymmetric standard model (MSSM). 
For example, in grand unified theories, large flavor violations
can be induced by running between the Planck and 
GUT scales \cite{hcr}.
Elaborate flavor symmetries may be imposed to limit
flavor violations with a Planck scale messenger 
sector \cite{flavorsym}.

Sensitivity to the far ultraviolet is removed in theories
with a messenger scale well below the Planck scale. 
In this case the soft terms are ``soft'' above the
messenger scale. 
Relations among the soft parameters are then not
``detuned'' by ultraviolet physics. 
In particular there can exist a sector which 
is responsible for the flavor structure of the Yukawa matrix.
This can arise from a hierarchy
of dynamically generated scales \cite{flavordyn}, or
from flavor symmetries, spontaneously broken
by a hierarchy of expectation values \cite{Yukvev}.
If the messenger sector for supersymmetry breaking is well
below the scale at which the Yukawa hierarchies are generated,
the soft terms can be insensitive to the flavor sector. 
Naturally small flavor violation can result without
specially designed symmetries. 

Gauge-mediated supersymmetry breaking gives an elegant 
realization of a messenger sector below the Planck scale
with ``soft'' soft terms and very small flavor violation. 
The direct flavor violation induced 
in the squark mass matrix at the messenger scale 
is GIM suppressed compared with
flavor conserving squark masses
by ${\cal O}(m_f^2/M^2)$, where $m_f$ is a fermion mass. 
The largest flavor violation is generated by renormalization
group evolution between the messenger and electroweak 
scales.
This experiences a GIM suppression of 
${\cal O}(m_f^2/\tilde{m}^2)\ln(M/\tilde{m})$,
where $\tilde{m}$ is a squark mass, and is 
well below current experimental bounds. 

If the messenger sector fields transforming under the 
standard model gauge group have the same quantum numbers
as visible sector fields, flavor violating
mixing can take place through Yukawa couplings. 
This generally leads to large flavor violating soft 
terms \cite{Yukflavor}.
These dangerous mixings are easily avoided by discrete
symmetries.  
For example, in the minimal model discussed in the next section,
if the messenger fields are even under $R$-parity,
no mixing occurs.
In more realistic models in which the messenger fields 
are embedded directly in the supersymmetry breaking
sector, messenger sector gauge symmetries responsible for
supersymmetry breaking forbid flavor violating mixings. 
The natural lack of flavor violation is a significant advantage
of gauge-mediated supersymmetry breaking.

\section{The Minimal Model of Gauge-Mediated Supersymmetry Breaking}

The standard model gauge interactions act as messengers of 
supersymmetry breaking if fields within the supersymmetry 
breaking sector transform under the standard model gauge group. 
Integrating out the messenger sector fields gives rise to 
radiatively generated soft terms within the visible sector,
as discussed below.  
The messenger fields should fall into vector representations
at the messenger scale 
in order to obtain a mass well above the electroweak scale. 
In order not to disturb 
the successful prediction of gauge coupling
unification within the MSSM \cite{dimgeo} at lowest order, 
it is sufficient (although not necessary \cite{martin})
that the messenger sector fields transform as complete 
multiplets of any grand unified gauge group which contains the
standard model. 
If the messenger fields remain elementary degrees of freedom
up to the unification scale, the further requirement of 
perturbative unification may be imposed on the 
messenger sector. 
For supersymmetry breaking at a low scale, 
these constraints allow up to four flavors of 
${\bf 5} + \overline{\bf 5}$ of $SU(5)$, 
a single ${\bf 10} + \overline{\bf 10}$ of $SU(5)$, or 
a single ${\bf 16} + \overline{\bf 16}$ of $SO(10)$. 
With the assumptions outlined above, these are the 
discrete choices for the standard model representations in the 
messenger sector. 

In the following subsection the minimal model of gauge-mediated
supersymmetry breaking is defined. 
In the subsequent subsection variations of the 
minimal model are introduced.

\subsection{The Minimal Model}

\label{minimalsection}

\sfig{gauginoloop}{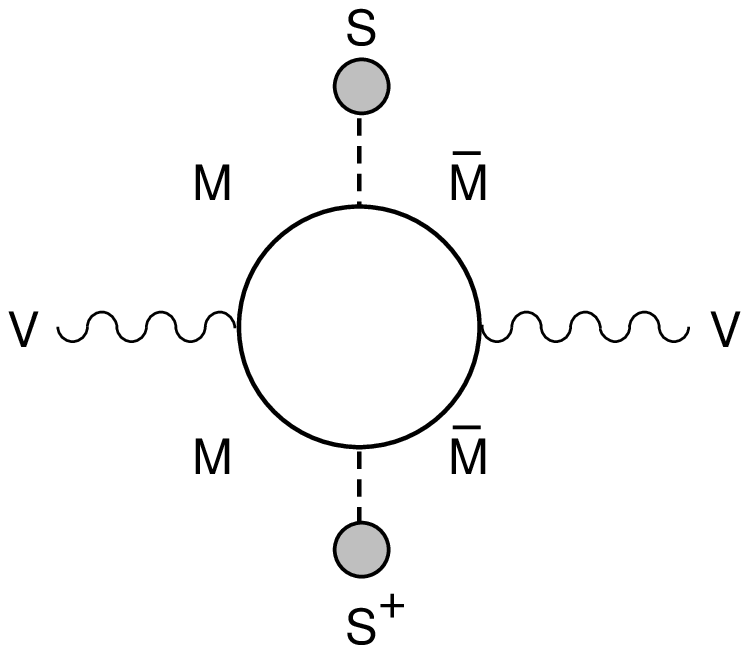}
{One-loop messenger sector supergraph which gives rise to visible sector
gaugino masses.}

The minimal model of gauge-mediated supersymmetry breaking
(which preserves the successful predictions of perturbative
gauge unification) consists of messenger fields which transform
as a single flavor of ${\bf 5} + \overline{\bf 5}$ of $SU(5)$, 
i.e. there are $SU(2)_L$ doublets 
$\ell$ and $\bar{\ell}$, and $SU(3)_C$ triplets $q$ and $\bar{q}$. 
In order to introduce supersymmetry breaking into the messenger
sector, these fields may be coupled to a gauge 
singlet spurion, $S$, through the superpotential
\beq
W = \lambda_2 S \ell \bar{\ell} + \lambda_3 S q \bar{q} 
\label{SQQbar}
\eq
The scalar expectation value of $S$ sets the overall scale
for the messenger sector, and the auxiliary component, $F$, 
sets the supersymmetry breaking scale. 
For $F \neq 0$ the messenger spectrum is not supersymmetric,
$$
m_b = M \sqrt{ 1 \pm {\Lambda \over M} }
$$
\beq
m_f = M
\eq
where $M = \lambda S$ and $\Lambda = F/S$. 
The parameter $\Lambda /M$ sets the scale for the fractional
splitting between bosons and fermions. 
Avoiding electroweak and color breaking in the messenger 
sector requires $M > \Lambda$. 

In the models of Ref. \cite{dnmodels} the field $S$ is an 
elementary singlet which couples through a secondary
messenger sector to the supersymmetry breaking sector. 
In more realistic models the messenger fields are
embedded directly in the supersymmetry breaking sector. 
This may be accomplished within a model of dynamical supersymmetry
breaking by identifying an unbroken global symmetry with the 
standard model gauge group. 
In the present context, 
the field $S$ should be thought of as a spurion which 
represents the dynamics which break supersymmetry. 
The physics discussed in this paper does not depend on the
details of the dynamics represented by the spurion. 
Because (\ref{SQQbar}) amounts to tree level breaking, the 
messenger spectrum satisfies the sum rule
${\cal S}Tr~m^2 = 0$.
With a dynamical supersymmetry breaking sector, this sum rule
need not be satisfied.
The precise value of ${\cal S}Tr~m^2$ in the messenger sector, however,
does not significantly affect the radiatively generated 
visible sector soft parameters discussed below.

\sfig{scalarloop}{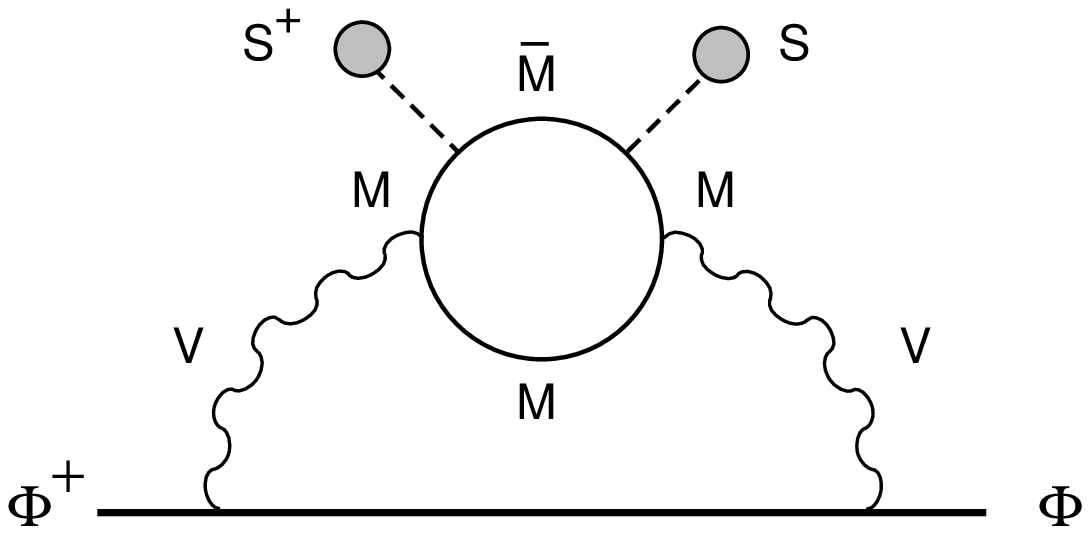}
{Two-loop messenger sector supergraph which gives rise to visible sector
scalar masses.  The one-loop subgraph gives rise to visible sector
gaugino wave function renormalization.  Other graphs related by
gauge invariance are not shown.}
Integrating out the non-supersymmetric messengers gives rise
to effective operators, which lead to supersymmetry breaking
in the visible sector. 
Gaugino masses arise at one-loop from the operator
\beq
\int d^2\theta~\ln S\, W^\alpha W_\alpha ~+~ h.c.
\label{SSWDV}
\eq
as shown in Fig. 1. 
In superspace this operator amounts to a shift of the 
gauge couplings in the presence of the background spurion.
Inserting a single spurion auxiliary component gives a
gaugino mass. 
For $F \ll \lambda S^2$ the gaugino masses 
are \cite{dnmodels}
\beq
m_{\lambda_i}(M) =  \frac{\alpha_i(M)}{4\pi}~ \Lambda 
\label{gauginomass}
\eq
where $\Lambda = F /S$ and 
GUT normalized gauge couplings are assumed 
($\alpha_1 = \alpha_2 = \alpha_3$
at the unification scale).\footnote{The standard model normalization
of hypercharge is related to the GUT normalization
by $\alpha^{\prime} = (3/5) \alpha_1$.}
The dominant loop momenta in Fig. 1 are ${\cal O}(M)$, so 
(\ref{gauginomass})
amounts to a boundary condition for
the gaugino masses at the messenger scale.
Visible sector scalar masses arise at two-loops from the
operator
\beq
\int d^4 \theta~ \ln(S^{\dagger} S) ~ \Phi^{\dagger} e^V \Phi
\label{SSPP}
\eq
as shown in Fig. \ref{scalarloop}.
In superspace this operator represents wave function
renormalization from the background spurion. 
Inserting two powers of the auxiliary component of the
spurion gives a scalar mass squared. 
For $F \ll \lambda S^2$ the scalar masses are \cite{dnmodels}
\beq
m^2(M) = 2 \Lambda^2~ \sum_{i=1}^3 ~ k_i 
  \left( \alpha_i(M) \over 4 \pi \right)^2
\label{scalarmass}
\eq
where the sum is over $SU(3) \times SU(2)_L \times U(1)_Y$, with 
$k_1 = (3/5) (Y/2)^2$ where the hypercharge is 
normalized as $Q = T_3 + {1 \over 2} Y$, 
$k_2 = 3/4$ for $SU(2)_L$ doublets and zero for singlets, 
and $k_3 = 4/3$ for $SU(3)_C$ triplets and zero for singlets. 
Again, the dominant loop momenta in Fig. \ref{scalarloop}
are ${\cal O}(M)$, so 
(\ref{scalarmass})
amounts to a boundary condition for
the scalar masses at the messenger scale.

It is interesting to note that for $F \ll \lambda S^2$ the soft masses 
(\ref{gauginomass}) and (\ref{scalarmass}) are independent
of the magnitude of the Yukawa couplings (\ref{SQQbar}). 
This is because the one-loop graph of Fig. \ref{gauginoloop}
has an infrared divergence, $k^{-2}$, which is cut off by the 
messenger mass $M=\lambda S$, thereby cancelling the $\lambda F$ dependence
in the numerator.  
The one-loop subgraph of Fig. (\ref{scalarloop}) 
has a similar infrared divergence which cancels the 
$\lambda$ dependence.
For finite $F/ (\lambda S^2)$ the corrections to (\ref{gauginomass})
and (\ref{scalarmass})
are small unless $M$ is very close to $\Lambda$ 
\cite{martin,dgp,yuritloop}.

Since the gaugino masses arise at one-loop and scalar masses
squared at two-loops, superpartners masses are generally
the same order for particles with similar gauge charges. 
If the messenger scale is well below the GUT scale, 
then $\alpha_3 \gg \alpha_2 > \alpha_1$, so the squarks and gluino 
receive mass predominantly from $SU(3)_C$ interactions, 
the left handed sleptons and $W$-ino from $SU(2)_L$ interactions, 
and the right handed sleptons and $B$-ino from $U(1)_Y$ interactions. 
The gaugino and scalar masses are then related at the messenger
scale by $m_3^2 \simeq {3 \over 8} m_{\tilde{q}}^2$, 
$m_2^2 \simeq {2 \over 3} m_{\lL}^2$, and 
$m_1^2 = {5 \over 6} m_{\lR}^2$. 
This also leads to a hierarchy in mass between electroweak and
strongly interacting states. 
The gaugino masses at the messenger scale are in the ratios
$m_1 : m_2 : m_3 = \alpha_1 : \alpha_2 : \alpha_3$,
while the scalar masses squared are in the approximate ratios
$m_{\tilde{q}}^2 : m_{\lL}^2 : m_{\lR}^2 \simeq  {4 \over 3} \alpha_3^2 : 
 {3 \over 4} \alpha_2^2 :  {3 \over  5} \alpha_1^2$.
The masses of particles with different gauge charges are tightly
correlated in the minimal model. 
These correlations are reflected in the constraints of electroweak
symmetry breaking on the low energy spectrum, as
discussed in section  \ref{RGEanalysis}.

The parameter $(\alpha / 4 \pi) \Lambda$ sets the scale for 
the soft masses. 
This should be of order the weak scale, implying 
$\Lambda \sim {\cal O}(100\tev)$. 
The messenger scale $M$ is, however, arbitrary in the minimal model,
subject to $M > \Lambda$.   
In models in which 
the messenger sector is embedded directly in a renormalizable
dynamical supersymmetry breaking sector \cite{SUquantum},
the messenger and 
effective supersymmetry breaking scales are
the same order, $M \sim \Lambda \sim {\cal O}(100\tev)$,
up to small hierarchies from messenger sector Yukawa couplings. 
This is also true of models
with a secondary messenger sector \cite{dnmodels}.
The messenger scale can, however, be well separated from the
supersymmetry breaking scale. 
This can arise in models with large ratios of dynamical scales.
Alternatively with non-renormalizable supersymmetry breaking,  
which vanishes in the flat space limit,
expectation values intermediate between the Planck and 
supersymmetry breaking 
scale can develop, leading to $M \gg \Lambda$. 

A noteworthy feature of the minimal messenger sector
is that it is invariant under charge conjugation and parity,
up to electroweak radiative corrections. 
This has the important effect of enforcing the vanishing
of the $U(1)_Y$ Fayet-Iliopoulos $D$-term 
at all orders in interactions that involve gauge interactions
and messenger fields only. 
This is crucial since a non-zero $U(1)_Y$ D-term at one-loop
would induce soft scalar masses much larger 
in magnitude than the
two-loop contributions (\ref{scalarmass}),
and lead to $SU(3)_C$ and $U(1)_Q$ breaking. 
This vanishing is unfortunately not an automatic feature of
models in which the messenger fields also transform
under a chiral representation of the gauge group responsible for
breaking supersymmetry. 
In the minimal model a $U(1)_Y$ $D$-term is generated only 
by gauge couplings to chiral standard model fields at three loops. 
The leading log contribution comes from renormalization group
evolution and is discussed in section~\ref{electroweaksection}.

The dimensionful parameters within the Higgs sector
\beq
W = \mu H_u H_d
\eq
and 
\beq
V = m_{12}^2 H_u H_d ~+~ h.c.
\eq
do not follow from the anzatz of gauge-mediated supersymmetry
breaking. 
These terms require additional interactions which violate
$U(1)_{PQ}$ and $U(1)_{R-PQ}$ symmetries. 
A number of models have been proposed to generate these
terms, including,
additional messenger quarks and 
singlets \cite{dnmodels}, singlets with an 
inverted hierarchy \cite{dnmodels}, and singlets with
an approximate global symmetry \cite{dgphiggs}.
In the minimal model the mechanisms for generating the 
parameters $\mu$ and $m_{12}^2$ are not specified, and 
they are taken as free parameters at the messenger scale. 
As discussed below, upon imposing electroweak symmetry breaking, 
these parameters may be eliminated in favor of 
$\tan \beta = v_u / v_d$ and $\mZ$.

\sfig{Aloop}{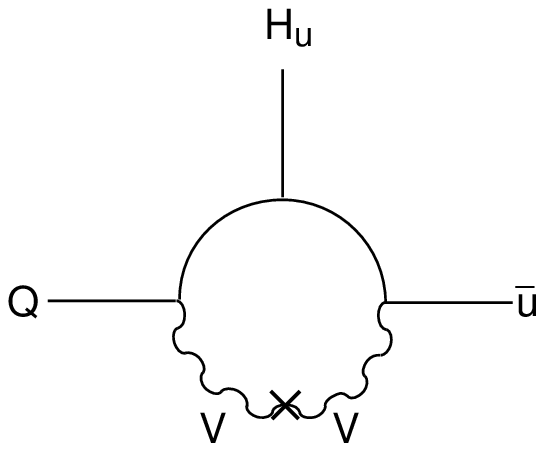}
{One-loop visible sector supergraph which contains both
logarithmic and finite contributions to visible sector
$A$-terms.  The cross on the visible sector gaugino line 
represents the gaugino mass insertion shown in Fig. 1.}
Soft tri-linear $A$-terms require interactions which 
violate both $U(1)_R$ and visible sector chiral 
flavor symmetries. 
Since the messenger sector does not violate visible sector
flavor symmetries, $A$-terms are not generated at one-loop.
However, two-loop contributions involving a visible sector
gaugino supermultiplet do give rise to 
operators of the form 
\beq
\int d^4 \theta~ \ln S ~ {D^2 \over \Dbox} Q H_u \bar{u} 
   ~+~h.c.
\label{SSQHu}
\eq
as shown in Fig. \ref{Aloop},
and similarly for down-type quarks and leptons. 
This operator is related by an integration by parts in superspace 
to a correction to 
the superpotential in the presence of the background spurion.
Inserting a single auxiliary component of the spurion 
(equivalent to a visible sector gaugino mass in the one-loop subgraph) 
gives a tri-linear $A$-term. 
The momenta in the effective one-loop graph shown in Fig. \ref{Aloop}
are equally weighted in logarithmic intervals between
the gaugino mass and messenger scale.  
Over these scales the gaugino mass is ``hard.''
The $A$-terms therefore effectively 
vanish at the messenger scale, and 
are generated
from renormalization group evolution below
the messenger scale, and finite contributions at 
the electroweak scale. 
At the low scale the $A$-terms have magnitude
$A \sim  (\alpha / 4 \pi) m_{\lambda}\ln(M/m_{\lambda})$.
Note that $A$ is small compared with the scale
of the other soft terms, 
unless the logarithm is large. 
As discussed in section \ref{sspectrum} the $A$-terms
do not have a qualitative effect on the superpartner
spectrum unless the messenger scale is large.

The general MSSM has a large number of $CP$-violating phases
beyond those of the standard model.
Since here the soft masses are flavor symmetric (up to very small
GIM suppressed corrections discussed in section (\ref{UVinsensitive})), 
only flavor symmetric phases are relevant. 
A background charge analysis \cite{dtphases} fixes the  
basis independent combination of flavor symmetric phases to be
${\rm Arg}(m_{\lambda} \mu (m_{12}^2)^*)$ and
${\rm Arg}(A^* m_{\lambda})$. 
Since the $A$-terms vanish at the messenger scale only the first
of these can arise in the soft terms. 
In the models of \cite{dnmodels} the auxiliary component of a
single field is the source for 
all soft terms, giving a correlation among the phases such
that ${\rm Arg}(m_{\lambda} \mu (m_{12}^2)^*) = 0$ mod $\pi$. 
In the minimal model, however, the mechanism for generating
$\mu$ and $m_{12}^2$ is not specified, and the phase is
arbitrary.

Below the messenger scale the particle content of the 
minimal model is just that of the MSSM, along with the 
gravitino discussed in section \ref{collidersection}
and appendix \ref{appgoldstino}.
At the messenger scale the boundary conditions for the 
visible sector 
soft terms are given by (\ref{gauginomass}) and 
(\ref{scalarmass}), $\mu$, $m_{12}^2$, and $A=0$. 
It is important to note that from the low energy
point of view the minimal model is just a set of 
boundary conditions specified at the messenger scale. 
These boundary conditions may be traded for the electroweak
scale parameters
\beq
( ~ \tan \beta~,~ \Lambda=F/S~,~{\rm Arg}~\mu~,~\ln M~)
\label{minpar}
\eq
The most important of these is $\Lambda$ which sets
the overall scale for the superpartner spectrum.
Since all the soft masses are related in the minimal model,
$\Lambda$ may be traded for any of these, such
as $m_{\tilde{B}}(M)$.
It may also be traded for a physical mass, such as 
$m_{\na}$ or $m_{\lL}$.
In addition, as discussed in section \ref{EWSB} 
$\tan \beta$ ``determines'' $m_{12}$ and $\mu$
in the low energy theory, and can have important effects
on the superpartner spectrum.


\subsection{Variations of the Minimal Model}

\label{subvariations}

The minimal model represents a highly constrained and 
very predictive theory for the soft supersymmetry 
breaking terms. 
It is therefore interesting to consider how 
the qualitative features discussed in the
remainder of the paper change under deformations
away from the minimal model. 

The most straightforward generalization is to 
the other messenger sector representations which are consistent
with perturbative unification discussed at the 
beginning of this section. 
The expressions for gaugino and scalar masses for a 
general messenger sector are given in appendix \ref{appgeneral}.
The gaugino masses grow like the quadratic index of 
the messenger sector matter, while the scalar masses
grow like the square root of the quadratic index.
Models with larger messenger sector representations generally
have gauginos which are relatively heavier, compared with the
scalars, than the minimal model. 
This can have important consequences for the standard model
superpartner spectrum. 
In particular, a scalar lepton can be the lightest 
standard model superpartner (as opposed to the lightest
neutralino for the minimal model).
This is the case for a range of parameters with two 
messenger generations of ${\bf 5} + \overline{\bf 5}$ of 
$SU(5)$, as discussed in section 
\ref{multiple}.

Another generalization is to introduce multiple spurions
with general scalar and auxiliary
components, and general Yukawa couplings. 
Unlike the case with a single spurion,
the scalar and fermion mass matrices in 
the messenger sector are in general not aligned.
Such a situation generally arises if the messengers receive
masses from a sector not associated with supersymmetry
breaking. 
This can occur in dynamical models with chiral messenger
representations which confine to, or are dual to, 
a low energy theory with vector representations.
The messengers can gain a mass at the 
confinement or duality scale, with
supersymmetry broken at a much lower scale \cite{itdual}. 
With multiple spurions, the infrared cancelations of the
messenger Yukawa couplings, described in the previous
subsection for the minimal model, no longer hold. 
This has the effect of removing the tight correlations 
in the minimal model
between masses of superpartners with different gauge charges.
As an example, a model is presented in appendix \ref{appnonmin}
with two generations of  ${\bf 5} + \overline{\bf 5}$ of 
$SU(5)$, and two singlet fields. 
One of the singlets is responsible for masses in the 
messenger sector, while the other breaks supersymmetry.
Even though the model can be embedded in a GUT theory, 
it yields a non-minimal spectrum. 

Soft scalar masses require supersymmetry breaking, while
gaugino masses require, in addition, breaking of $U(1)_R$ 
symmetry.
If $U(1)_R$ symmetry is broken at a lower scale
than supersymmetry, the gauginos can be lighter
than in the minimal model. 
This may be represented in the low energy
theory by a parameter $\Rslash$ which is the ratio of 
a gaugino to scalar mass at the messenger 
scale, relative to that in the minimal model.
The general definition of $\Rslash$ in theories with 
a single spurion is given in appendix \ref{appgeneral}.
With multiple spurions $\Rslash < 1$ is generally obtained
since the messenger scalar and fermion mass matrices do not 
necessarily align. 
Since the gauginos have little influence on electroweak 
symmetry breaking, as discussed in section \ref{appUR}, 
the main effect of $\Rslash < 1$ is simply to make
the gauginos lighter than in the minimal model. 
The non-minimal model given in appendix \ref{appnonmin}
has, in one limit, 
an approximate $U(1)_R$ symmetry and light gauginos.

Additional interactions in the messenger sector 
are required in order to generate $\mu$ and $m_{12}^2$. 
It is likely that these interactions also contribute
to the Higgs boson soft masses.\footnote{We thank Gian Giudice
for this important observation.}
Therefore, even though Higgs bosons and lepton doublets
have the same electroweak quantum numbers, their soft
masses at the messenger scale can be different. 
In the low energy theory this may be parameterized by
the quantities 
$$
\Delta_{+}^2 \equiv 
m_{H_d}^2 + m_{H_u}^2 - 2 m_{l_L}^2
$$
\beq
\Delta_{-}^2 \equiv 
m_{H_d}^2 - m_{H_u}^2  
\label{splithiggs}
\eq
where $m_{\lL}^2$ is the gauge-mediated left handed
slepton mass,
and all masses are evaluated at the messenger scale.
In the minimal model $\Delta_{\pm}^2=0$. 
Since the Higgs soft masses affect electroweak symmetry 
breaking, these splittings can potentially
have significant effects on the superpartner
spectrum. 
However, as discussed in section \ref{addhiggs}, unless the non-minimal
contributions are very large, the general form of the 
spectrum is largely unaltered.

Finally, the $U(1)_Y$ $D$-term, $D_Y$, can have a non-zero 
expectation value at the messenger scale. 
This leads to a shift of the soft scalar masses at the messenger
scale proportional to hypercharge, $Y$, 
\beq
\delta m^2 (M) = g' Y ~ D_Y(M)
\label{Dmass}
\eq
where $g' = \sqrt{3/5} g_1$ is the $U(1)_Y$ coupling.
Note that for the Higgs bosons, this yields a shift 
$\Delta_-^2 = -2 g' D_Y(M)$. 
As discussed in the previous subsection, $D_Y(M)=0$ 
in the minimal model as the result of a parity symmetry in 
the messenger sector. 
In non-minimal models $D_Y(M)$ need not vanish. 
In general, an unsupressed $U(1)_Y$ $D$-term generated at 
one-loop in the messenger sector destabilizes the electroweak
scale, and leads to $SU(3)_C$ and $U(1)_Q$ breaking. 
However, it is possible for $D_Y(M)$ to be generated
at a level which gives rise to soft masses of the same order
as the gauge mediated masses. 
For example, the one-loop contribution to $D_Y(M)$ is 
suppressed if there is an approximate parity symmetry in 
the messenger sector, and further suppressed if the messenger
couplings are unified at the GUT scale \cite{appparity}.
As another example, if the messengers sector transforms under
an additional $U(1)_{\tilde{Y}}$, gauge kinetic mixing between $U(1)_Y$ and 
$U(1)_{\tilde{Y}}$ couples $D_Y$ and $D_{\tilde{Y}}$ \cite{kineticmix}. 
Renormalization group evolution gives a one-loop 
mixing contribution
$D_Y(M) \simeq {\cal O}((g' \tilde{g} / 16 \pi^2) \ln(M_c / M)
D_{\tilde{Y}})$ where $M_c$ is the GUT or compactification scale. 
If the messengers in this case were embedded directly in a 
renormalizable dynamical supersymmetry breaking sector, 
$\tilde{g} D_{\tilde{Y}} \sim (\lambda^2 / \tilde{g}^2) F$, 
where $\lambda$ is a Yukawa coupling in the dynamical sector.
For $\lambda \ll \tilde{g}$ the one-loop contribution to 
$D_Y(M)$ is then suppressed. 
With a non-renormalizable model $D_{\tilde{Y}}$ is suppressed by 
powers of the dynamical scale over the Planck scale.

Given the above variations, 
we consider the following parameters, in addition
to those of the minimal model (\ref{minpar})
\beq
(~N~,~\Rslash~ , ~\Delta^2_+ ~ , ~\Delta^2_-~, ~D_Y~)
\eq
where $N$ is the number of generations of 
${\bf 5} + \overline{\bf 5}$ in the messenger sector. 
In the remainder of the paper we consider the minimal 
case, $N=1$, $\Rslash=1$, $\Delta_{\pm}^2=0$, 
and $D_Y=0$, and 
discuss in detail the effects on the superpartner spectrum 
for $N=2$, $\Rslash \neq 1$, $\Delta_-^2 \neq 0$, and 
$D_Y \neq 0$ in 
section \ref{varcon}.

\section{Renormalization Group Analysis}

\label{RGEanalysis}

The general features of the superpartner spectrum 
and resulting phenomenology are 
determined by the boundary conditions at the messenger scale. 
Renormalization group evolution of the parameters
between the messenger and electroweak scales 
can have a number of important effects. 
Electroweak symmetry breaking results from the negative 
evolution of the up-type Higgs mass squared from the large
top quark Yukawa coupling. 
Imposing electroweak symmetry breaking gives 
relations among the Higgs sector mass parameters. 
Details of the sparticle spectrum can also be affected by
renormalization. 
In particular, changes in the 
splittings among the light states can have an important impact
on several collider signatures, as discussed in section 
\ref{collidersection}.
In addition, general features of the spectrum, such as the
splitting between squarks and left handed sleptons, can 
be logarithmically sensitive to the messenger scale.

In this section the effects of renormalization group evolution 
on electroweak symmetry breaking and the superpartner spectrum in 
the minimal model are presented. 
Gauge couplings, gaugino masses, and third generation
Yukawa couplings are evolved at two-loops. 
Scalar masses, tri-linear $A$-terms, the $\mu$ parameter, 
and $m_{12}^2$ are evolved at one-loop. 
For the scalar masses, the $D$-term contributions to the 
$\beta$-functions 
are included (these are sometimes neglected in the literature).
Unless stated otherwise, the top quark pole mass is 
taken to be $m_t^{\rm pole} = 175$ GeV. 
The renormalization group analysis is similar to that of 
Ref. \cite{CMSSM}, being modified for the boundary conditions and 
arbitrary messenger scale of the minimal model of gauge-mediated
supersymmetry breaking.

As discussed in the previous section, 
the boundary conditions at the high scale are 
$m_{\lambda_i}, m^2_i, \mu, m_{12}^2$, and
${\rm sgn}~\mu$. 
The precise scale at which a  
soft term is defined depends on the 
messenger field masses. 
In the minimal model the mass of the messenger doublets and
triplets are determined by the messenger Yukawa couplings, 
$M_{2,3} = \lambda_{2,3} S$, which can in principle differ. 
Even if $\lambda_2 = \lambda_3$ at the GUT scale these couplings
are split under renormalization group evolution \cite{hitoshi}. 
In a full model, the supersymmetry breaking dynamics 
also in part determine the values of $\lambda_2$ and $\lambda_3$
through renormalization group evolution.  
Since we are interested primarily in effects which follow from the
anzatz of gauge-mediated supersymmetry breaking, and not 
on details of the mechanism of supersymmetry breaking, 
we will neglect any splitting between doublet and triplet masses. 
This would in fact be the case, up to small gauge coupling
corrections, if the messengers are embedded 
in a supersymmetry breaking sector near a strongly coupled 
fixed point. 
All soft terms are therefore assumed to be specified
at a single messenger scale, $M$.
The range of allowed $M$ is taken to be $\Lambda \leq M \leq M_{GUT}$. 
In the minimal model, $M$ precisely equal to 
$\Lambda$ is unrealistic since some
messenger sector scalars become massless at this point. 
In what follows
the limit $M = \Lambda$ 
is not to be taken literally in the minimal model, but
should be thought of as indicative
of a realistic model with a single dynamical scale, and 
no small parameters \cite{SUquantum}.
In many of the specific examples given in the following subsections
we take $\Lambda = 76,118,163$ TeV, with $\Lambda=M$, which
gives a $B$-ino mass at the messenger scale of 
$m_{\tilde B}(M) = 115, 180, 250$ GeV,
with the other soft masses related by the boundary conditions
(\ref{gauginomass}) and (\ref{scalarmass}).

\jfig{sfig7n}{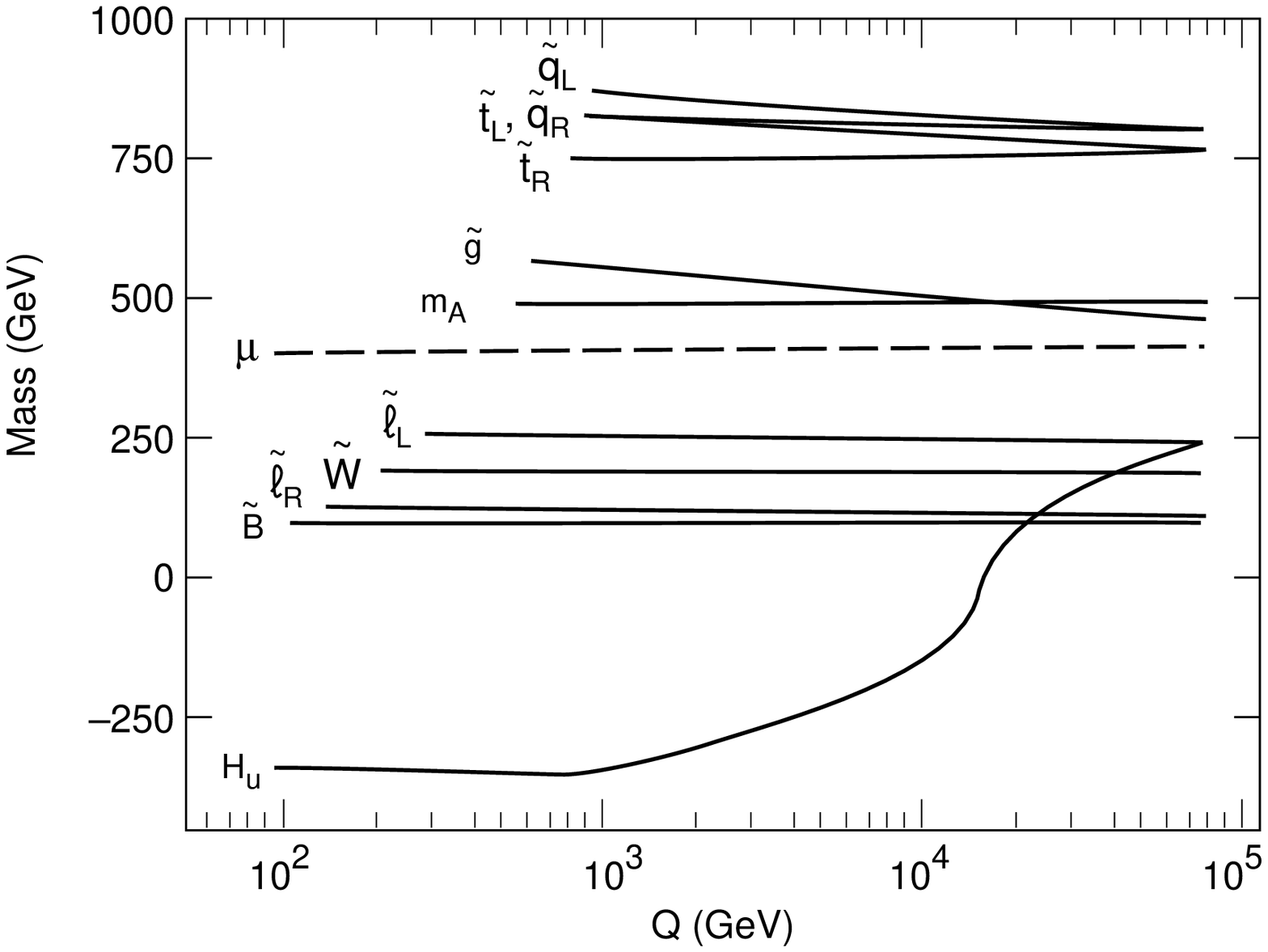}{Renormalization group evolution of the
$\overline{\rm DR}$ mass parameters with MGM boundary conditions.  The
messenger scale is $M=76\tev$, $m_{\tilde B}(M)=115\gev$,
$\tan\beta =3$, ${\rm sgn}(\mu) = +1$,
and $m_t^{\rm pole} = 175$ GeV.
The masses are plotted as $m \equiv {\rm sgn}(m) (|m|)^{1/2}$.
}
The renormalization of the $\overline{\rm DR}$ mass parameters
according to the one- and two-loop $\beta$ functions described
above 
for the minimal model 
with $M = \Lambda = 76$ TeV, 
$m_{\tilde{B}}(M) = 115$ GeV, $\tan \beta =3$,
and ${\rm sgn}(\mu) = +1$, 
is shown in Fig. \ref{sfig7n}.
As can be seen, renormalization has an effect 
on the mass parameters. 
In the following subsections the 
constraints imposed
by electroweak symmetry breaking, and the 
form of the low energy spectrum resulting from renormalization group
evolution are discussed.

\subsection{Electroweak Symmetry Breaking}

\label{EWSB}

The most significant effect of renormalization group evolution
is the negative contribution to the 
up-type Higgs boson mass squared from the large
top quark Yukawa coupling. 
As can be seen in Fig. \ref{sfig7n} this leads to a negative
mass squared for $H_u$. 
The $\beta$-function for $\mHu^2$ is dominated by the 
heavy stop mass
\beq
\frac{dm^2_{H_u}}{dt} \simeq 
\frac{1}{16\pi^2} (6h^2_t (m^2_{\tilde{t}_L} +m^2_{\tilde{t}_R} 
  + \mHu^2) +\cdots )
\eeq
where $t=\ln(Q)$, with a small 
${\cal O}(g_2^2 m_2^2 / h_t^2 m^2_{\tilde{t}})$
correction from gauge interactions. 
For the parameters given in Fig. \ref{sfig7n}
the full $\beta$-function for $\mHu^2$
is approximately constant above the stop 
thresholds,
differing by less than 1\% between $M$ and $m_{\tilde t}$. 
The evolution of $\mHu^2$ is therefore approximately
linear in this region
(the non-linear feature in Fig. \ref{sfig7n} is a square-root
singularity because the quantity plotted is 
${\rm sgn}(\mHu) \sqrt{ | \mHu |^2}$).
It is worth noting that 
for $\tan \beta$ not too large, and 
$M$ not too much larger than $\Lambda$,
$\mHu^2(m_{\tilde t})$ is then well approximated by 
\beq
\mHu^2(m_{\tilde t}) \simeq \mHu^2(M) - {3 \over 8 \pi^2} h_t^2
     ( m^2_{\tilde{t}_L} +m^2_{\tilde{t}_R} ) \ln(M/ m_{\tilde t}) 
\label{Huapp}
\eq
(although throughout we use numerical integration of the 
full renormalization group equations). 
For the parameters of Fig. \ref{sfig7n} 
the approximation (\ref{Huapp}) 
differs from the full numerical integration by 2\%
(using the messenger scale values of $h_t$,
$m_{\tilde t_L}$, and $m_{\tilde t_R}$).
The magnitude of the small positive gauge contribution 
can be seen below the stop thresholds in Fig. \ref{sfig7n}.

The negative value of $\mHu^2$ leads to electroweak symmetry
breaking in the low energy theory. 
The mechanism of radiative symmetry breaking \cite{dnmodels}
is similar to that for high scale supersymmetry breaking with universal
boundary conditions. 
With high scale supersymmetry breaking, $\mHu^2 < 0$ develops
because of the large logarithm. 
Here $\mHu^2 < 0$ results not because the logarithm is large, 
but because the stop masses are large \cite{dnmodels}.
Notice in Fig. \ref{sfig7n} that $\mHu^2$ turns negative
in less than a decade of running. 
The negative stop correction 
effectively amounts to an 
${\cal O}((\alpha_3 / 4 \pi)^2 
  (h_t/4 \pi)^2 \ln(M / m_{\tilde t}))$
three-loop contribution which is larger than the 
${\cal O}((\alpha_2 / 4 \pi)^2 )$ two-loop contribution
\cite{dnmodels}.
Naturally large stop masses which lead automatically to 
radiative electroweak symmetry breaking are one of the
nice features of low scale 
gauge-mediated supersymmetry breaking.

Imposing electroweak symmetry breaking gives relations
among the Higgs sector mass parameters. 
In the approach taken here we solve for the electroweak
scale values of $\mu$ and $m_{12}^2$ 
in terms of $\tan \beta$ and $\mZ$
using the minimization
conditions
\bea
|\mu|^2+\frac{\mZ^2}{2} & = & 
 \frac{(m^2_{H_d}+\Sigma_d)-(m^2_{H_u}+\Sigma_u)\tan^2\beta}
{\tan^2\beta -1} 
   \label{mincona} \\ 
   \label{minconb}  
  \sin 2\beta & = & 
  \frac{-2m^2_{12}}{(m^2_{H_u}+\Sigma_u)+(m^2_{H_d}+\Sigma_d)+2|\mu|^2}
\eea
where 
$\Sigma_{u,d}$ represent finite one-loop corrections 
from gauge interactions and top and bottom Yukawas
\cite{finitehiggs}.
These corrections are
necessary to reduce substantially the scale dependence 
of the minimization conditions. 
In order to minimize the stop contributions to the finite corrections,
the renormalization scale is taken to be the geometric mean
of the stop masses, $Q^2 = m_{\tilde{t}_1} m_{\tilde{t}_2}$. 
The finite corrections to the one-loop effective potential
make a non-negligible contribution
to (\ref{mincona}) and (\ref{minconb}).
For example, with the parameters of Fig. \ref{sfig7n}, 
minimization of the tree level conditions with the renormalization
scheme given above gives $\mu = 360$ GeV, 
while inclusion of the finite corrections results in $\mu = 395$ GeV. 
The minimization conditions (\ref{mincona}) and (\ref{minconb})
depend on 
the value of the top quark mass, $m_t$, mainly through
the running of $\mHu^2$, and also through the finite corrections. 
For the parameters of Fig. \ref{sfig7n},
a top mass range in the range $m_t = 175 \pm 15$ GeV gives a 
$\mu$ parameter in the range $\mu = 395 \pm 50$ GeV.

\jfig{sfig16n}{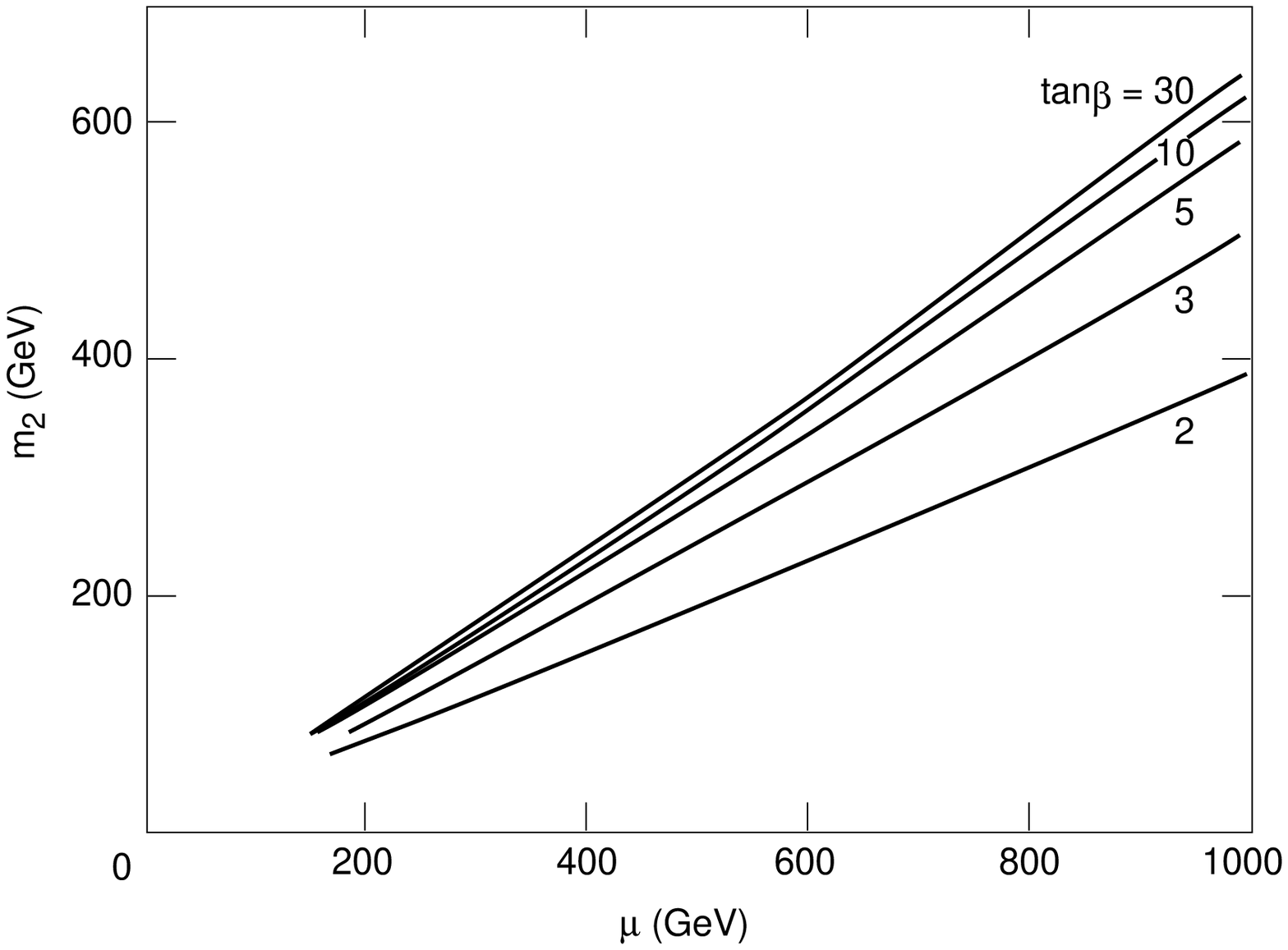}
{The relation between $m_2$ and $|\mu|$
imposed by electroweak symmetry breaking with MGM boundary conditions
for $\tan\beta =2,3,5,10,30$, and $\Lambda=M$.}
The correlation between $\mu$ and $m_{12}^2$ can be obtained from 
(\ref{mincona}) and (\ref{minconb})
in terms of $\tan \beta$ and $\mZ$. 
The relation between the $W$-ino mass, $m_2$,
and $\mu$ (evaluated at the renormalization scale) 
imposed by electroweak symmetry breaking 
in the minimal model is shown 
in Fig. \ref{sfig16n} for 
$\tan \beta = 2,3,5,10$, and $30$, with $\Lambda = M$ and
${\rm sgn}(\mu)=+1$. 
The actual correlation is of course between the Higgs sector
mass parameters.
The parameter $m_2$ is plotted just as a representative mass 
of states transforming under $SU(2)_L$, and of the overall scale
of the soft masses
(the gaugino masses directly affect electroweak
symmetry breaking only through very small higher order corrections
to renormalized Higgs sector parameters). 
The $\mu$ parameter typically lies in the range
$ {3 \over 2} m_2 \lsim |\mu| \lsim 3 m_2$ or 
$ m_{\lL} \lsim |\mu| \lsim 2 m_{\lL}$,
depending on the precise values of 
$\tan \beta$ and $\ln M$.

The correlation between $\mu$ and the 
overall scale of the soft masses arises because 
the stop masses set the scale for $\mHu^2$ at the electroweak
scale, and therefore the depth of the Higgs potential. 
For $\tan \beta \gg 1$ 
the conditions (\ref{mincona}) reduce to 
$|\mu|^2 \simeq -\mHu^2 - {1 \over 2} \mZ^2$.
In this limit, for $|\mu|^2 \gg \mZ^2$, 
$|\mu| \simeq (- \mHu^2)^{1/2}$, with the small difference determining
the electroweak scale. 
At moderate $\tan \beta$ the corrections to this approximation
increase $\mu$ for a fixed overall scale. 
At fixed $\tan \beta$, and $M$ not too far 
above $\Lambda$, $\mHu^2$ at the renormalization scale
is approximately a linear function of 
the overall scale $\Lambda$, as can be seen from Eq. (\ref{Huapp}).
The very slight non-linearity in Fig. \ref{sfig16n} arises
from $\ln M$ dependence, and ${\cal O}(\mZ^2 / \mu^2)$
effects in the minimization conditions.
The limit $|\mu|^2, |\mHu|^2 \gg \mZ^2$ of course represents
a tuning among the Higgs potential parameters
in order to obtain proper electroweak symmetry breaking.

\jfig{sfig11n}{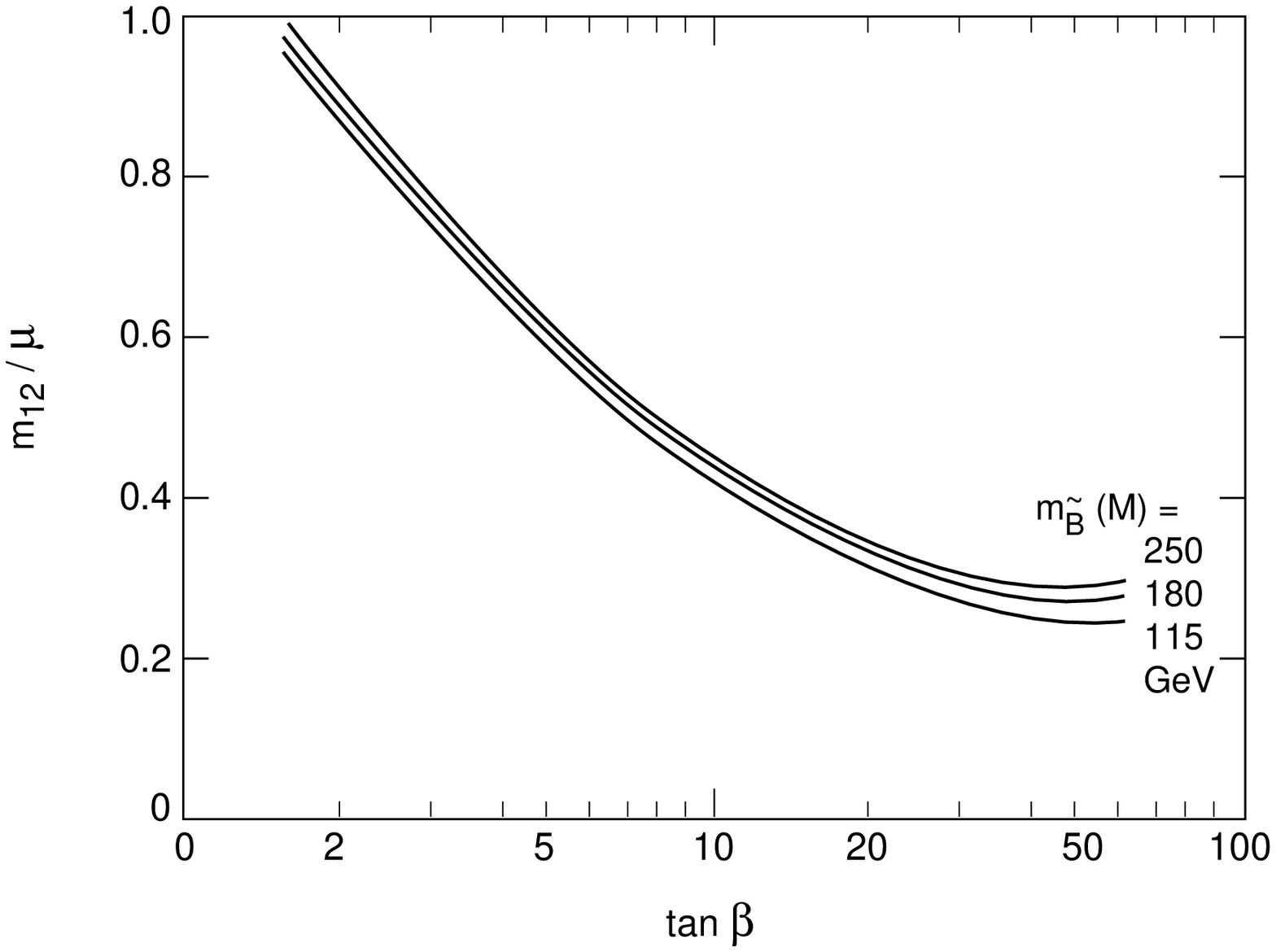}
{The ratio $m_{12}/\mu$ as a function of
$\tan\beta$ imposed by electroweak symmetry breaking for 
$m_{\tilde B}(M)=115, 180,250$ GeV and $\Lambda=M$.}
The ratio $m_{12} / \mu$ at the renormalization scale is
plotted in Fig. \ref{sfig11n}
for $m_{\tilde B}(M)=115, 180,250$ GeV and $\Lambda=M$.
Again, the tight correlation, approximately independent of the
overall scale, arises because all soft terms are related to 
a single scale, $\Lambda$. 
The small splitting between the three cases shown 
in Fig. \ref{sfig11n} arises from $\ln M$ dependence, and  
${\cal O}(\mZ^2 / \mu^2)$
effects in the minimization conditions.
Ignoring corrections from the bottom Yukawa, 
$m_{12} / \mu \rightarrow 0$
for $\tan \beta \gg 1$.  
The saturation of $m_{12} / \mu$ at large $\tan \beta$ is 
due to bottom Yukawa effects in the renormalization 
group evolution and finite corrections to $\mHd^2$. 
Any theory for the origin of $\mu$ and $m_{12}^2$ with
minimal gauge mediation, and 
only the MSSM degrees of freedom at low energy, would
have to reproduce (or at least be compatible with) the
relation given in Fig. \ref{sfig11n}.
Note that all the low scale Higgs sector mass parameters are quite
similar in magnitude 
over essentially all the parameter space of the MGM.

\subsection{Sparticle Spectroscopy}

\label{sspectrum}

\begin{table}
\begin{center}
\begin{tabular}{cc}
\hline \hline 
Particle &  Mass (GeV) \\ \hline
$\tilde{u}_L, \tilde{d}_L$  &  869, 871 \\
$\tilde{u}_R, \tilde{d}_R$  &  834, 832 \\
$\tilde{t}_1, \tilde{t}_2$  &  765, 860 \\
$\tilde{g}$  &  642 \\
$A^0, H^0, H^{\pm}$ &   506, 510, 516 \\
$\chi_3^0, \chi_2^{\pm}, \chi_4^0$  &  404, 426, 429 \\
$\tilde{\nu}_L, \tilde{l}_L$   &  260, 270 \\
$ {\chi_1^{\pm}}, {\chi_2^0}$  &  174, 175 \\
$\tilde{l}_R$   &  137 \\
$h^0$  &  104 \\
${\chi_1^0}$  & 95 \\
\hline \hline
\end{tabular}
\caption{Superpartner physical spectrum for the parameters given in 
Fig. 4.}
\end{center}
\end{table}

The gross features of the superpartner spectrum are determined 
by the boundary conditions at the messenger scale. 
Renormalization group evolution can modify these somewhat, 
and electroweak symmetry breaking imposes relations which 
are reflected in the spectrum. 
Mixing and $D$-term contributions also shift 
some of the states slightly. 
The physical spectrum resulting from the renormalized parameters
given in Fig. \ref{sfig7n} is presented in Table 1. 
In the following subsections the spectroscopy of the electroweak
states,
strongly interacting states, and Higgs bosons are discussed. 
We also consider the 
dependence of the spectrum on the messenger scale,
and discuss quantitative relations among the superpartner masses
which test the hypothesis of gauge-mediation and can be 
sensitive to the messenger scale.

\subsubsection{Electroweak States}

\label{electroweaksection}

The physical charginos, $\chi_i^{\pm}$, and neutralinos, 
$\chi_i^0$, are mixtures
of the electroweak gauginos and Higgsinos.
As discussed in the previous subsection,
imposing electroweak symmetry breaking with MGM boundary
conditions implies 
${3 \over 2} m_2  \lsim |\mu| \lsim 3 m_2 $ 
over all the allowed
parameter space. 
With this inequality, in the limit 
$\mu^2 - m_2^2 \gg \mW^2$ the lightest chargino is mostly
gaugino. 
Likewise, in the limit $\mu^2 - m_1^2 \gg \mZ^2$, the lightest
two neutralinos are mostly gaugino. 
Under renormalization group evolution both $m_1$ 
and $m_2$ are slightly decreased.  
For the parameters of Fig. \ref{sfig7n} this amounts to a
$-15$ GeV shift for the $B$-ino and a $-10$ GeV shift for the 
$W$-ino.  
At the electroweak scale $m_2 \simeq 2 m_1$. 
The lightest neutralino is therefore mostly $B$-ino. 
For example, with the parameters given in Fig. \ref{sfig7n},
the $\na$ eigenvectors are 
$N_{1 \tilde{B}} = 0.98$,
$N_{1 \tilde{W}} = - 0.09$,
$N_{1d} = 0.14$, and
$N_{1u} = - 0.07$. 
Expanding in $\mZ^2/ (\mu^2-m_1^2)$,
the $\na$ mass is given by~\cite{ssconstraints}
\beq
m_{\na} \simeq m_1 - { \mZ^2 \sin^2 \theta_W \
    ( m_1 + \mu \sin2 \beta) \over |\mu|^2 - m_1^2 } .
\label{binoshift}
\eq
Note that the shift in the physical 
$\na$ mass relative to the $B$-ino mass parameter $m_1$
depends on ${\rm sgn}(\mu)$.  
For the parameters in Fig. \ref{sfig7n} with ${\rm sgn}(\mu) = +1$ 
this amounts to a $-5$ GeV shift. 
Except for very large $\tan \beta$ discussed below, 
$\na$ is the lightest standard model superpartner.

The lightest chargino and second lightest neutralino 
are mostly $W$-ino, and form an approximately degenerate
triplet of $SU(2)_L$, 
$(\chi_1^+, \chi_2^0, \chi_1^-)$,
as can be seen in Table 1. 
This approximate triplet is very degenerate, with splittings 
arising only at 
${\cal O}(\mZ^4 / \mu^4)$. 
Expanding in $\mW^2/(\mu^2 - m_2^2)$,
the triplet mass is given by~\cite{ssconstraints}
\beq
m_{\nb,\ca} \simeq m_2 - { \mW^2     
  ( m_2 + \mu \sin2 \beta) \over |\mu|^2 - m_2^2 } .
\label{winoshift}
\eq
Again, the shift in the physical mass relative to the $W$-ino
mass parameter $m_2$ is anticorrelated with ${\rm sgn}(\mu)$.
For the parameters of Fig. \ref{sfig7n} this amounts
to a $-19$ GeV shift. 

The heavier chargino and two heaviest neutralinos
are mostly Higgsino, and form an approximately degenerate
singlet and triplet of $SU(2)_L$,
all with mass set by the $\mu$ parameter. 
This is also apparent in Table 1, where $\chi_3^0$ is
the singlet, and $(\chi_2^{+}, \chi_4^0, \chi_2^-)$ form the triplet.
    The splitting between the singlet and triplet 
is ${\cal O}(\mZ^2 / \mu^2)$ while the 
splitting among the triplets are ${\cal O}(\mZ^4 / \mu^4)$.
All these splittings may be verified by an effective operator analysis 
which shows that splittings within a triplet require
four Higgs insertions, while splitting between a singlet
and triplet requires only two.

The right handed sleptons are lighter than the other scalars 
since they only couple to the messenger sector through 
$U(1)_Y$ interactions. 
The low scale value of $m_{\lR}$ is shifted by a number
of effects. 
First, renormalization due to gaugino interactions increases 
$m_{\lR}$ in proportion to the gaugino mass.
In addition, the radiatively generated $U(1)_Y$ $D$-term,
proportional to $S \equiv {1 \over 2}{\rm Tr}(Y \tilde{m}^2)$
where ${\rm Tr}$ is over all scalars,
also contributes to renormalization of scalar masses \cite{alvarez}.
With gauge-mediated boundary conditions, $S =0$ at the 
messenger scale as the result of anomaly cancelation, 
${\rm Tr}(Y \{T^a,T^b \})=0$, where 
$T^a$ is any standard model gauge generator. 
The $\beta$-function for $S$ is homogeneous and 
very small in magnitude
(below the messenger scale and above all sparticle thresholds 
$\beta_S = (66 / 20 \pi) \alpha_1 S$ \cite{Sref})
so $S \simeq 0$ in the absence of
scalar thresholds. 
The largest contribution comes below the squark thresholds.
The ``image'' squarks make a large negative contribution to $S$ in 
this range.
Although not visible with the resolution in Fig. \ref{sfig7n},
the slope of $m_{\lR}(Q)$ has a kink at the 
squark thresholds from this effect. 
Finally, the classical $U(1)_Y$ $D$-term also increases 
the physical mass in the presence of electroweak symmetry breaking,
$\tilde{m}^2_{\lR} = {m}^2_{\lR} - \sin^2 \theta_W 
   \cos 2 \beta \mZ^2$, where $\cos 2 \beta < 0$.
For the parameters of Fig. \ref{sfig7n} 
the gauge and $U(1)_Y$ $D$-term contributions to renormalization,
and the classical $U(1)_Y$ $D$-term, contribute a 
positive shift to $\tilde{m}_{\lR}$ of 
$+2$, $+3$, and $+6$ GeV respectively.
As discussed in section \ref{collidersection}, 
the sum of all these small shifts,
and the $m_{\na}$ shift (\ref{binoshift}), can have
an important effect on signatures at hadron colliders.

The left handed sleptons receive mass from both $SU(2)_L$ and
$U(1)_Y$ interactions. 
Under renormalization $m_{\lL}$ is increased slightly
by gaugino interactions. 
The most important shift is the splitting between  $m_{\lL}$ and
$m_{\nL}$ arising from $SU(2)_L$ classical $D$-terms 
in the presence of electroweak symmetry breaking,
$m_{\lL}^2 - m_{\nL}^2 = - \mW^2 \cos 2 \beta$. 
For the parameters of Fig. \ref{sfig7n} this amounts to 
a 10 GeV splitting between $\lL$ and $\nL$.

\jfig{sfig17n}{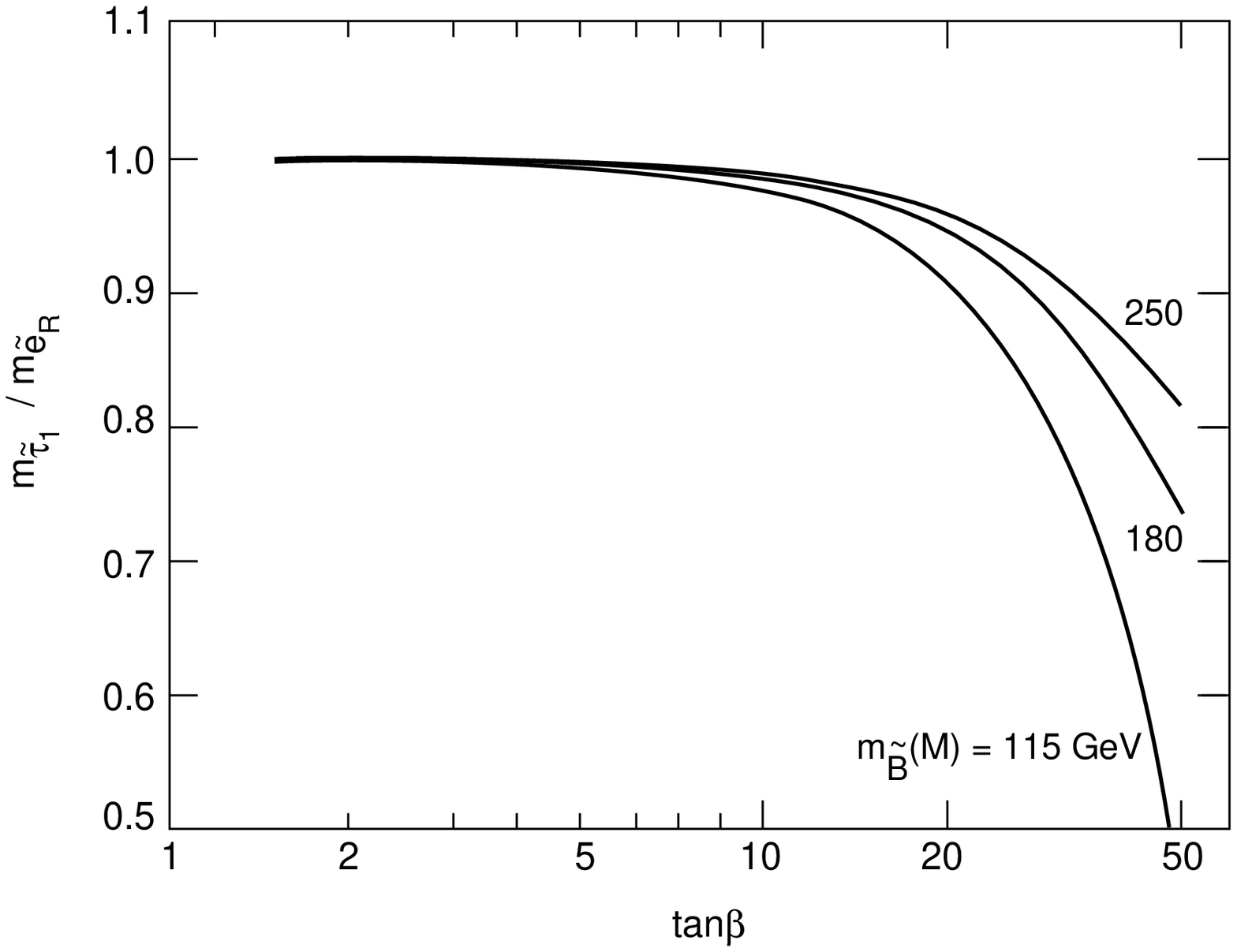}
{The ratio $\mstauR / \meR$ as a function of $\tbeta$ for 
$\mbino(M) = 115, 180,250$ GeV, and $\Lambda=M$.}
Because of the larger Yukawa coupling, the $\stau$ 
slepton masses
receive other contributions, beyond those
for $\tilde{e}$ and $\tilde{\mu}$ discussed above. 
The $\tau$ Yukawa gives a negative contribution 
to the renormalization group evolution of $m_{\stau_L}$ 
and $m_{\stau_R}$. 
In addition the left and right handed $\stau$ are mixed
in the presence of electroweak symmetry breaking
\beq
m_{\stau}^2 = \left(  \begin{array}{cc}
 m_{{\stau}_L}^2 + m_{\tau}^2 + \Delta_{{\stau}_L}  &  
     m_{\tau} ( A_{\stau} - \mu \tan \beta)  \\
     m_{\tau} ( A_{\stau} - \mu \tan \beta)  & 
  m_{\tilde{\tau}_R}^2 + m_{\tau}^2 + \Delta_{\tilde{\tau}_R} 
 \end{array}
  \right)
\label{staumatrix}
\eq
where $\Delta_{\stau_L} = (-{1 \over 2} + \sin^2 \theta_W) 
\mZ^2 \cos 2 \beta$ and
$\Delta_{\stau_R} = - \sin^2 \theta_W \cos 2 \beta$ are 
classical $D$-term contributions. 
As discussed in section \ref{minimalsection},
$A_{\stau}$ is only generated by renormalization group evolution
in proportional to $m_2$.
It is therefore small, and does not contribute significantly
to the mixing terms. 
For the parameters of Fig. \ref{sfig7n} $A_{\stau} \simeq -25$ GeV,
and remains small for all $\tan \beta$. 
\jfig{sfig10n}{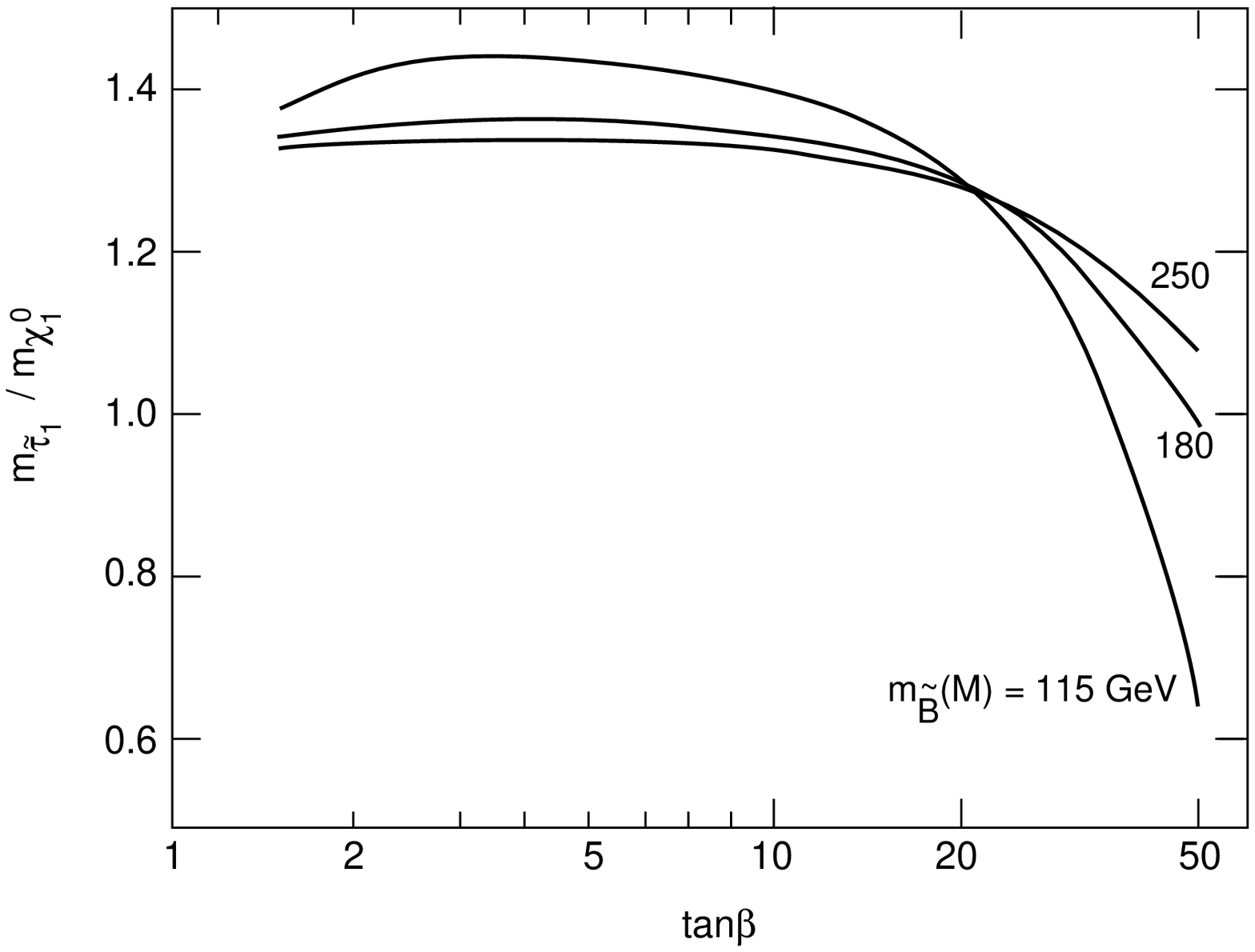}
{The ratio $m_{\tilde \tau_1}/m_{\chi^0_1}$ as a function of 
$\tan\beta$ for $m_{\tilde B}(M) = 115,180,250$ GeV.
For large $\tan\beta$ the 
$\tilde \tau_1$ becomes the lightest standard model superpartner.}
For large $\tan \beta$ the $\tau$ Yukawa coupling becomes large, 
and the mixing terms cause level repulsion which lowers
the $\stau_1$ mass below $\lR$. 
The ratio $\mstauR / \meR$ is shown in Fig. \ref{sfig17n}
as a function of $\tan \beta$ for 
$\mbino(M) = 115, 180,250$ GeV, and $\Lambda=M$.
For large $\tan \beta$ the $\stau_1$ can be significantly
lighter than $\eR$ and $\tilde{\mu}_R$. 
The negative contributions to $m_{\stau_1}$ 
in this regime comes partially from 
renormalization group evolution, but mostly from mixing. 
For example, 
with $\mbino(M)=115$ GeV, $\Lambda=M$, and $\tan \beta = 40$,
the Yukawa renormalization and mixing contributions to 
$m_{\stau_1}$ amount to $-10$ GeV and $-36$ GeV shifts 
with respect to $m_{\eR}$.
For these parameters $m_{\eR} = 137$ GeV and 
$m_{\stau_1} = 91$ GeV.
As the messenger scale is increased the relative contribution
to the mass shift due to renormalization group evolution
increases.

The negative shift of the lightest $\stau$ can even be large enough
to make $\stau_1$ the lightest standard model superpartner. 
The ratio $m_{\tilde \tau_1}/m_{\chi^0_1}$ is plotted as 
a function of $\tan \beta$ in Fig. \ref{sfig10n}.
For $\mbino(M)=115$ GeV and $\Lambda=M$ the $\stau_1$
becomes lighter than $\na$ at $\tan \beta = 38 \pm 4$ for
$m_t^{\rm pole} = 175 \pm 15$ GeV. 
The negative shift from mixing is an 
${\cal O}(m_{\tau} \mu \tan \beta / (m_{\stau_L}^2 - m_{\stau_R}^2))$
correction to the lightest eigenvalue,
and therefore becomes smaller as the overall 
scale of the soft masses is increased,
as can be seen in Fig. \ref{sfig10n}.
In addition, $\tan \beta$ is bounded from above if the 
$b$ quark Yukawa coupling remains perturbative up to a large scale. 
We find that for 
$\mbino(M) \gsim 200$ GeV the $\stau_1$ is never lighter
than $\na$ in the minimal model without $h_b$ becoming
non-perturbative somewhere below the GUT scale. 
The slight decrease in $m_{\stau_1} / m_{\na}$ 
which can be seen in Fig. \ref{sfig10n}
at small $\tan \beta$
is due partly to the increase in $\mu$
which decreases the $\na$ eigenvalue 
for ${\rm sgn}(\mu)=+1$, 
and partly to classical $U(1)_Y$ $D$-terms which increase 
$\meR$. 
Both these effects are less important for a larger overall
scale. 
The negative shift of $\stau_1$ 
relative to the other sleptons and $\na$ at large $\tan \beta$ 
can have important implications
for collider signatures, as discussed in section
\ref{collidersection}.

The $\tau$ Yukawa also gives a negative shift in the mass of the
$\nu_{\stau}$ sneutrino under renormalization group evolution. 
The magnitude of this 
shift is however smaller than for that of $\stau_R$ by 
${\cal O}(m_{\lR}^2 / m_{\lL}^2)$. 
For example, 
with $\mbino(M)=115$ GeV, $\Lambda=M$, and $\tan \beta = 40$,
the Yukawa renormalization contribution amounts to a $-2$ GeV shift
in $m_{\nu_{\stau}}$ with respect to $m_{\nu_{\tilde e}}$.

\subsubsection{Strongly Interacting States}

\label{strongsection}

The gluino and squarks receive mass predominantly
from $SU(3)_C$ interactions with the messenger sector. 
The gluino mass increases under renormalization
in proportion to $\alpha_3$ at lowest order,
$m_3 = m_3(M) ( \alpha_3(m_3) / \alpha_3(M))$.
The physical pole mass of the gluino is related to the 
renormalized $\overline{\rm DR}$ mass by finite corrections
\cite{finitegluino}.
With very heavy squarks, the general expression 
for the finite corrections given in Ref. 
\cite{finitegluino} reduces to 
\beq
m_{\tilde{g}}^{\rm pole} \simeq m_3 \left[ 1 + {\alpha_3 \over 4 \pi} 
  \left( 15 + 12 I(r) \right) \right]
\label{gluinopole}
\eq
where $m_3$ and $\alpha_3$ are the renormalized $\overline{\rm DR}$
parameters evaluated at the scale $m_3$. 
The first term in the inner parenthesis is from QCD corrections, 
and the second from squark-quark couplings, where 
the loop function is 
$I(r) = {1 \over 2} \ln r + {1 \over 2} (r-1)^2 \ln(1-r^{-1}) 
 + {1 \over 2} r -1$ for $r \geq 1$
where $r = m_{\tilde{q}}^2 / m_3^2$. 
For $r \gg 1$ $I(r) \to {1 \over 2} \ln r$, $I(2)=0$, 
and $I(1)=-{1 \over 2}$. 
The largest corrections to (\ref{gluinopole}) are
${\cal O}(( m_t^2 / m_3^2)\alpha_3 / 4 \pi)$. 
In the minimal model $r \lsim {8 \over 3}$,
which happens to be near the zero of $I(r)$.
For example, with the parameters of Fig. \ref{sfig7n}, 
$I(r)\simeq 0.035$. 
The corrections to the gluino pole mass are then dominated by QCD.
For the parameters of Fig. \ref{sfig7n} the finite corrections
amount to a $+70$ GeV contribution to the physical mass.

The squark masses receive a small increase under renormalization
in proportion to the gluino mass at lowest order. 
Since the gluino and squark masses are related within
gauge mediation by 
$m_3^2(M) \simeq {3 \over 8} m_{\tilde{q}}^2(M)$, 
this can be written as a multiplicative shift
of the squark masses
by integrating the one-loop $\beta$-function
\beq
m_{\tilde{q}} \simeq  m_{\tilde{q}}(M)
  \left[ 1 + { \Rslash^2 \over 3} 
  \left( { \alpha_3^2(m_{\tilde{q}}) \over \alpha_3^2(M)}  -1 \right)
  \right]^{1/2}
\label{squarkrenorm}
\eq
where $\Rslash$ is the gaugino masses parameter defined in section
\ref{subvariations}.
The 
${\cal O}((m_i^2 / m_3^2)
(\alpha_i^2(m_{\tilde{q}}) / \alpha_i^2(M)-1))$, $i=1,2$, 
renormalization group 
corrections to (\ref{squarkrenorm}) are quite small since
$\alpha_2$ runs very slowly, and $\alpha_1$ is small.
For the minimal model with $\Rslash=1$ and $\Lambda=M$ 
renormalization amounts to a 7\% upward shift in the squark masses. 
For a messenger scale not too far above $\Lambda$, the squark
masses are determined mainly by $\alpha_3(M)$.

The left and right handed squarks are split mainly 
by $SU(2)_L$ interactions with the messenger sector
at ${\cal O}(m_{\lL}^2 / m_{\tilde{q}}^2)$.
The much smaller splitting between up and down 
type left handed squarks is due to 
classical $SU(2)_L$ $D$-terms at ${\cal O}(\mW^2 / m_{\tilde{q}}^2)$.
Finally, the small splitting 
between up and down
right handed sleptons is from classical $U(1)_Y$ $D$-terms
at ${\cal O}(\mZ^2 / m_{\tilde{q}}^2)$
and $U(1)_Y$ interactions with the messenger sector
at ${\cal O}(m_{\lR}^2 / m_{\tilde{q}}^2)$. 
The magnitude of these small splittings can be seen in Table 1. 
The gluino is lighter than all the first and second generation 
squarks for any messenger scale in the minimal model.


The stop squarks receive additional contributions because
of the large top quark Yukawa. 
For a messenger scale not too much larger than $\Lambda$,
the positive renormalization group contribution from 
gauge interactions is largely offset by a negative 
contribution from the top Yukawa. 
In addition the left and right handed stops are
mixed in the presence of electroweak symmetry
breaking
\beq
m_{\tilde t}^2 = \left(  \begin{array}{cc}
 m_{\tilde{t}_L}^2 + m_t^2 + \Delta_{\tilde{t}_L}  &  
     m_t ( A_t - \mu \cot \beta)  \\
     m_t ( A_t - \mu \cot \beta)  & 
  m_{\tilde{t}_R}^2 + m_t^2 + \Delta_{\tilde{t}_R} 
 \end{array}
  \right)
\label{stopmatrix}
\eq
where 
$\Delta_{\tilde{t}_L}=({1 \over 2} - {2 \over 3} \sin^2 \theta_W)
\cos 2 \beta \mZ^2$ and 
$ \Delta_{\tilde{t}_R}={2 \over 3} \sin^2 \theta_W \cos 2 \beta 
\mZ^2$ 
are classical $D$-term contributions. 
The radiatively generated $A$-terms for squarks are 
somewhat larger
than for the electroweak scalars because of the larger gluino 
mass. 
\jfig{sfig12n}{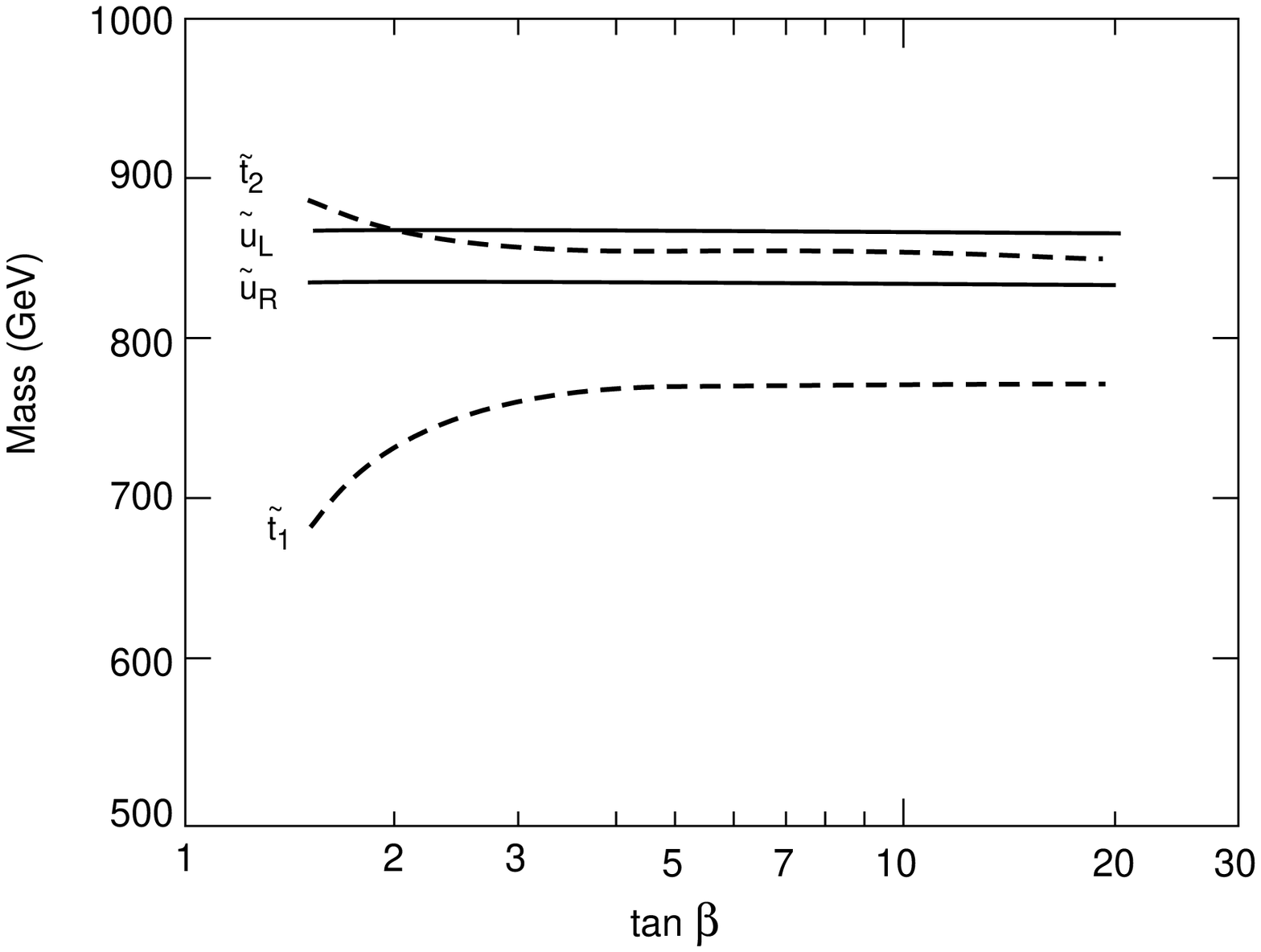}{Squark spectrum as a function of
$\tan\beta$ for $m_{\tilde B}(M)=115$ GeV and $\Lambda=M$.}
For the parameters of Fig. \ref{sfig7n} $A_{\tilde{t}} \simeq -250$ GeV,
and does not vary significantly over all $\tan \beta$.
At large $\tan \beta$ mixing induces only an 
${\cal O}(m_t A_t / m_{\tilde{t}}^2)$
correction to the lightest eigenvalue.
Because of the large squark masses, 
$A$-terms therefore do not contribute significantly to mixing.
At small $\tan \beta$ the top Yukawa and $\mu$ become large,
and $\tilde{t}_1$ can be pushed down. 
The squark spectrum is shown
in Fig. \ref{sfig12n} as a function of $\tan \beta$ for 
$\mbino(M)=115$ GeV and $\Lambda = M$. 
In contrast to the $\stau$, most of the negative shift in the 
stop masses
relative to the first two generations comes from renormalization 
group evolution (except for small $\tan \beta$).
In Fig. \ref{sfig12n}, for large  $\tan \beta$, the renormalization
and mixing contributions to $m_{\tilde{t}_1}$ are
$-50$ GeV and $-5$ GeV respectively. 
Because of the large overall squark masses, and relatively small
$\mu$ and $A$-terms, a light stop is never obtained in the MGM
parameter space.


\subsubsection{Higgs Bosons}

\label{higgssection}

The qualitative features of the Higgs boson spectrum are
determined by the pseudo-scalar mass
$\mA^2 = 2 m_{12}^2 / \sin 2 \beta$. 
The pseudo-scalar mass is shown in Fig. \ref{sfig14n} as a function 
$m_{\na}$, for
$\tan\beta =2,3,5,30$, and $\Lambda=M$.
\jfig{sfig14n}{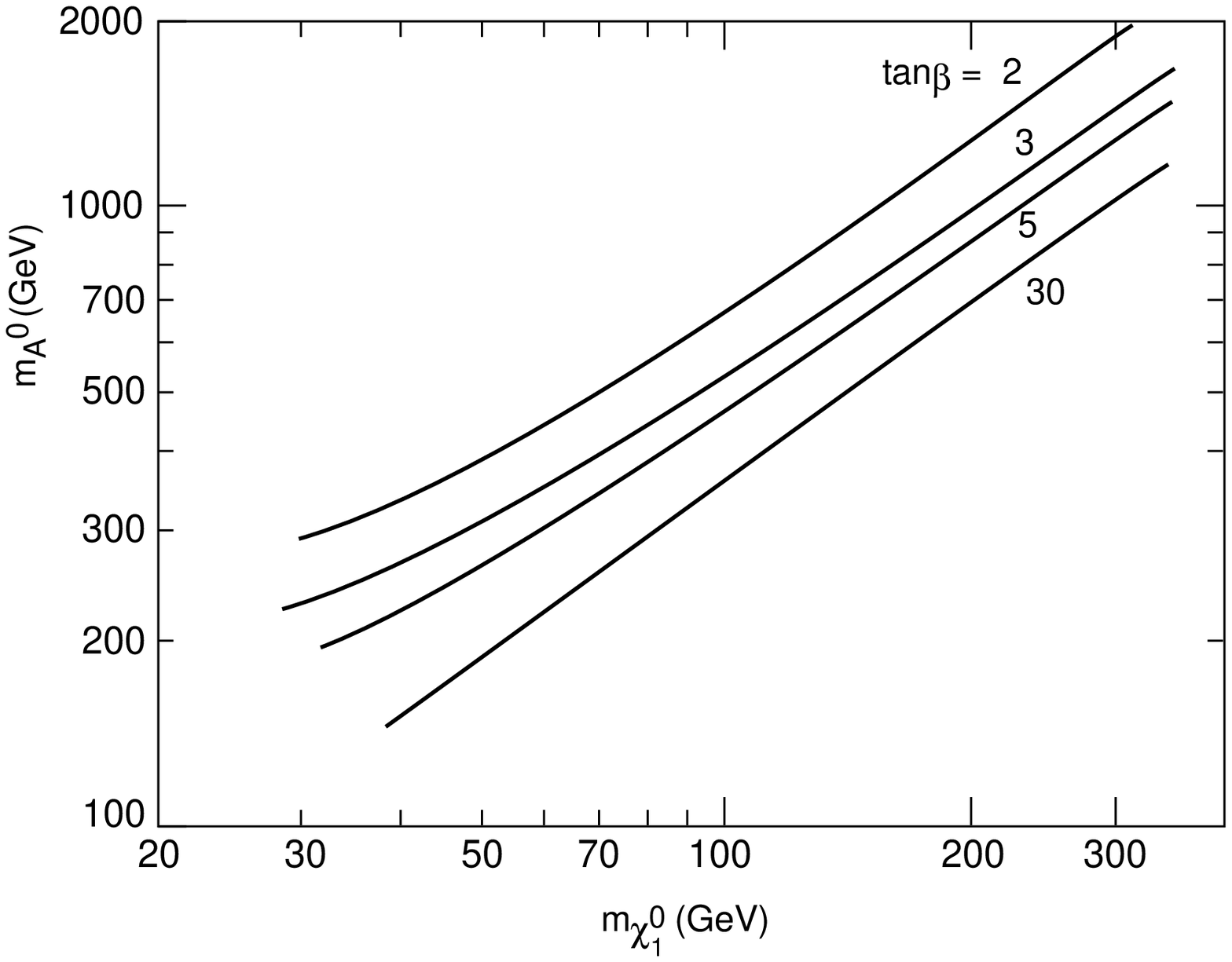}{The pseudo-scalar Higgs mass, $\mA$, as
a function of the lightest neutralino mass, $m_{\chi^0_1}$, for
$\tan\beta =2,3,5,30$, and $\Lambda=M$.}
The lightest neutralino mass is plotted in Fig. \ref{sfig14n}
as representative of the overall scale of the superpartner 
spectrum. 
Using the minimization conditions (\ref{mincona}) and (\ref{minconb})
the pseudo-scalar mass may be written, for $\tan \beta \gg 1$, 
as $\mA^2 \simeq |\mu|^2 + (\mHd^2 + \Sigma_d) - {1 \over 2} \mZ^2$.
For moderate values of $\tan \beta$ this gives the inequality
$\mA \gsim |\mu|$ over all the allowed parameter space. 
Since electroweak symmetry breaking implies $3 m_1 \lsim |\mu| 
\lsim 6 m_1$, $\mA \gg m_{\na}$ in this range. 
For small $\tan \beta $ the corrections from (\ref{mincona}) 
to this approximate relation make $\mA$ even larger. 
For $\tan \beta \gsim 35$ the negative contribution of the bottom
Yukawa to the renormalization group evolution of $\mHd^2$,
and finite corrections, allow $\mA \lsim |\mu|$. 
Also note that 
since $\mu$ is determined by the overall scale of the superpartner
masses, $\mA$ scales linearly with $m_{\na}$ for $|\mu|^2 \gg \mZ^2$. 
This scaling persists for moderate values of $\tan \beta$,
as can be seen in Fig. \ref{sfig14n}. 
The non-linear behavior at small $\mA$ is due to 
${\cal O}(\mZ^2 / \mu^2)$ contributions to the mass.

Over essentially all the allowed parameter space $\mA \gg \mZ$. 
In this case the Higgs decoupling limit is reached in which
$A^0$, $H^0$ and $H^{\pm}$ form an approximately
degenerate complex doublet of $SU(2)_L$,
with fractional splittings of ${\cal O}(\mZ^2 / \mA^2)$.
This limit is apparent in Table 1. 
Since $\mA \gsim |\mu|$ over most of the parameter space, the
heavy Higgs bosons are heavier than the Higgsinos, except
for $\tan \beta$ very large.

\jfig{sfig13n}{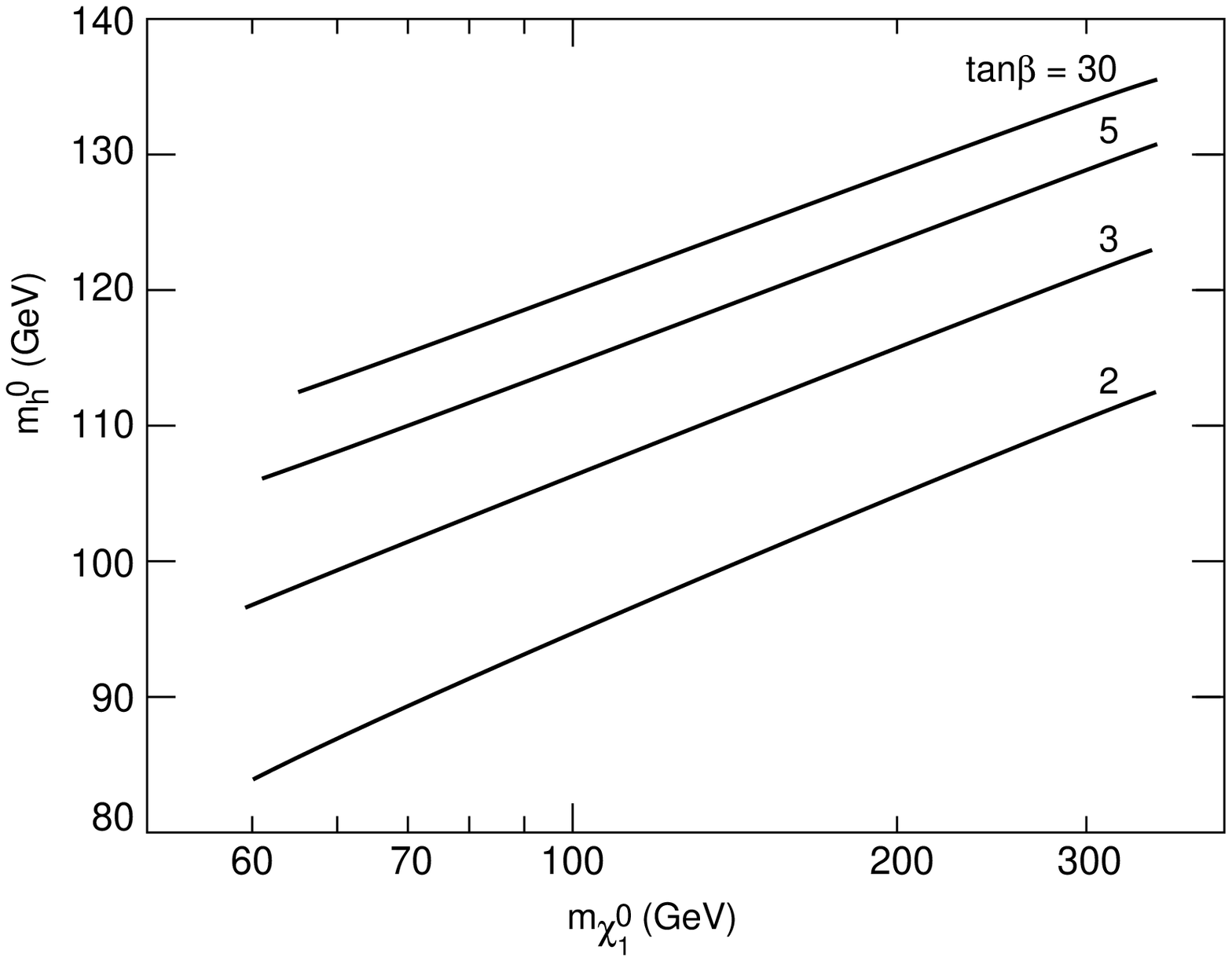}{The lightest Higgs mass, $m_{h^0}$, as a 
function of the lightest neutralino mass, $m_{\chi^0_1}$, for
$\tan\beta =2,3,5,30$, and $\Lambda=M$.}
In the decoupling limit the light Higgs, $h^0$, remains light
with couplings approaching standard model values. 
The radiative corrections to $\mh$ are sizeable~\cite{higgsrad} since
the stop squarks are so heavy with MGM boundary conditions \cite{riot}.
The physical $h^0$ mass is shown in Fig. \ref{sfig13n} 
as a function of $m_{\na}$ for $\tan \beta = 2,3,5,30$, 
and $\Lambda=M$. 
Since $m_{\tilde t_{1,2}},m_{A^0}\gg \mZ$ and stop
mixings are small, as discussed in section \ref{strongsection},
$\mh$ is well approximated by the
leading log correction to the tree level mass in the decoupling limit
\beq
m^2_{h^0} \simeq \cos^2 2 \beta \mZ^2 
 + \frac{3g^2m^4_t}{8 \pi^2m^2_W}\ln 
 \left( \frac{m_{\tilde t_1}m_{\tilde t_2}}{m^2_t} \right).
\label{higgsmass}
\eeq
For moderate values of $\tan \beta$ (\ref{higgsmass}) 
overestimates the full one-loop mass shown
in Fig. \ref{sfig13n} by $4-5\gev$.  
The $\tan \beta$ dependence of $\mh$ in Fig. \ref{sfig13n}
comes mainly from the tree level contribution. 
In general, the one-loop corrections are
largely independent of $\tan \beta$ and depend mainly on the overall
scale for the superpartners through the stop masses.  
This is
apparent from the log dependence of $m_{h^0}$ on $m_{\na}$ 
in Fig.~\ref{sfig13n},
and in the approximation (\ref{higgsmass}). 
Note that for $m_{\na} < 100$ GeV, $m_{h^0} \lsim 120$ GeV.


\subsubsection{Messenger Scale Dependence}

Much of the spectroscopy discussed above assumed a low messenger
scale 
$M \sim \Lambda \sim {\cal O}(100\tev)$. 
However, in principle $M$ can be anywhere between $\Lambda$ 
and $M_{GUT}$. 
The physical spectrum for $\mbino(M) = 115$ GeV and $\tan \beta =3$
is shown in Fig. \ref{sfig15n} as a function of the messenger scale. 
\jfig{sfig15n}{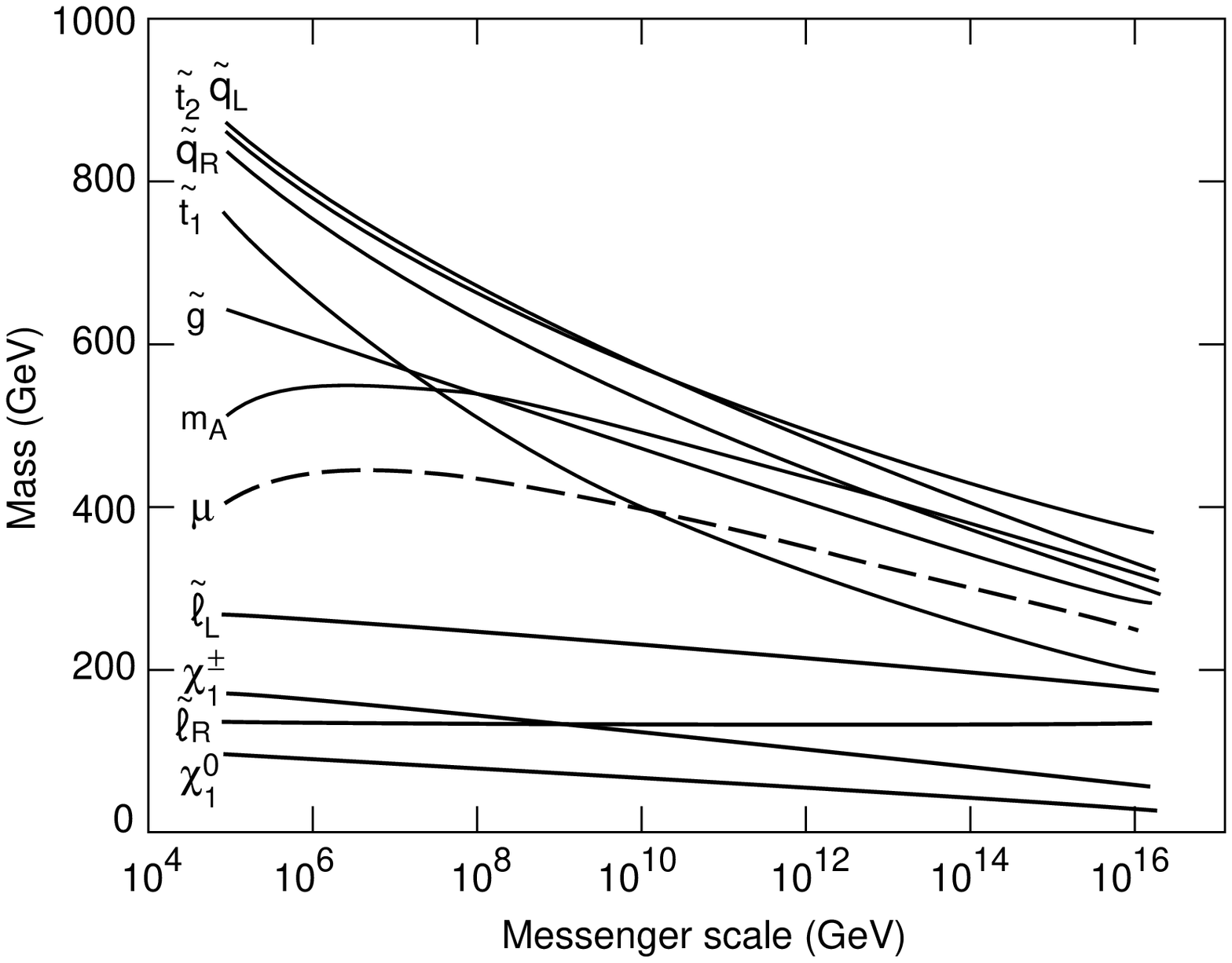}
{The physical spectrum as a function of
the messenger scale for $m_{\tilde B}(M)=115$ GeV and $\tan\beta =3$.}
The scalar masses are sensitive to the gauge couplings at 
the messenger scale. 
For a fixed $B$-ino mass at the messenger scale
(proportional to $\alpha_1(M)$),
the squarks become lighter as the messenger scale is increased
because $\alpha_3 / \alpha_1$ is smaller at higher scales. 
Conversely the right handed sleptons become heavier as $M$
is increased because
of the larger contribution from renormalization group
evolution. 
The gauge coupling $\alpha_2$ increases more slowly than 
$\alpha_1$ as the scale is increased. 
For fixed $\mbino(M)$, as
in Fig. \ref{sfig15n}, the left handed slepton masses therefore
become smaller as the messenger scale is increased. 
The sensitive dependence of the squark masses on $\alpha_3(M)$
provides a logarithmically sensitive probe of the messenger
scale, as discussed in the next section.

For larger messenger scales the spread among the superpartner
masses becomes smaller. 
This is simply because all soft masses are proportional to 
gauge couplings squared, and the gauge couplings 
converge at larger scales. 
The boundary conditions for the scalar masses with $M = M_{GUT}$
satisfy the relations
$m_{\eR}^2 : m_{\eL}^2 : m_{\tilde{Q}_L}^2 : m_{\tilde{u}_R}^2 
  : m_{\tilde{d}_R}^2 = 
 (3/5) : (9/10) : 3 : (8/5) : (21/15)$.
These do not satisfy GUT relations because only 
$SU(3)_C \times SU(2)_L \times U(1)_Y$ interactions are included. 
If the full $SU(5)$ gauge interactions are included
$m_{\bar{\bf 5}}^2 : m_{\bar{\bf 10}}^2 = 2 : 3$ 
where $\eL, \tilde{d}_R \in \overline{\bf 5}$ and 
$\eR, \tilde{Q}_L, \tilde{u}_R \in \overline{\bf 10}$.
Of course, for a messenger scale this large, gravitational 
effects are also important. 

For a messenger scale slightly above $\Lambda$,
$\mHu^2$ is driven to more negative values by the top Yukawa
under renormalization group evolution. 
Obtaining correct electroweak symmetry breaking therefore requires
larger values of $\mu$ and $m_{12}^2$. 
This can be seen in Fig. \ref{sfig15n}
as an increase in $\mu$ and $\mA$ for $M \gsim \Lambda$.
For larger messengers scales the increase in the magnitude of $\mHu^2$
from more running is eventually offset by the smaller stop masses.
This can be seen in Fig. \ref{sfig15n} as a decrease in $\mu$
and $\mA$ for $M \gsim 10^7$ GeV. 
The spectra as a function of the messenger scale for different
values of $\tan \beta$ are essentially identical to Fig. \ref{sfig15n}
aside from $\mu$, $\mA$, and $m_{\tilde{t}}$. 
This is because $\tan \beta$ only affects directly the Higgs sector
parameters, which in turn  
influence the mass of the other states 
only through two-loop corrections
(except for the third generation scalars discussed in 
sections \ref{electroweaksection} and \ref{strongsection}).

\subsubsection{Relations Among the Superpartner Masses}

\label{relations}

The minimal model of gauge-mediated supersymmetry breaking
represents a very constrained theory of the soft terms.
In this section we present some quantitative 
relations among the superpartner masses. 
These can be used to distinguish the MGM from other theories
of the soft terms and within the MGM can be logarithmically
sensitive to the messenger scale. 

The gaugino masses at the messenger scale are in proportion
to the gauge couplings squared, 
$m_1 : m_2 : m_3 = \alpha_1 : \alpha_2 : \alpha_3$.
Since $\alpha_i m_{\lambda_i}^{-1}$ is a renormalization group
invariant at one loop, this relation is preserved to
lowest order at the electroweak scale, 
where 
$m_{\lambda_i}$ are the $\overline{DR}$ masses. 
The MGM therefore yields, in leading log approximation, the 
same ratios of gaugino masses as high scale supersymmetry breaking
with universal gaugino boundary conditions.
``Gaugino unification'' is a generic feature of any 
weakly coupled gauge-mediated messenger sector 
which forms a representation of any GUT group and 
which has a single spurion.  
The gaugino mass ratios are independent of $\Rslash$. 
However, as discussed in section \ref{subvariations}, 
with multiple sources of supersymmetry breaking and/or messenger
fermion masses the gaugino masses can be sensitive to 
messenger Yukawa couplings. 
``Gaugino unification'' therefore does {\it not} follow 
just from the anzatz
of gauge-mediation, even for messenger sectors which can 
be embedded in a GUT theory. 
An example of such a messenger sector is given in appendix 
\ref{appnonmin}.
Of course, a messenger sector which forms an incomplete GUT
multiplet (and modifies gauge coupling unification 
unification at one-loop under renormalization
group evolution) does not in general yield 
``gaugino unification'' \cite{alon}.



With gauge-mediated supersymmetry breaking
the scalar and gaugino masses are related at the messenger scale.
For a messenger sector well below the GUT scale, 
$\alpha_3 \gg \alpha_2 > \alpha_1$, so the most important 
scalar-gaugino correlations
are between squarks and gluino, left handed sleptons and 
$W$-ino, and right handed sleptons and $B$-ino. 
\jfig{sfig4x}{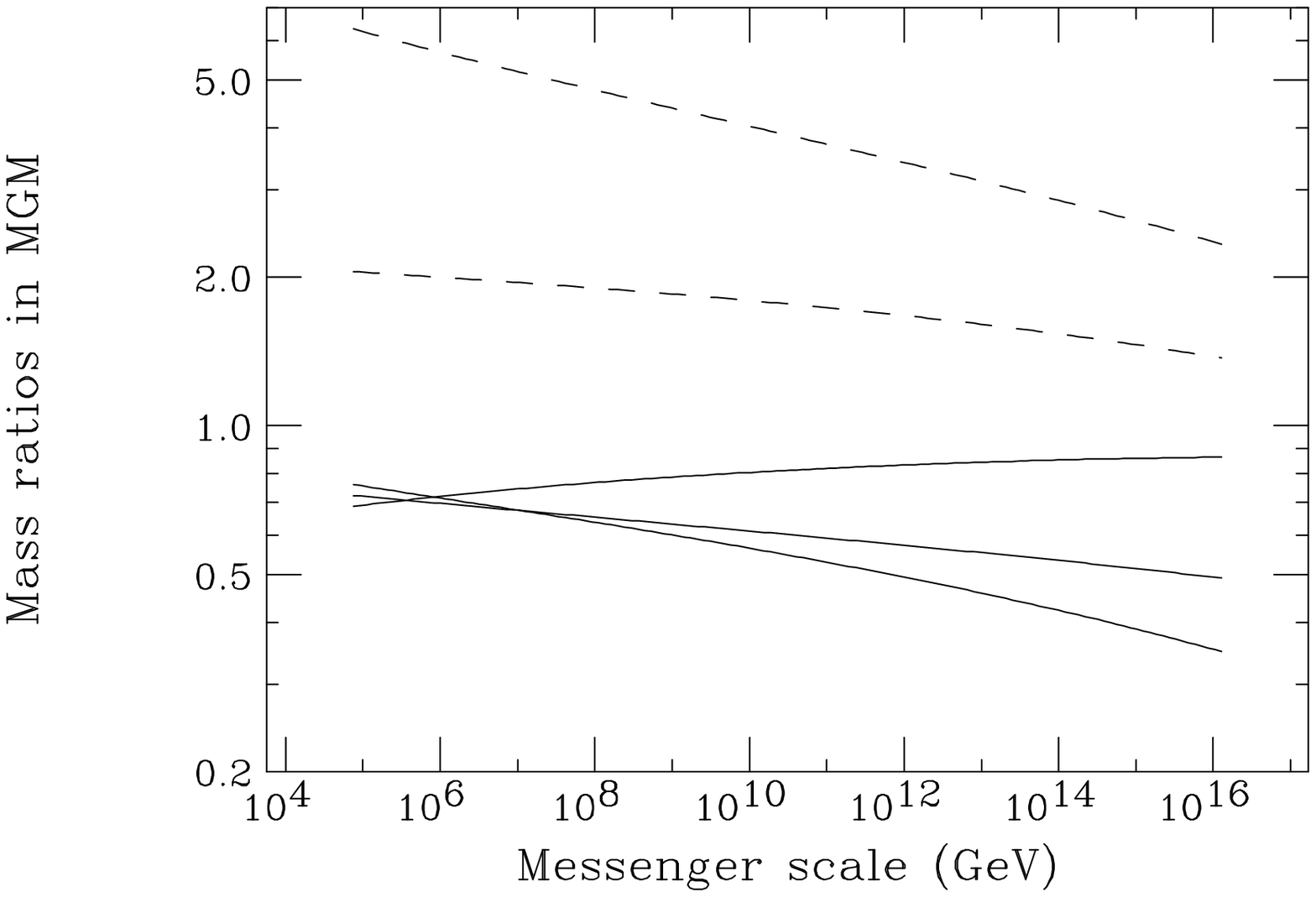}
{Ratios of $\overline{\rm DR}$ mass parameters
with MGM boundary conditions as a function
of the messenger scale: 
$m_{\tilde{q}_R} / m_{\lR}$ (upper dashed line), 
$m_{\lL} / m_{\lR}$ (lower dashed line), 
$m_3 / m_{\tilde{q}_R}$ (upper solid line),
$m_2 / m_{\lL}$ (middle solid line), and 
$m_1 / m_{\lR}$ (lower solid line).}
The ratios are of course proportional to $\Rslash$ which 
determines the overall scale of the gaugino masses at
the messenger scale,
and are modified by renormalization group evolution 
to the low scale. 
Ratios of the $\overline{\rm DR}$ masses 
$m_3 / m_{\tilde{q}_R}$,
$m_2 / m_{\lL}$, and 
$m_1 / m_{\lR}$ in the minimal model
are shown in Fig. \ref{sfig4x} as a function
of the messenger scale. 
As discussed in the next section, these ratios can be altered 
with non-minimal messenger sectors.

In the minimal model, with $\Rslash=1$, a measurement 
of any ratio $m_{\lambda_i}/ m $ gives a logarithmically
sensitive measurement of the messenger scale. 
Because of the larger magnitude of the $U(1)_Y$
gauge $\beta$-function the ratio
$m_{\lR} / m_1$ is most sensitive to the messenger scale. 
Notice also that $m_3 / m_{\tilde{q}}$ is larger 
for a larger messenger scale, while 
$m_{\lL} / m_2$ and 
$m_{\lR} / m_1$ decrease with the messenger scale. 
Because of this disparate sensitivity,
within the anzatz of minimal gauge-mediation,  
both $\Rslash$ and $\ln M$ could be extracted from a precision 
measurement of all three ratios.

For $\Rslash \leq 1$ the ratio of scalar mass 
to associated gaugino mass is always $ \geq 1$ 
for any messenger scale. 
Observation of a first or second generation scalar
lighter than the associated gaugino is therefore a 
signal for $\Rslash >1$. 
As discussed in section \ref{multiple},
$\Rslash >1$ is actually possible with larger messenger
sector representations. 
In fact, as discussed in appendix \ref{appgeneral} in models
with a single spurion in the messenger sector, 
$\Rslash$ is senstive to the index of the messenger sector
matter. 
Additional matter which transforms under the standard model
gauge group between the electroweak and messenger scales
would of course modify these relations slightly through
renormalization group evolution contributions.

Ratios of 
scalar masses at the messenger scale are related by ratios
of gauge couplings squared. 
These ratios are reflected in the low energy spectrum. 
In particular, since $\alpha_3 \gg \alpha_1$ if the messenger
scale is well below the GUT scale, the ratio
$m_{\tilde{q}} / m_{\lR}$ is sizeable. 
Ratios of the $\overline{\rm DR}$ masses 
$m_{\tilde{q}_R} / m_{\lR}$ and 
$m_{\lL} / m_{\lR}$ are shown in Fig. \ref{sfig4x} as a function
of the messenger scale. 
For $\Lambda = M$, $m_{\tilde{q_R}} / m_{\lR} \simeq 6.3$.
Notice that $m_{\tilde{q}_R} / m_{\lR}$ is smaller for larger
messenger scales, 
and is fairly sensitive to $\ln M$.
This is because 
$\alpha_3$ decreases rapidly at larger scales,
while $\alpha_1$ increases. 
This sensitivity allows an indirect measure of $\ln M$. 
The ratio $m_{\lL} / m_{\lR}$ is also fairly sizeable
but not as sensitive to the messenger scale. 
For $\Lambda = M$, $m_{\lL} / m_{\lR} \simeq 2.1$. 
It is important to note that with 
$SU(3)_C \times SU(2)_L \times U(1)_Y$
gauge-mediated supersymmetry
breaking, any parity and charge conjugate invariant
messenger sector 
which forms a representation of any GUT group and 
which has a single spurion
yields, at leading order, the same 
scalar mass ratios as in the minimal model.
These mass ratios therefore represent a fairly generic
feature of minimal gauge-mediation.

The sizeable hierarchy which arises in gauge-mediated supersymmetry 
breaking between 
scalar masses of particles with different
gauge charges  generally does not arise with universal
boundary conditions with a large overall scalar mass. 
With gravity mediated supersymmetry breaking and universal
boundary conditions the largest hierarchy results
for the no-scale boundary condition $m_0=0$. 
In this case the scalar masses are ``gaugino-dominated,''
being generated in proportion to the gaugino masses under
renormalization group evolution. 
The scalar mass 
ratios turn out to be just slightly smaller 
than the maximum gauge-mediated ratios.  
With no-scale boundary conditions at $M_{GUT}$,
$m_{\tilde{q}_R} / m_{\eR} \simeq 5.6$ and
$m_{\lL} / m_{\lR} \simeq 1.9$.  
However, the scalars in this case 
are just slightly lighter than the 
associated gauginos, in contrast to the MGM with $\Rslash=1$,
in which they are heavier. 
It is interesting to note, 
however, that for $M \sim 1000$ TeV
and $N=2$ or 
$\Rslash \simeq \sqrt{2}$, gauge mediation coincidentally
 gives almost identical
mass ratios as high scale supersymmetry breaking with 
the no-scale boundary condition at the GUT scale.

With gauge-mediation,
scalar masses at the messenger scale receive contributions
proportional to gauge couplings squared. 
Splitting among squarks with different gauge charges 
can therefore be related to 
right and left handed slepton masses (cf. Eq. \ref{scalarmass}). 
This can be quantified in the form of sum rules which involve
various linear combinations of 
all the first generation scalar masses squared \cite{martin}.
The splitting due to $U(1)_Y$ interactions with the messenger
sector can be quantified by 
${\rm Tr}(Ym^2)$, where ${\rm Tr}$ is over 
first generation sleptons and squarks. 
As discussed in section \ref{electroweaksection}
this quantity vanishes with gauge-mediated boundary conditions
as the result of anomaly cancelation. 
It is therefore interesting to consider the low scale quantity 
$$
M_Y^2 =
{1 \over 2} \left( m_{\tilde{u}_L}^2 +  m_{\tilde{d}_L}^2 \right)
  -2 m_{\tilde{u}_R}^2 +  m_{\tilde{d}_L}^2 
 - {1 \over 2} \left(  m_{\tilde{e}_L}^2 +  m_{\tilde{\nu}_L}^2 
     \right) 
 +  m_{\tilde{e}_R}^2  
$$
\beq
 + {10 \over 3} \sin^2 \theta_W \cos 2 \beta \mZ^2
\label{Ysumrule}
\eq
where the the sum of the $m^2$ terms is 
${1 \over 2}{\rm Tr}(Ym^2)$ over the first generation, and  
the ${\cal O}(\mZ^2)$ term is a correction for 
classical $U(1)_Y$ $D$-terms. 
The contribution of the gaugino masses to $M_Y^2$ under 
renormalization group evolution cancels at one-loop. 
So this quantity is independent of the gaugino spectrum. 
In addition, the $\beta$-function for ${\rm Tr}(Ym^2)$ is homogeneous 
\cite{alvarez}
and independent of the Yukawa couplings at one-loop, even
though the individual masses are affected. 
So if 
$M_Y^2=0$ at the messenger scale,
it is not generated above scalar thresholds. 
It only receives very small contributions below the squark 
thresholds of ${\cal O}((\alpha_1 / 4 \pi) m_{\tilde{q}}^2
 \ln( m_{\tilde{q}} / m_{\tilde{l}} )) $.
The relation $M_Y^2 \simeq 0$ tests the assumption that
splittings within the squark and slepton spectrum
are related to $U(1)_Y$ quantum numbers. 
The quantity $M_Y^2$ also vanishes in any model in which 
soft scalar masses are univeral within GUT multiplets. 
This is because ${\rm Tr} Y=0$ over any GUT multiplet. 
Within the anzatz of gauge-mediation, a violation of 
$M_Y^2 \simeq 0$ can result from a number of sources. 
First, the messengers might not transform under 
$U(1)_Y$. 
In this case the $B$-ino should also be very light. 
Second, a large $U(1)_Y$ $D$-term can be generated 
radiatively if the messenger sector is not parity and 
charge conjugate invariant. 
Finally, the squarks and/or sleptons might
transform under additional gauge interactions 
which couple with the messenger sector
so that ${\rm Tr}(Ym^2)$ does not vanish over
any generation. 
This implies the existence of additional
electroweak scale matter in order to cancel the 
${\rm Tr}(Y \{T^a, T^b \})$ anomaly, where $T^a$ is a 
generator of the extra gauge interactions. 


Unfortunately, sum rules which involve near cancelation among squark 
and slepton 
masses squared, such as $M_Y^2 =0$, if in fact satisfied, are 
often not particularly useful experimentally. 
This is because the squark masses are split only at 
${\cal O}(m_{\tilde{l}}^2 / m_{\tilde{q}}^2)$
by $SU(2)_L$ and $U(1)_Y$ interactions with the messenger 
sector, and at ${\cal O}(\mZ^2 / m_{\tilde{q}}^2)$ from 
classical $SU(2)_L$ and $U(1)_Y$ $D$-terms. 
Testing such sum rules therefore requires, 
in general, measurements of
squark masses at the sub-GeV level, as can be determined
from the masses given in Table 1.
It is more useful to 
consider sum rules, such as the ones given below, 
which isolate the dominant splitting arising from $SU(2)_L$ interactions,
and are only violated by $U(1)_Y$ interactions. 
These violations 
are typically smaller than the experimental resolution. 
The sum rules may then be tested with somewhat less precise 
determinations of squark masses.


The near degeneracy among squarks may be quantified by
the splitting between right handed squarks 
\beq
\Delta_{\tilde{q}_R}^2 =  m_{\tilde{u}_R}^2 -  m_{\tilde{d}_R}^2 .
\label{sumruleright}
\eq
Ignoring $U(1)_Y$ interactions, 
this quantity is a renormalization group invariant. 
It receives non-zero contributions at  
${\cal O}(m_{\eR}^2/ m_{\tilde{q}}^2 )$ 
from $U(1)_Y$ interactions with the messenger sector and  
renormalization group contributions from the $B$-ino mass,
and
${\cal O}(\mZ^2/ m_{\tilde{q}}^2 )$
from classical $U(1)_Y$ $D$-terms at the low scale.
Numerically 
$\Delta_{\tilde{q}_R}^2 / ( m_{\tilde{u}_R}^2+m_{\tilde{d}_R}^2)
\simeq 0$
to better than 0.3\% with MGM boundary conditions. 
The near degeneracy between right handed squarks 
is a necessary condition if squarks receive
mass mainly from $SU(3)_C$ interactions.
The quantity $\Delta_{\tilde{q}_R}^2$ 
also vanishes to the same order with universal boundary
conditions, but need not even approximately 
vanish in theories in which 
the soft masses are only universal within GUT multiplets.  

An experimentally  more interesting measure 
which quantifies the splitting between left and 
right handed squarks is
\beq
M_{L-R}^2 = 
 m_{\tilde{u}_L}^2 +  m_{\tilde{d}_L}^2 -  
  \left(  m_{\tilde{u}_R}^2 +  m_{\tilde{d}_R}^2 \right)
- \left( m_{\lL}^2 + m_{\tilde{\nu}_L}^2 
  \right)
\eq
This quantity is also a renormalization group invariant ignoring
$U(1)_Y$ interactions. 
It formally vanishes at the same order as (\ref{sumruleright}).
Numerically 
$M_{L-R}^2 / ( m_{\tilde{u}_R}^2 +  m_{\tilde{d}_R}^2)
\simeq 0$ 
to better than 1\% with 
MGM boundary conditions. 
Without the left handed slepton contribution, 
$M_{L-R}^2 / ( m_{\tilde{u}_R}^2 +  m_{\tilde{d}_R}^2)
\simeq 0$ 
can be violated by up to 10\%.
This relation tests the assumption that the splitting between
the left and right handed squarks is due mainly to 
$SU(2)_L$ interactions within the messenger sector.
The splitting is therefore correlated with the left handed slepton
masses, which receive masses predominantly from the same source. 

If the squarks and sleptons receive mass predominantly from 
gauge interactions with the messenger sector, 
the masses depend only on gauge quantum numbers, and 
are independent of generation up to very small 
${\cal O}(m_f^2 / M^2)$ corrections at the messenger scale, where 
$m_f$ is the partner fermion mass. 
However, third generation masses are modified by Yukawa
contributions under renormalization group evolution and 
mixing. 
Mixing effects can be eliminated by considering the quantity
${\rm Tr}(m_{LR}^2)$ where $m_{LR}^2$ is the left-right 
scalar mass squared 
matrix. 
In addition, it is possible to choose linear combinations
of masses which are independent of Yukawa couplings
under renormalization group evolution at 
one-loop,  
$m_{\tilde{u}_L}^2 + m_{\tilde{u}_R}^2 - 3 m_{\tilde{d}_L}^2$,
and similarly for sleptons
\cite{ssconstraints}.
The quantities
\beq
M^2_{\tilde{t} - \tilde{q}} = 
m_{\tilde{t}_1}^2 + m_{\tilde{t}_2}^2 - 2 m_t^2 
 - 3 m_{\tilde{b}_2}^2
 - \left( m_{\tilde{u}_L}^2 + m_{\tilde{u}_R}^2 
  - 3 m_{\tilde{d}_L}^2  \right)
\label{thirdsquark}
\eq
\beq
M^2_{\tilde{\tau} - \tilde{e}} = 
m_{\tilde{\tau}_1}^2 + m_{\tilde{\tau}_2}^2 
 - 3 m_{\tilde{\nu}_{\tau}}^2
 - \left( m_{\tilde{e}_L}^2 + m_{\tilde{e}_R}^2 
  - 3 m_{\tilde{\nu}_e}^2  \right)
\label{thirdslepton}
\eq
only receive contributions at two-loops under renormalization,
and in the case 
of $M^2_{\tilde{t} - \tilde{q}}$ from $\tilde{b}$ mixing effects
which are negligible unless $\tan \beta$ is very large. 
The relations $M^2_{\tilde{t} - \tilde{q}} \simeq 0$
and $M^2_{\tilde{\tau} - \tilde{e}} \simeq 0$ 
test the assumption that scalars with different
gauge quantum numbers have a flavor independent mass 
at the messenger scale. 
They vanish in any theory of the 
soft terms with flavor independent masses at the messenger scale, 
but need not vanish in theories in which alignment of the 
squark mass matrices with the quark masses
is responsible for the lack of supersymmetric
contributions to flavor changing neutral currents. 
Within the anzatz of gauge-mediation, violations of these relations
would imply additional flavor dependent interactions with the
messenger sector. 


If the quantities (\ref{thirdsquark}) and (\ref{thirdslepton})
are satisfied, implying the masses are generation independent
at the messenger scale, it is possible to extract the
Yukawa contribution to the renormalization group evolution. 
The quantities 
\beq
M_{h_t}^2 = m_{\tilde{t}_1}^2 +  m_{\tilde{t}_2}^2 - 2 m_t^2 
 - \left(  m_{\tilde{u}_L}^2 +  m_{\tilde{u}_R}^2 \right)
\eq
\beq
M_{h_{\tau}}^2 = m_{\tilde{\tau}_1}^2 +  m_{\tilde{\tau}_2}^2 
 - \left(  m_{\tilde{e}_L}^2 +  m_{\tilde{e}_R}^2 \right)
\eq
are independent of third generation mixing effects. 
Under renormalization group evolution 
$M_{h_t}^2$ receives an ${\cal O}((h_t / 4 \pi)^2 m_{\tilde{t}}^2
  \ln(M/m_{\tilde{t}}) )$ 
negative contribution from the top Yukawa. 
For moderate values of $\tan \beta$ this amounts to a 
14\% deviation from 
$M_{h_t}^2 / (m_{\tilde{t}_1}^2 + m_{\tilde{t}_1}^2)=0$ 
for $M = \Lambda$ and 
grows to 29\% for $M= 10^5 \Lambda$. 
Given an independent measure of $\tan \beta$ to fix the value
of $h_t$, this quantity gives an indirect probe of $\ln M$. 
Unfortunately it requires a fairly precise measurement of the
squark and stop masses, but is complimentary to the 
$\ln M$ dependence of the mass ratios of scalars with different 
gauge charges discussed above.  
The quantity $M_{h_{\tau}}^2$ is only significant if $\tan \beta$
is very large. 
If this is the case, the splitting between $\tilde{\nu}_{\tau}$
and $\tilde{\nu}_e$, $\Delta^2_{\tilde{\nu}_{\tau} - \tilde{\nu}_e} = 
  m^2_{\tilde{\nu}_{\tau}} -  m^2_{\tilde{\nu}_{e}}$, 
gives an independent check of the renormalization contribution 
through the relation 
\beq
M_{h_{\tau}}^2 = 3 \Delta^2_{\tilde{\nu}_{\tau} - \tilde{\nu}_e}
\eq


\section{Variations of the Minimal Model}

\label{varcon}

The results of the renormalization group analysis given above 
are for the minimal model of gauge-mediated 
supersymmetry breaking. 
In this section we discuss how variations of the minimal
model affect the 
form of the superpartner spectrum and the constraints
imposed by electroweak symmetry breaking.

\subsection{Approximate $U(1)_R$ Symmetry}

\label{appUR}

Soft scalar masses require supersymmetry breaking, while
gaugino masses require both supersymmetry and $U(1)_R$
breaking, as discussed in section \ref{subvariations}.
It is therefore possible that the scale for gaugino
masses is somewhat different than that for the scalar masses,
as quantified by the parameter $\Rslash$. 
An example of a messenger sector with an approximate $U(1)_R$
symmetry is given in appendix \ref{appnonmin}.
The gaugino masses affect the scalar masses only through
renormalization group evolution. 
For $\Rslash < 1$ the small positive contribution to scalar masses  
from gaugino masses is slightly reduced. 
The scalar mass relations discussed in section 
\ref{relations} are not affected by this renormalization and
so are not altered. 
The main effect of $\Rslash < 1$ is simply to lower the overall
scale for the gauginos relative to the scalars. 
This also does not affect the relation among gaugino masses.

\subsection{Multiple Messenger Generations}

\label{multiple}

The minimal model contains a single messenger generation of 
${\bf 5} + \overline{\bf 5}$ of $SU(5)$. 
This can be extended to any vector 
representation of the standard model gauge group.
Such generalizations may be parameterized by the equivalent number of
${\bf 5} + \overline{\bf 5}$ messenger generations,
$N = C_3$, where $C_3$ is 
defined in Appendix \ref{appgeneral}.
For a ${\bf 10} + \overline{\bf 10}$ of $SU(5)$ $N=3$.
{}From the general expressions given in appendix \ref{appgeneral}
for gaugino and scalar masses, it is apparent that
gaugino masses grow like $N$ while scalar masses grow 
like $\sqrt{N}$ \cite{signatures}.
This corresponds roughly to the gaugino mass parameter
$\Rslash = \sqrt{N}$. 
Messenger sectors with larger matter representations 
therefore result
in gauginos which are heavier relative to the scalars than 
in the minimal model. 

\jfig{sfig4n}{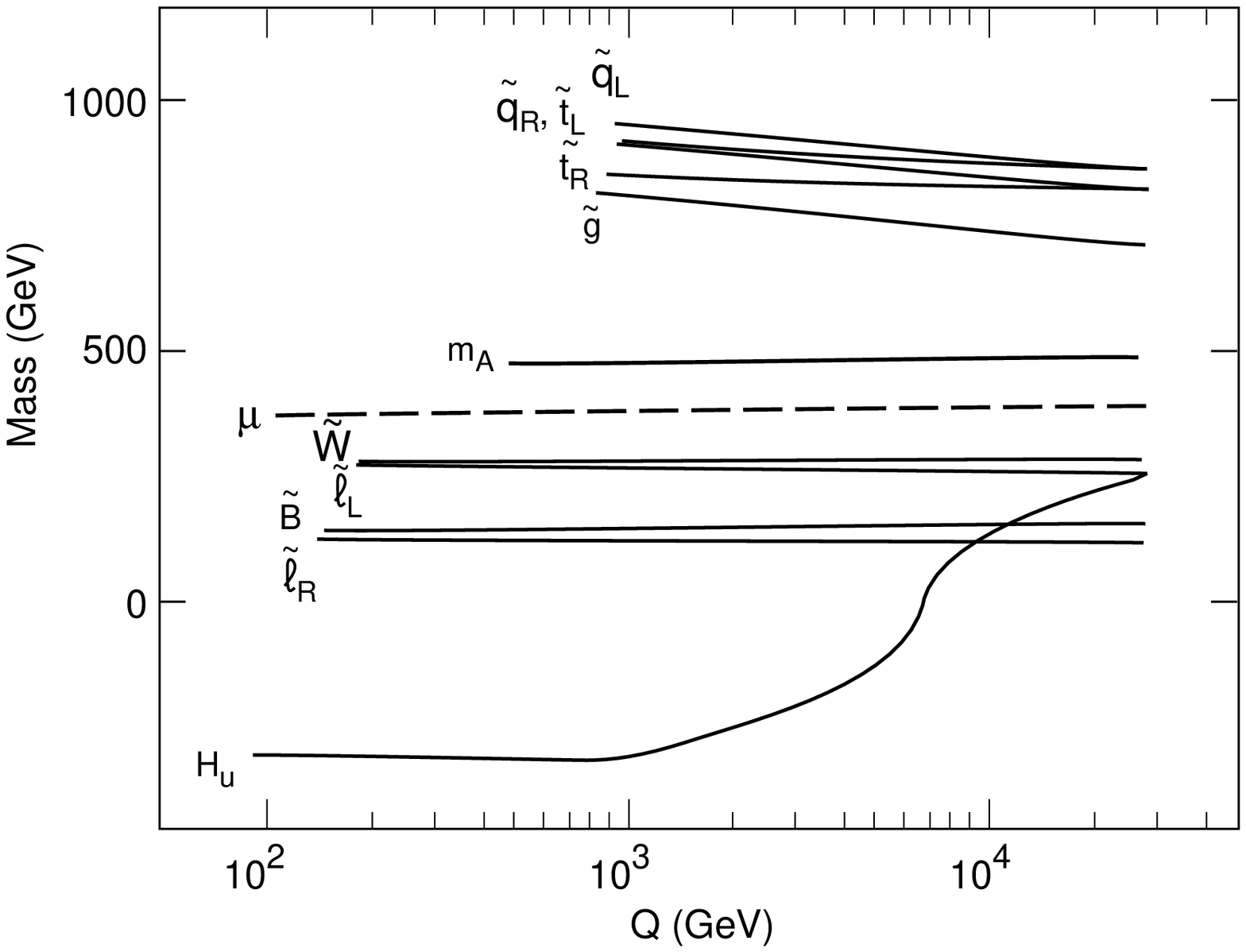}
{Renormalization group evolution of the 
${\overline{\rm DR}}$ mass parameters with boundary conditions
of the two generation messenger sector. 
The messenger scale is  $M = 54$ TeV, $\Lambda=M$, 
$\mbino(M) = 163$ GeV, and $\tan \beta = 3$.} 
The renormalization group evolution of the $\overline{\rm DR}$ 
parameters for $N=2$ with a messenger scale of 
$M=54$ TeV, $\Lambda = M$,
$\mbino(M) = 163$ GeV, and $\tan \beta =3$ is 
shown in Fig. \ref{sfig4n}.   
The renormalization group contribution to the scalar masses
proportional to the gaugino masses is slightly larger than 
for $N=1$.
Notice that at the low scale the renormalized 
right handed slepton masses are slightly smaller than the 
$B$-ino mass.
The physical slepton masses, however, receive a positive
contribution from the classical $U(1)_Y$ $D$-term, while
the physical $\na$ mass receives a negative contribution from 
mixing with the Higgsinos
for ${\rm sgn}(\mu)=+1$. 
With $N=2$ and the messenger scale not too far above $\Lambda$, 
the $\lR$ and $\na$ are therefore very close in mass. 
For the parameters of Fig. \ref{sfig4n}
$m_{\na} = 138$ GeV and $m_{\lR} = 140$ GeV,
so that $\na$ remains the lightest standard model superpartner.  
The $D$-term and Higgsino mixing contributions become smaller
for a larger overall scale. 
For $M=60$ TeV, $\Lambda=M$, and $\tan \beta =3$, the 
$\na$ and $\lR$ masses cross 
at $m_{\na} = m_{\lR} \simeq 153$ GeV.
Since the $B$-ino mass decreases while the right handed slepton
masses increase under renormalization, 
$\meR > m_{\na}$ for a messenger scale well above $\Lambda$.
The near degeneracy of $\lR$ and $\na$ is just a coincidence
of the numerical boundary conditions for $N=2$
and ${\rm sgn}(\mu)=+1$. 
For messenger sectors with $N \geq 3$ a right handed slepton 
is naturally the lightest standard model superpartner.

The heavier gauginos which result for $N \geq 2$ only 
slightly modify electroweak symmetry breaking through
a larger positive renormalization group 
contribution to the Higgs soft masses, 
and finite corrections at the low scale. 
The negative contribution to the $\stau_1$ mass 
relative to $\eR$ and $\tilde{\mu}_R$ from 
mixing and Yukawa contributions to renormalization 
are therefore also only slightly modified. 
For a given physical scalar mass at the low scale, the
ratio $m_{\stau_1} / m_{\eR}$ is very similar to the $N=1$ 
case. 
For $N \geq 3$, and the regions of the $N=2$ parameter space
in which $m_{\lR} < m_{\na}$, the $\stau_1$
is the lightest standard model superpartner. 
As discussed in section \ref{collidersection}, collider signatures for 
these cases are much different than for the MGM with $N=1$
with $\na$ as the lightest standard model superpartner.

\subsection{Additional Soft Terms in the Higgs Sector}

\label{addhiggs}

The Higgs sector parameters $\mu$ and $m_{12}^2$ require
additional interactions
with the messenger sector beyond the standard model gauge 
interactions. 
In the minimal model the precise form of these interactions is
not specified, and $\mu$ and $m_{12}^2$ are taken 
as free parameters.
The additional interactions which couple to the Higgs sector
are likely to contribute to the Higgs soft masses 
$\mHu^2$ and $\mHd^2$, and split these from 
the left handed sleptons, $m_{\lL}^2$.
Splittings of ${\cal O}(1)$ are not unreasonable since
the additional interactions must generate $\mu$ and $m_{12}^2$
of the same order. 
The Higgs splitting may be parameterized by $\Delta_{\pm}^2$,
defined in Eq. (\ref{splithiggs}) of section \ref{subvariations}.

It is possible that other scalars also receive additional 
contributions to soft masses.
The right handed sleptons 
receive a gauge-mediated mass only from 
$U(1)_Y$ coupling, and are therefore most 
susceptible to a shift in mass from additional 
interactions.
Right handed sleptons represent the potentially
most sensitive probe for such interactions \cite{tevatron}.
Note that additional messenger sector interactions do not modify
at lowest order 
the relations among gaugino masses. 
Since additional interactions {\it must} arise in the Higgs sector,
we focus in this section on 
the effect of
additional contributions to the Higgs soft masses
on electroweak
symmetry breaking and the superpartner spectrum.
We also consider the possibility that $m_{12}^2$ is generated
entirely from renormalization group evolution \cite{frank}.

Additional contributions to Higgs sector masses can in principle
have large effects on electroweak symmetry breaking. 
With the Higgs bosons split from the left handed sleptons 
the minimization condition (\ref{mincona}) is modified 
to 
\beq
\label{newEWSB}
|\mu|^2+\frac{m_Z^2}{2} =  
 \frac{(m^2_{H_d,0}+\Sigma_d)+
    (m^2_{H_u,0}+\Sigma_u)\tan^2\beta}{\tan^2\beta -1} 
 -\frac{\Delta^2_+}{2}+\frac{\Delta^2_-}{2}\left( 
     \frac{\tan\beta^2 +1}{\tan\beta^2-1}\right)
\label{minsplit}
\eq
where all quantities are evaluated at the minimization scale, 
and 
$m^2_{H_u,0}$ and $m^2_{H_d,0}$ are the gauge-mediated contributions
to the soft masses. 
\jfig{sfig5n}{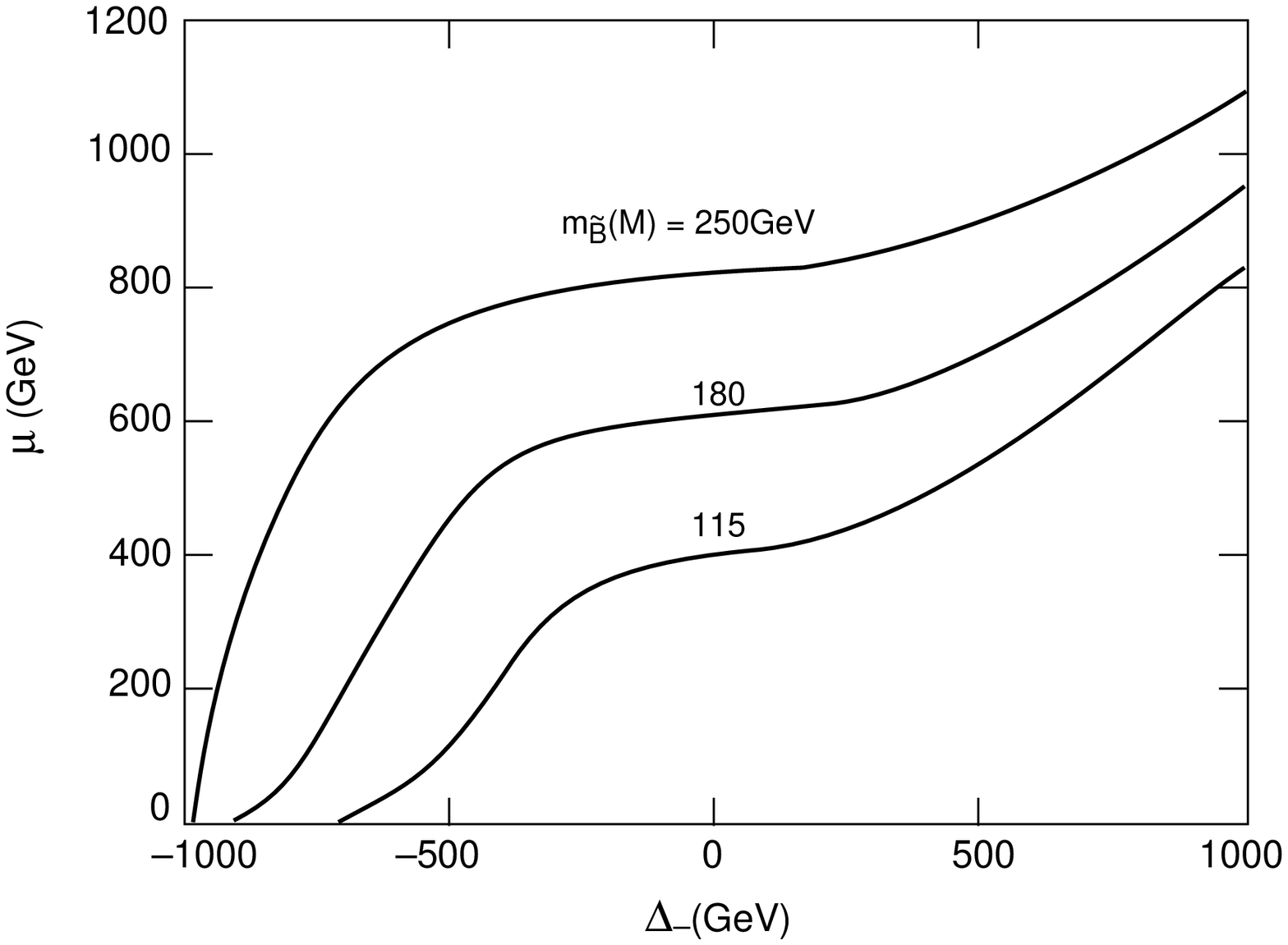}
{The relation between the low scale $|\mu|$ parameter and 
$\Delta_-\equiv {\rm sgn}(\Delta^2_-(M))(|\Delta^2_-(M)|)^{1/2}$ 
at the messenger scale 
imposed by electroweak symmetry breaking for
$\Delta_+(M)=0$, 
$m_{\tilde B}(M)=115,180,250$, GeV $\tan\beta =3$, 
and $\Lambda=M$.}
The relation between $\mu$ at the minimization scale and
$\Delta_-\equiv {\rm sgn}(\Delta^2_-(M) )\sqrt{|\Delta^2_-(M)|}$
at the messenger scale is shown in 
Fig. \ref{sfig5n} for 
$\Delta_+(M)=0$, 
$m_{\tilde B}(M)=115,180,250$, GeV, $\tan\beta =3$, 
and $\Lambda=M$.
For moderate values of $\Delta_-$, the additional Higgs splittings
contribute in quadrature 
with the gauge-mediated contributions,
and only give ${\cal O}(\Delta^2_- / \mu^2)$ corrections 
to the minimization condition (\ref{minsplit}).
This is the origin of the shallow plateau in Fig. 
\ref{sfig5n} along which $\mu$ does not significantly vary.
The plateau extends over the range $|\Delta_-^2| \lsim |m_{H_u,0}^2|$
at the messenger scale. 
For $\tan \beta \gg 1$ the minimization condition
(\ref{minsplit}) becomes 
$|\mu|^2 \simeq -\mHu^2 + {1 \over 2}
  (\Delta_-^2 - \Delta_+^2  - \mZ^2)$. 
For very large $(\Delta_-^2 - \Delta_+^2)$ this reduces to 
$\sqrt{2}|\mu| \simeq  (\Delta_-^2 - \Delta_+^2)^{1/2}$.
This linear correlation between $\mu$ and $\Delta_-$ 
for $\Delta_-$ large and $\Delta_+=0$ is apparent
in Fig. \ref{sfig5n}.
The non-linear behavior at small $\Delta_-$ 
arises from ${\cal O}(\mZ^2 / \mu^2)$ contributions to the
minimization condition (\ref{minsplit}). 

The physical correlation between $\mu$ and $\Delta_{\pm}$ is
easily understood in terms of $\mHu^2$ at the messenger scale.  
For $\Delta_+=0$ and $\Delta_- > 0$, $\mHu^2$ 
is more negative than in the minimal model, 
leading to a deeper minimum
in the Higgs potential. 
In fact, for the $\mbino(M)=115$ GeV case shown in Fig. 
\ref{sfig5n}, $\mHu^2 <0$ already at the messenger scale
for $\Delta_- \gsim 260$ GeV. 
Obtaining correct electroweak symmetry breaking for $\Delta_- > 0$
therefore requires 
a larger value of $|\mu|$, 
as can be seen in Fig. \ref{sfig5n}.
Conversely, 
for $\Delta_+=0$ and $\Delta_- < 0$, $\mHu^2$ 
is less
negative than in the minimal model, leading to a more shallow minimum
in the Higgs potential. 
Obtaining correct electroweak symmetry breaking in this limit 
therefore requires
a smaller value of $|\mu|$, 
as can also be seen in Fig. \ref{sfig5n}.
Eventually, 
for $\Delta_-$ very negative, 
$\mHu^2$ at the messenger scale is large enough that 
the negative 
renormalization group evolution from the top Yukawa is
insufficient to drive electroweak symmetry breaking. 
In Fig. \ref{sfig5n} this corresponds to $|\mu| < 0$. 

With $\Delta_-=0$ and $\Delta_+ > 0$ both $\mHu^2$ and $\mHd^2$
are larger at the messenger scale than in the minimal model,
leading to a more shallow minimum in the Higgs potential.
This results in smaller values of $\mu$, and conversely
larger values of $\mu$ for $\Delta_+ < 0$. 
Again, there is only a significant effect for 
$|\Delta_+^2| \gsim |\mHu^2|$.

The pseudo-scalar Higgs mass also depends on additional
contributions to the Higgs soft masses,
$\mA^2 = 2 |\mu|^2 + (m^2_{H_u,0} + \Sigma_u) + 
 (m^2_{H_d,0} + \Sigma_d) + \Delta_+^2$. 
For large $\tan \beta$ the minimization condition (\ref{minsplit})
gives
$\mA^2 \simeq  - (m^2_{H_u,0} + \Sigma_u) +  (m^2_{H_d,0} + \Sigma_d)
 + \Delta_-^2$. 
Again, for $|\Delta_-^2| \lsim |m^2_{H_u,0}|$ the pseudo-scalar
mass is only slightly affected, but can be altered significantly
for $\Delta_-$ very large in magnitude. 
Notice that $\mA$ is independent of $\Delta_+$ in this limit. 
This is because in the contribution 
to $\mA^2$ the change in $|\mu|^2$ induced by $\Delta_+^2$
is cancelled by a 
compensating change in $\mHd^2$.
This approximate independence of $\mA$ on $\Delta_+$ persists
for moderate values of $\tan \beta$. 
For example, 
for the parameters of Fig. \ref{sfig5n} with 
$\mbino(M) = 115$ GeV and $\Delta_-=0$, $\mA$ only varies
between 485 GeV and 525 GeV for 
$-500$ GeV $< \Delta_+ < $ $500$ GeV, while $\mu$ varies from 510 GeV to 
230 GeV over the same range. 

The additional contributions to the Higgs soft masses
can, if large enough, change the form of the superpartner
spectrum. 
The charginos and neutralinos are affected mainly through
the value of $\mu$ implied by electroweak symmetry breaking. 
For very large $|\mu|$, the approximately degenerate 
singlet $\chi_3^0$ and triplet $(\chi_2^{+}, \chi_4^0, \chi_2^-)$
discussed in section \ref{electroweaksection} are mostly 
Higgsino, and have mass $\mu$. 
For $\mu \lsim m_2$ the charginos and neutralinos are
a general mixture of gaugino and Higgsino. 
A value of $\mu$ in this range,
as evidenced by a sizeable Higgsino component of 
$\chi_1^0$, $\chi_2^0$, or $\chi_1^{\pm}$, or 
a light $\chi_3^0$ or $\chi_2^{\pm}$,
would be strong evidence for
deviations from the minimal model in the Higgs sector.

The heavy Higgs masses are determined by $\mA$.
Since $\mA^2$ is roughly independent of $\Delta_+^2$, while
$|\mu|$ is sensitive to $( \Delta_-^2 - \Delta_+^2)$, 
the relative shift between
the Higgsinos and heavy Higgses
is sensitive to the individual
splittings of $\mHu^2$ and $\mHd^2$ from the left handed sleptons,
$m_{\lL}^2$.  
Within the MGM,
given an independent measure of $\tan \beta$ (such as from 
left handed slepton - sneutrino splitting, 
$m_{\lL}^2 - m_{\nu_L}^2 = -\mW^2 \cos 2 \beta$)
the mass of the Higgsinos and heavy Higgses 
therefore provides an indirect probe
for additional contributions to the Higgs soft masses.

\jfig{sfig6n}{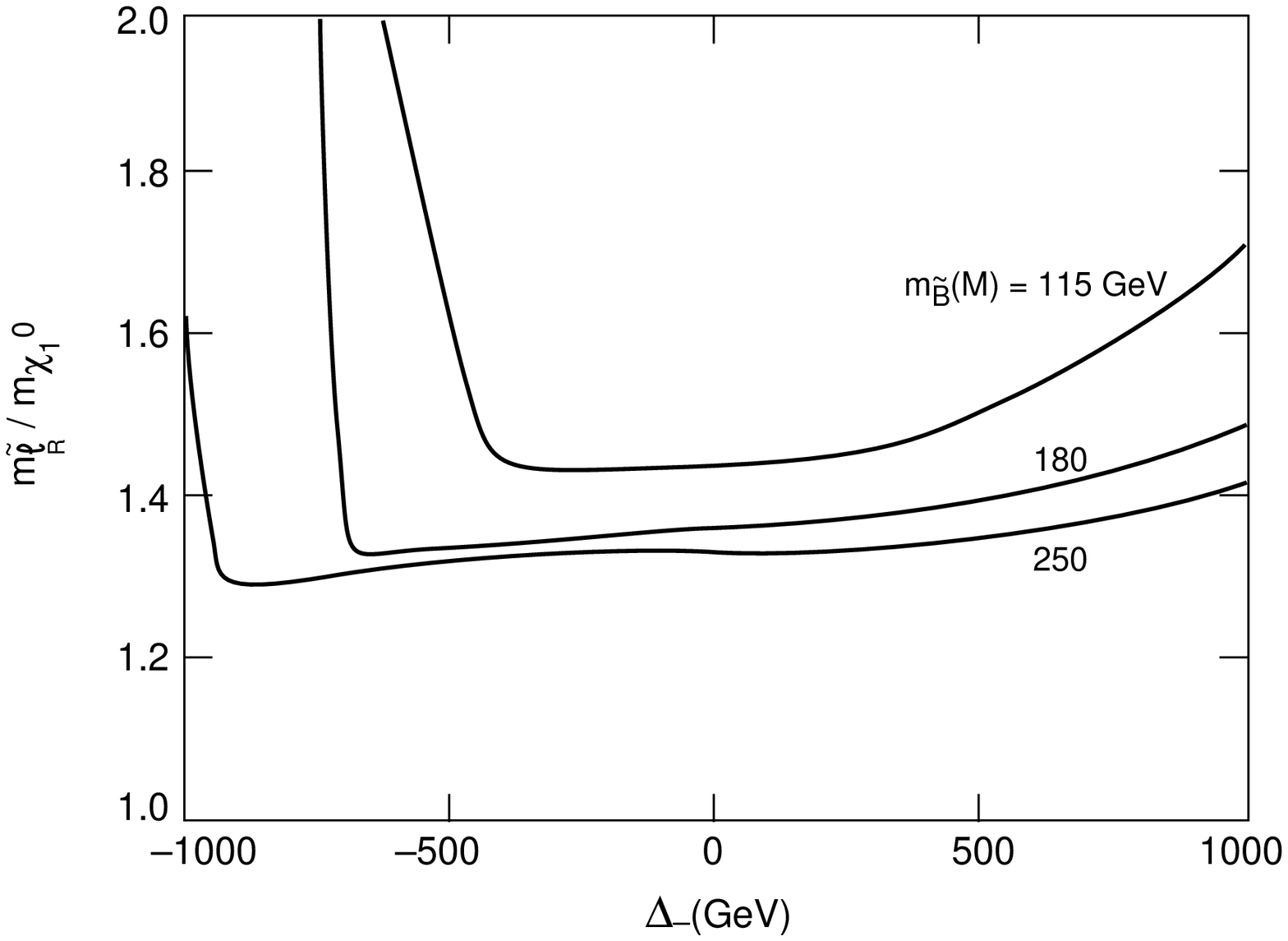}
{The ratio $m_{\tilde l_R}/m_{\chi^0_1}$ as a
function of $\Delta_-$ at the messenger scale 
for $m_{\tilde B}(M)=115,180,250\gev$, $\tan\beta =3$, 
${\rm sgn}(\mu)=+1$, and $\Lambda=M$.}
Non-minimal contributions to Higgs soft masses can 
also affect the other scalar masses through renormalization 
group evolution. 
The largest effect comes from the radiative contribution to 
the $U(1)_Y$ $D$-term, which is generated in 
proportion to $S = {1 \over 2}{\rm Tr}(Ym^2)$. 
In the minimal model the Higgs contribution to $S$ 
vanishes at the messenger scale because the Higgses 
are degenerate and have opposite hypercharge. 
For $\Delta_- > 0$ they are no longer degenerate and 
give a negative contribution to $S$.
This increases the magnitude of the contribution in the minimal
model from running below the squark thresholds. 
To illustrate the effect of the Higgs contribution to $S$ on 
the scalar masses, 
$m_{\tilde l_R}/m_{\chi^0_1}$ is shown 
in Fig. \ref{sfig6n} as a
function of $\Delta_-$ at the messenger scale 
for $m_{\tilde B}(M)=115,180,250\gev$, $\tan\beta =3$, 
${\rm sgn}(\mu)=+1$, and $\Lambda=M$.
For $\Delta_-$ very large and 
positive the radiatively generated
$U(1)_Y$ $D$-term contribution to right handed slepton masses
increases the ratio $m_{\tilde l_R}/m_{\chi^0_1}$. 
For $\Delta_-$ very negative, 
the rapid increase in  $m_{\tilde l_R}/m_{\chi^0_1}$
occurs because $|\mu|$ is so small that $\chi_1^0$ becomes
mostly Higgsino with mass $\mu$.

All these modifications of the 
form of the superpartner spectrum are significant 
only if the Higgs bosons receive 
additional contributions to the soft masses which are 
roughly larger in magnitude than the gauge-mediated contribution. 

The $\mu$ parameter is renormalized multiplicatively while 
$m_{12}^2$ receives renormalization group contributions proportional
to $\mu m_{\lambda}$, where $m_{\lambda}$ is the $B$-ino
or $W$-ino mass.
As suggested in Ref. \cite{frank}, it is therefore interesting
to investigate the possibility that $m_{12}^2$ is generated 
only radiatively below the messenger scale, with 
the boundary condition $m_{12}^2(M)=0$.
Most models of the Higgs sector interactions actually suffer 
from $m_{12}^2 \gg \mu^2$ \cite{dnmodels,dgphiggs}, but
$m_{12}^2(M)=0$ represents a potentially interesting, and 
highly constrained subspace of the MGM. 
\jfig{sfig8n}{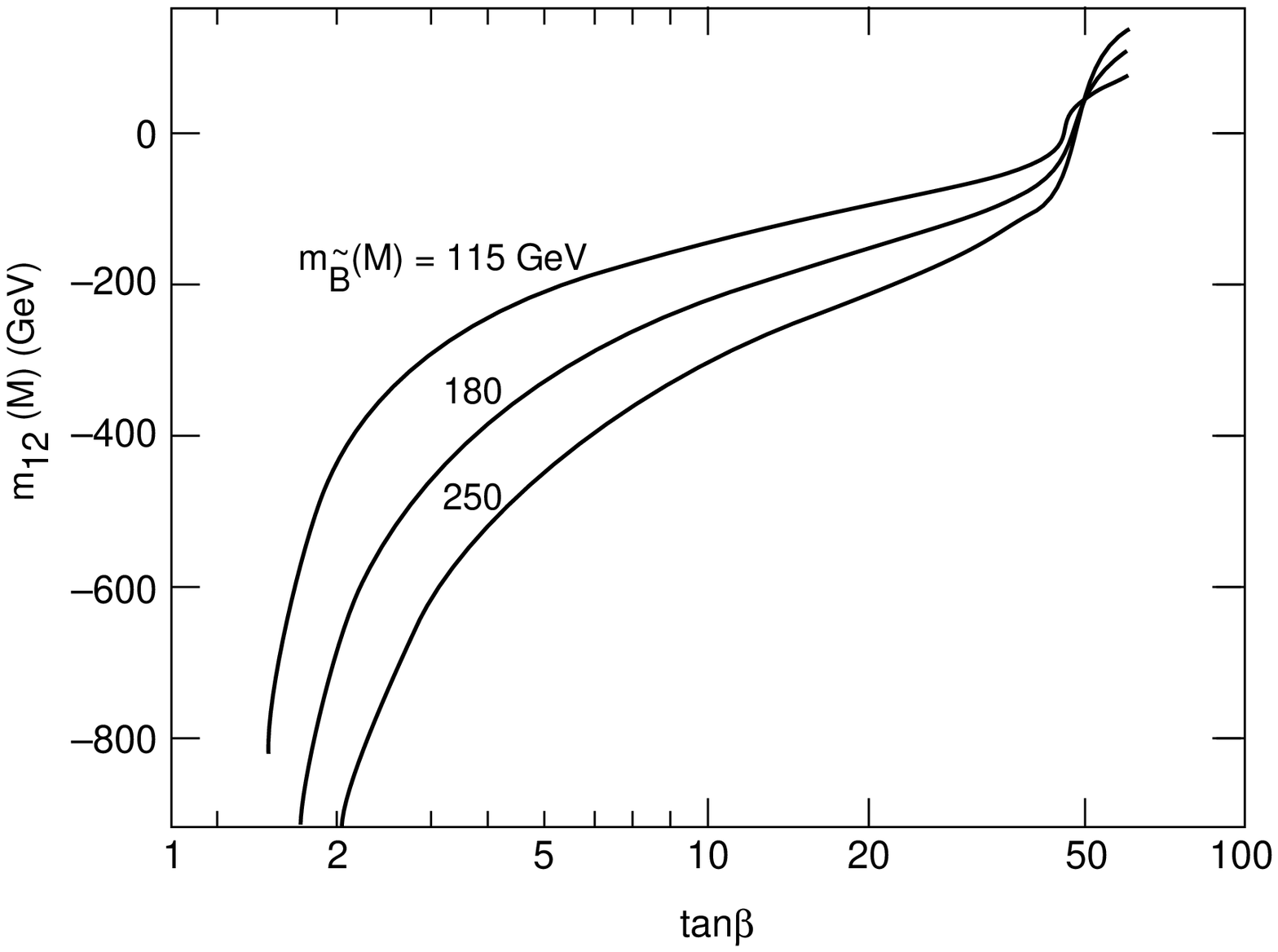}
{The relation between $m_{12}(M)$ and $\tan\beta$ imposed by electroweak
symmetry breaking for 
$m_{\tilde B}(M) =115,180,250$ GeV, and $\Lambda=M$.}
In order to illustrate what constraints this boundary condition
implies, the relation between $m_{12}(M)$ and 
$\tan \beta$ imposed by electroweak symmetry breaking is
shown in Fig. \ref{sfig8n} for 
$m_{\tilde B}(M) =115,180,250$ GeV and $\Lambda=M$.
The non-linear feature at $m_{12}(M)\simeq 0$ is a square root
singularity since the $\beta$-function for $m_{12}^2$
is an implicit function of $\tan \beta$ only through the slow
dependence of $\mu$ on $\tan \beta$.
The value of $\tan \beta$ for which $m_{12}(M)=0$ is almost
entirely independent of the overall scale of the superpartners.  
This is because to lowest order the minimization condition 
(\ref{minconb}) fixes $m_{12}^2$ at the low scale
to be a homogeneous function of the overall superpartner scale
(up to $\ln(m_{\tilde{t}_1} m_{\tilde{t}_2} / m_t^2)$ finite 
corrections)
$m_{12}^2 \simeq f(\alpha_i,\tan \beta) (\alpha / 4 \pi)^2 \Lambda^2$.
If $m_{12}$ vanishes at any scale, then the function $f$ vanishes
at that scale, thereby determining $\tan \beta$. 
For $m_{12}(M)=0$ and $\Lambda=M$ we find $\tan \beta \simeq 46$. 

With the boundary condition $m_{12}(M)=0$, the resulting
large value of $\tan \beta$ is natural. 
This is because
$m_{12}^2(Q)$ at the minimization scale, $Q$, is small. 
With the parameters given above $m_{12}(Q) \simeq -80$ GeV.
For $m_{12}(Q) \rightarrow 0$, $H_d$ does not participate
in electroweak symmetry breaking, and $\tan \beta \to \infty$. 
As discussed in section \ref{electroweaksection}, 
at large $\tan \beta$,
$m_{\stau_1}$ receives a large negative contribution 
from the $\tau$ Yukawa due to renormalization
group evolution and mixing. 
For the values of $\tan \beta$ given above 
we find $m_{\stau_1} \lsim m_{\na}$. 
It is important to note that for such large values of 
$\tan \beta$, physical quantities, such as 
$m_{\stau_1} /  m_{\na}$, depend sensitively on 
the precise value of the 
$b$ Yukawa through renormalization group and finite
contributions to the Higgs potential.


\subsection{$U(1)_Y$ $D$-term}

The $U(1)_Y$ $D$-term can be non-vanishing at the messenger scale, 
as discussed in section 2.2. 
This gives an additional contribution to the soft scalar masses
proportional to the $U(1)_Y$ coupling, as given in 
Eq. (\ref{Dmass}). 
This splits $\mHu$ and $\mHd$, and has the same affect on electroweak
symmetry breaking as $\Delta_-$ discussed in the previous subsection. 
The right handed sleptons have the smallest gauge-mediated
contribution to soft masses, and are therefore most 
susceptible to $D_Y(M)$. 
The biggest effect 
on the scalar spectrum 
is therefore a modification of the splitting
between left and right handed sleptons. 
This splitting can have an important impact on the relative
rates of 
$p \bar{p} \to l^+ l^- \gamma \gamma + \EmissT$ and 
$p \bar{p} \to l^{\pm} \gamma \gamma + \EmissT$ at
hadron colliders \cite{tevatron} as compared with the minimal
model discussed in section \ref{misssig}.

\section{Phenomenological Consequences}

Since the parameter space of the MGM is so constrained
it is interesting to investigate what phenomenological 
consequences follow. 
In the first subsection below
we discuss virtual effects, with
emphasis on the constraints within the MGM from $b \rightarrow s \gamma$. 
In the second subsection we discuss the collider signatures
associated with the gauge-mediated supersymmetry breaking. 
These can differ significantly from the standard MSSM with
$R$-parity conservation and  
high scale supersymmetry breaking.
This is because first, the lightest
standard model superpartner can decay within the detector
to its partner plus the Goldstino,
and second, the lightest standard model superpartner
can be either either $\na$ or $\lR^{\pm}$.

\subsection{Virtual Effects}

Supersymmetric theories can be probed indirectly by virtual
effects on low energy, high precision, processes \cite{dpf}.
Among these are precision electroweak measurements,
electric dipole moments, and flavor changing neutral currents.
In the minimal model of gauge-mediation, supersymmetric
corrections to electroweak observables are unobservably
small since the charginos, left handed sleptons, 
and squarks are too heavy. 
Likewise, the effect on 
$R_b = \Gamma(Z^0 \to b \bar{b})/ \Gamma(Z^0 \to {\rm had})$
is tiny since the Higgsinos and both stops are heavy. 
Electric dipole moments can arise from the single $CP$-violating
phase in the soft terms, discussed in section \ref{minimalsection}.
The dominant contributions to the dipole moments of atoms
with paired or unpaired electrons, and the neutron, 
come from one-loop chargino processes, just as 
with high scale supersymmetry breaking.
The bounds on the phase are therefore comparable to those
in the standard MSSM,
${\rm Arg}(m_{\lambda} \mu (m_{12}^2)^*) \lsim 10^{-2}$ 
\cite{edm,edma}.
It is important to note that in some schemes for
generating the Higgs sector parameters $\mu$ and $m_{12}^2$,
the soft terms are $CP$ conserving \cite{dnmodels},
in which case electric dipole moments are unobservably small. 
This is also true for the boundary condition $m_{12}^2(M)=0$
since 
$(m_{\lambda} \mu (m_{12}^2)^*)$ vanishes in this case.

Contributions to flavor changing neutral currents can 
come from two sources in supersymmetric theories. 
The first is from flavor violation in the squark or 
slepton sectors. 
As discussed in section \ref{UVinsensitive} this source
for flavor violation is naturally small with gauge-mediated
supersymmetry breaking. 
The second source is from second order electroweak virtual
processes which are sensitive to flavor violation in the
quark Yukawa couplings. 
At present the most sensitive probe for contributions of this
type beyond those of the standard model is $b \to s \gamma$.
In a supersymmetric theory one-loop 
$\chi^{\pm}-\tilde{t}$ and $H^{\pm}-t$ contributions 
can compete with the standard model
$W^{\pm}-t$ one-loop effect. 
The standard model effect is dominated by the transition 
magnetic dipole operator which arises from the electromagnetic penguin, 
and the tree level charged current operator, which contributes
under renormalization group evolution. 
The dominant supersymmetric contributions 
are through the transition dipole operator. 
It is therefore convenient to parameterize the 
supersymmetric contributions as 
\beq
R_7 \equiv { C^{\rm MSSM}_7(\mW) \over  C^{\rm SM}_7(\mW) } -1
\eq
where $C_7(\mW)$ is the coefficient of the dipole
operator at a renormalization scale $\mW$, and
$ C^{\rm MSSM}_7(\mW)$ contains the entire MSSM contributions
(including the $W^{\pm}-t$ loop).
In the limit of 
decoupling the supersymmetric states and heavy Higgs bosons
$R_7 =0$. 
\jfig{sfig9n}{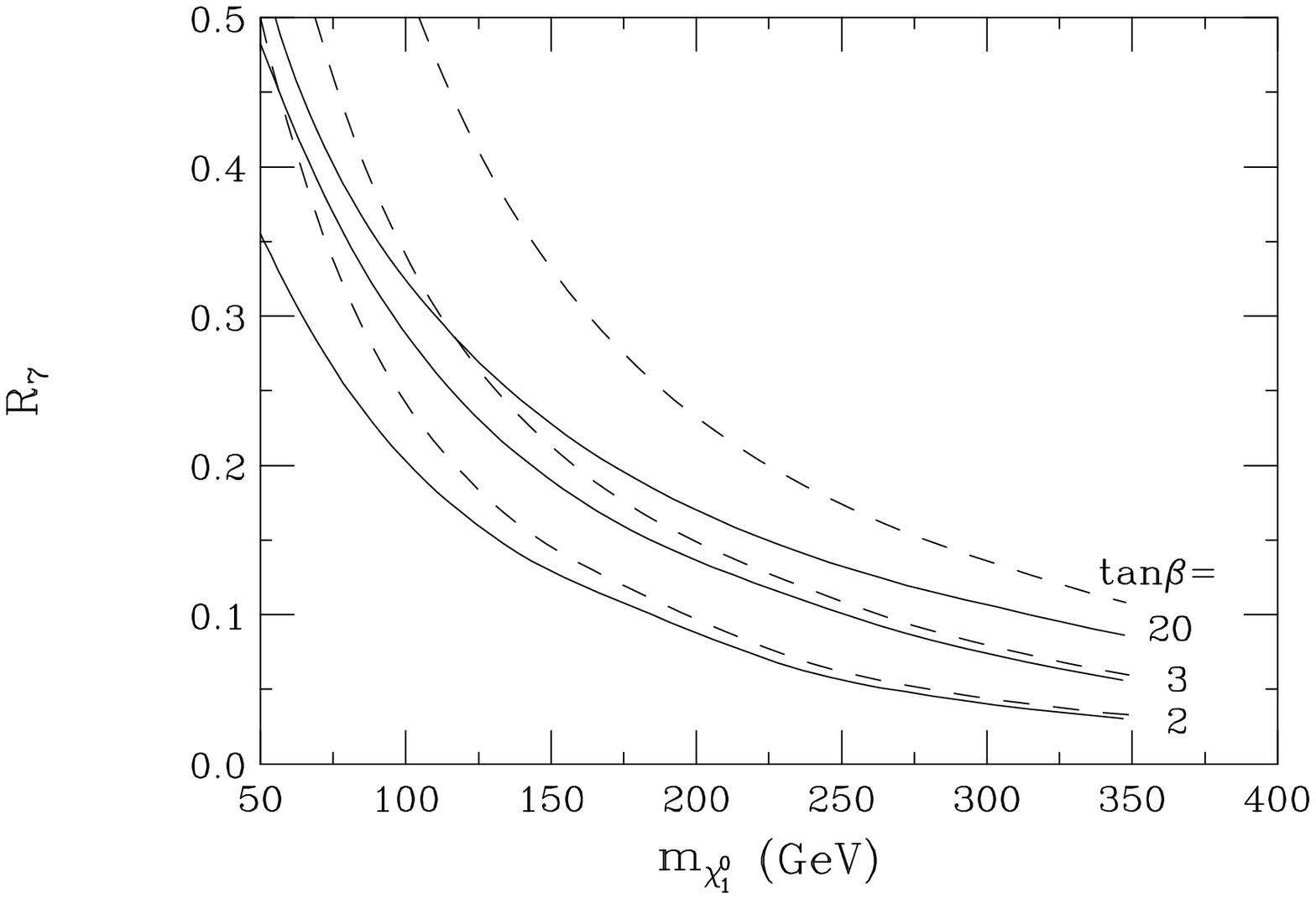}
{The parameter 
$R_7\equiv C^{\rm{MSSM}}_7(\mW)/C^{\rm{SM}}_7(\mW)-1$ as a function
of the lightest neutralino mass, $m_{\chi^0_1}$, for
$\tan\beta =2,3,20$, and $\Lambda=M$.
The solid lines are for $\mu >0$ and the dashed lines for $\mu <0$. 
}
The parameter $R_7$ is shown in Fig. \ref{sfig9n} as a function
of the lightest neutralino mass, $m_{\chi^0_1}$, for
both signs of $\mu$, 
$\tan\beta =2,3,20$, and $\Lambda=M$ \cite{bsgammaref}.
The $\chi_1^0$  mass is plotted in Fig. \ref{sfig9n}
as representative of the overall scale of the superpartner
masses. 
The dominant contribution comes from the 
$H^{\pm}-t$ loop which adds constructively to the 
standard model $W^{\pm}-t$ loop.
The $\chi^{\pm}-\tilde{t}$ loop gives a  
destructive contribution which is smaller in magnitude 
because the stops are so heavy. 
The ${\rm sgn}~\mu$ dependence of $R_7$ results from this small
destructive contribution mainly because the Higgsino
component of the lightest chargino is larger(smaller) for 
${\rm sgn}~\mu = +(-)$ (cf. Eq. \ref{winoshift}). 
The $\chi^{\pm}-\tilde{t}$ loop amounts to roughly a  
$-$15(5)\% contribution compared with the $H^{\pm}-t$ loop for 
${\rm sgn}~\mu=+(-)$. 
The non-standard model contribution to $R_7$ decreases for small
$\tan \beta$ since $m_{H^{\pm}} \simeq \mA$ increases
in this region.

In order to relate $R_7$ to ${\rm Br}( b \to s \gamma)$ 
the dipole and tree level charged current operators must
be evolved down to the scale $m_b$.
Using the results of Ref. \cite{buras}, which include
the leading QCD contributions to the anomalous dimension
matrix, we find
\beq
\frac{{\rm Br}^{\rm{MSSM}}(b\to s\gamma)}{{\rm Br}^{\rm{SM}}
    (b\to s\gamma)}  \simeq |1+0.45 ~R_7(\mW)|^2.
\eq
for $m_t^{\rm pole} = 175$ GeV.
For this top mass
${\rm Br}^{\rm SM}(b \to s \gamma) \simeq (3.25 \pm 0.5)\times 10^{-4}$
where the uncertainties are estimates of the theoretical 
uncertainty coming mainly from $\alpha_s(m_b)$ and 
renormalization scale 
dependence \cite{greub}.
Using the ``lower'' theoretical value and the 95\% CL
experimental 
upper limit of ${\rm Br}(b \to s \gamma) < 4.2 \times 10^{-4}$
from the CLEO measurement \cite{CLEO}, we find
$R_7 < 0.5$.\footnote{
This is somewhat more conservative than the bound
of $R_7 < 0.2$ suggested in Ref. \cite{cho}.}
This bound assumes that the non-standard model effects
arise predominantly in the dipole operator, and are constructive
with the standard model contribution.
In the MGM 
for $\mu > 0$,
$\tan \beta =3$, and $\Lambda=M$, this bound corresponds to 
$m_{\na} \gsim 45$ GeV, or a charged Higgs mass of 
$m_{H^{\pm}} \gsim 300$ GeV.

The present experimental limit does not severely constrain
the parameter space of the MGM. 
This follows from the fact that the charged Higgs is very
heavy over most of the allowed parameter space.
Except for very large $\tan \beta$ 
$m_{H^{\pm}} \gsim |\mu|$, and imposing electroweak symmetry 
breaking implies 
$ 3 m_{\na} \lsim |\mu| \lsim 6 m_{\na}$, 
as discussed 
in sections \ref{EWSB} and \ref{higgssection}.
For example, with the parameters of Table 1 
$m_{H^{\pm}} \simeq 5.4 m_{\na}$.
Note that since the stops are never light in the minimal
model there is no region of the parameter space for which the
$\ca - \tilde{t}$ loop can cancel the $H^{\pm} -t$ loop.

Precise measurements of ${\rm Br}(b \to s \gamma)$ at future
$B$-factories, and improved calculations of the anomalous
dimension matrix and finite contributions at the scale $m_b$, will
improve the uncertainty in $R_7$ to $\pm 0.1$ \cite{postb}.
Within the MGM, even for $\tan \beta =2$ and $\mu > 0$, 
a measurement of ${\rm Br}(b \to s \gamma)$ consistent with the standard
model would give a bound on
the charged Higgs mass of $m_{H^{\pm}} \gsim 1200 $ GeV,
or equivalently an indirect bound on the chargino mass of
$\ca$ mass of $m_{\ca} \gsim 350 $ GeV.
Such an indirect bound on the chargino mass  
is more stringent than the direct bound that 
could be obtained at the main injector
upgrade at the Tevatron \cite{tevatron}, and significantly
better than the direct bound that will be available at LEP II.

\subsection{Collider Signatures}

\label{collidersection}

Direct searches for superpartner production at high energy 
colliders represent the best probe for supersymmetry. 
Most searches assume that $R$-parity is conserved and that
the lightest standard model 
superpartner is a stable neutralino.
Pair production of supersymmetric states then takes place
through gauge or gaugino interactions, with cascade decays
to pairs of neutralinos. 
The neutralinos escape the detector leading to the classic
signature of missing energy. 
With gauge-mediated supersymmetry breaking the collider
signatures can be much different in some circumstances. 
First, for a messenger scale well below the Planck scale, 
the gravitino is naturally the lightest supersymmetric particle. 
If the supersymmetry breaking scale is below a 
few 1000 TeV,
the lightest standard model superpartner can decay to its partner 
plus the Goldstino component of the gravitino 
inside the detector \cite{sbtalk,signatures}.
The Goldstino, and associated decay rates, are discussed in 
appendix \ref{appgoldstino}.
Second, as discussed in sections \ref{electroweaksection}
and \ref{multiple} it is possible that the lightest standard 
model superpartner is a slepton \cite{sbtalk,signatures}. 
If the supersymmetry breaking scale is larger than a few 1000 
TeV, the signature for supersymmetry is then a pair of heavy
charged particles plowing through the detector,
rather than missing energy.

The form of the superpartner spectrum has an important impact
on what discovery modes are available at a collider. 
With gauge-mediation, all the strongly interacting states, 
including the stops, are generally too heavy to be relevant to discovery
in the near future.
In addition, the constraints of electroweak symmetry breaking
imply that the heavy Higgs bosons and mostly Higgsino
singlet $\chi_3^0$ and triplet $(\chi_2^+, \chi_4^0, \chi_2^-)$
are also too heavy.
The mostly $B$-ino $\chi_1^0$, mostly $W$-ino triplet
$(\chi_1^+, \chi_2^0, \chi_1^-)$, right handed sleptons
$\lR^{\pm}$, and lightest Higgs boson, $h^0$, are the accessible
light states. 

In this section we discuss the collider signatures of
gauge-mediated supersymmetry breaking associated with the
electroweak supersymmetric states.  
In the next two subsections the signatures associated
with either a neutralino or slepton as the lightest standard
model superpartner are presented.

\subsubsection{Missing Energy Signatures}

\label{misssig}

The minimal model has a conserved 
$R$-parity by assumption.
At moderate $\tan \beta$, $\chi_1^0$ is the lightest
standard model superpartner. 
If decay to the Goldstino takes place well outside the detector
the classic signature of missing energy results. 
However, 
the form of the low lying spectrum largely dictates 
the modes which can be observed. 
The lightest charged states are the right handed sleptons, 
$\lR^{\pm}$. 
At an $e^+e^-$ collider the most relevant mode
is then $e^+ e^- \to \lR^+ \lR^-$
with $\lR^{\pm} \to l^{\pm} \na$. 
For small $\tan \beta$ all the the sleptons are essentially 
degenerate so the rates 
to each lepton flavor should be essentially identical. 
For large $\tan \beta$ the $\stau_1$ can become
measurably lighter than $\tilde{e}_R$ and 
$\tilde{\mu}_R$ (cf. Fig. \ref{sfig17n}).
If sleptons receive masses at the messenger scale
only from standard model gauge interactions, the only
source for splitting of $\tau_1$ from $\eR$ and $\tilde{\mu}_R$
is the $\tau$ Yukawa in renormalization group evolution and 
mixing.
As discussed in section \ref{electroweaksection} the largest
effect is from $\stau_L - \stau_R$ 
mixing proportional to $\tan \beta$. 
A precision measurement of $m_{\stau_1}$ therefore
provides an indirect probe of 
whether 
$\tan \beta$ is large or not.

\jfig{sfig2x}{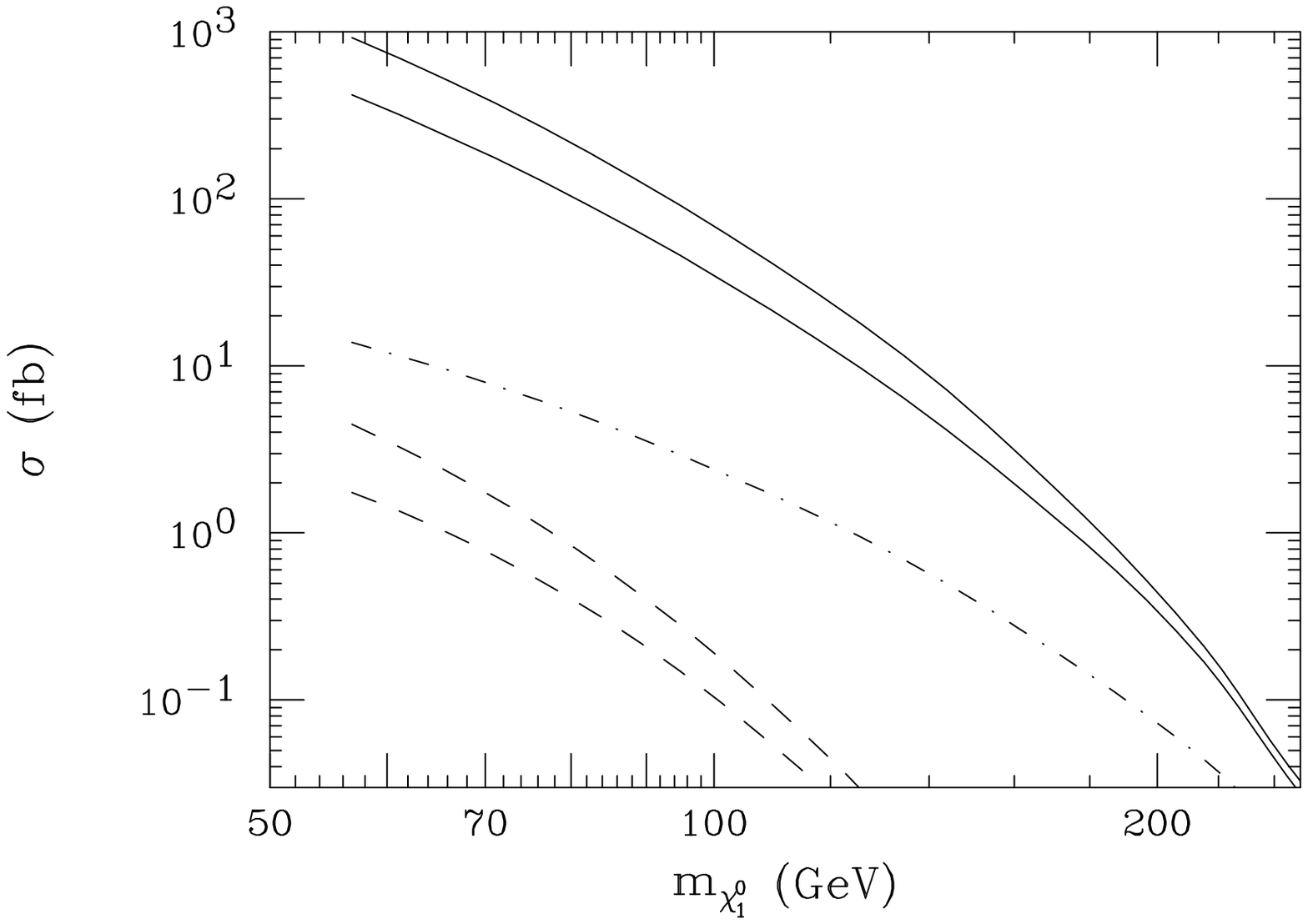}
{Production cross sections (fb) for $p \bar{p}$ initial state to the
final states $\chi_1^{\pm} \chi_2^0$ (upper solid line),
$\chi_1^+ \chi_1^-$ (lower solid line),
$\lR^+ \lR^-$ (dot-dashed line),
$\nL \lL^{\pm}$ (upper dashed line), and
$\lL^+ \lL^-$ (lower dashed line).
Lepton flavors are not summed. 
The center of mass energy is 2 TeV, ${\rm sgn}(\mu)=+1$, 
and $\Lambda=M$.} 
At a hadron collider both the mass and gauge quantum numbers determine
the production rate for supersymmetric states. 
The production cross sections for electroweak states in
$p \bar{p}$ collisions at $\sqrt{s} = 2$ TeV
(appropriate for the main injector upgrade at the Tevatron)
are shown in Fig. \ref{sfig2x} as a function of 
$m_{\na}$ for MGM boundary conditions with $\Lambda=M$ 
and ${\rm sgn}(\mu)=+1$.  
The largest cross section is for pairs of the mostly $W$-ino
$SU(2)_L$ triplet 
$(\chi_1^+, \chi_2^0, \chi_1^-)$ through off-shell $W^{\pm *}$ and
$Z^{0*}$. 
Pair production of $\lR^+ \lR^-$ is relatively
suppressed even though $m_{\lR} < m_{\chi_1^{\pm}}$ because
scalar production suffers a $\beta^3$ suppression near threshold, 
and the right handed sleptons couple only through $U(1)_Y$
interactions via off-shell $\gamma^*$ and $Z^{0*}$.
However, 
as the overall scale of the superpartner masses is increased
$\lR^+ \lR^-$ production becomes relatively more important
as can be seen in Fig. \ref{sfig2x}.
This is because production of the more massive $\ca \na$ and 
$\chi_1^+ \chi_1^-$ is reduced by the rapidly falling 
parton distribution functions. 
Pair production of $\lL^+ \lL^-$, $\lL^{\pm} \nL$, and 
$\nL \nL$ through off-shell $\gamma^*$, $Z^{0*}$, and 
$W^{\pm*}$ is suppressed relative to $\lR^+ \lR^-$ 
by the larger left handed slepton 
masses. 

The renormalization group and classical $U(1)_Y$ $D$-term
contributions which slightly increase $m_{\lR}$, 
and the renormalization group contribution which decreases
$m_{\na}$, have an impact on the 
relative importance $\lR^+ \lR^-$ production. 
These effects,  along with the radiatively generated $U(1)_Y$ $D$-term,
``improve'' the kinematics of the leptons arising
from $\lR^{\pm} \to l^{\pm} \na$ 
since $m_{\lR} - m_{\na}$ is increased \cite{frank}. 
However, the overall rate is simultaneously reduced 
to a fairly insignificant level \cite{tevatron}.
For example, with ${\rm sgn}(\mu)=+1$
an overall scale which would give an average of one 
$\tilde{l}_R^+ \tilde{l}_R^-$ event 
in 100 pb$^{-1}$ of integrated luminosity, would result
in over 80 
chargino events. 
As discussed in section \ref{electroweaksection}, the shift
in the triplet $(\chi_1^+, \chi_2^0, \chi_1^-)$ mass 
from mixing with the Higgsinos is anti-correlated with
${\rm sgn}(\mu)$.
For ${\rm sgn}(\mu)=-1$ the splitting between the right handed
sleptons and triplet is larger, thereby reducing slightly
chargino production.  
For example, with ${\rm sgn}(\mu)=-1$, a 
single $\tilde{l}_R^+ \tilde{l}_R^-$ event 
in 100 pb$^{-1}$ of integrated luminosity, would result
in 30 chargino events. 
The relative rate of the $\lR^+ \lR^-$ initial state
is increased in the minimal model for $\Rslash > 1$.
However, as discussed in Ref. \cite{tevatron}, 
obtaining a rate comparable to $\chi_1^{\pm} \chi_2^0$
results in ``poor'' kinematics, in that the leptons
arising from $\lR^{\pm} \to l^{\pm} \na$ are fairly
soft since 
$m_{\lR} - m_{\na}$ is reduced.
Note that for $\Rslash < 1$ chargino production becomes even more
important than $\lR^+ \lR^-$ production.


In the minimal model pair production of 
$\ca \nb$ and $\chi_1^+ \chi_1^-$ are the most important modes at a 
hadron collider. 
The cascade decays of $\ca$ and $\nb$
are largely fixed by the form of the
superpartner spectrum and couplings. 
If open, $\ca$ decays predominantly through its 
Higgsino components to the Higgsino components of $\na$ by 
$\ca \to \na W^{\pm}$.
Likewise, $\nb$ can also decay by $\nb \to \na Z^0$. 
However, if open $\nb \to h^0 \na$ is suppressed by
only a single Higgsino component in 
either $\nb$ or $\na$, and represents the dominant 
decay mode for $m_{\nb} \gsim \mh + m_{\na}$. 
The decay $\nb \to \lR^{\pm} l^{\mp}$ is suppressed by
the very small $B$-ino component of $\nb$, and is only important
if the other two-body modes given above are closed. 
If the two body decay modes for $\ca$ are closed, it decays
through three-body final states predominantly 
through off-shell $W^{\pm*}$. 
Over much of the parameter space the minimal model therefore
gives rise to the signatures 
$p \bar{p} \to W^{\pm} Z^0 + \EmissT$, 
$W^{\pm} h^0 + \EmissT$, and 
$W^+ W^- + \EmissT$. 
If decay to the Goldstino takes place well outside the
detector, the minimal model yields 
the ``standard'' chargino signatures at a hadron collider 
\cite{charginoref}.

If the intrinsic supersymmetry breaking scale is below a few
1000 TeV, the lightest standard model superpartner can 
decay to its partner plus the Goldstino within the 
detector \cite{sbtalk,signatures}.
For the case of $\na$ as the lightest standard model superpartner,
this degrades somewhat the missing energy, but leads to 
additional visible energy.
The neutralino $\na$ decays by 
$\na \to \gamma + G$ and if kinematically accessible
$\na \to (Z^0, h^0, H^0, A^0) + G$. 
In the minimal model $\mA, \mH > m_{\na}$ so the only two 
body final states potentially open are  
$\na \to (\gamma, Z^0, h^0) + G$.
However,  as discussed in section 
\ref{electroweaksection}, with MGM boundary conditions,
electroweak symmetry breaking implies 
that $\na$ is mostly $B$-ino,
and therefore decays predominantly to the 
gauge boson final states.
The decay $\na \to h^0 + G$ 
takes place only through the small Higgsino components.
In appendix \ref{appgoldstino}
the decay rate to the $h^0$ final state is 
shown to be suppressed by
${\cal O}(\mZ^2 m_{\na}^2 / \mu^4)$
compared with the
gauge boson final states, and is therefore 
insignificant in the minimal model. 
Observation of the decay $\na \to h^0 + G$ would 
imply non-negligible Higgsino components in $\na$,
and be a clear signal for deviations from the minimal
model in the Higgs sector. 
\jfig{sfig1x}{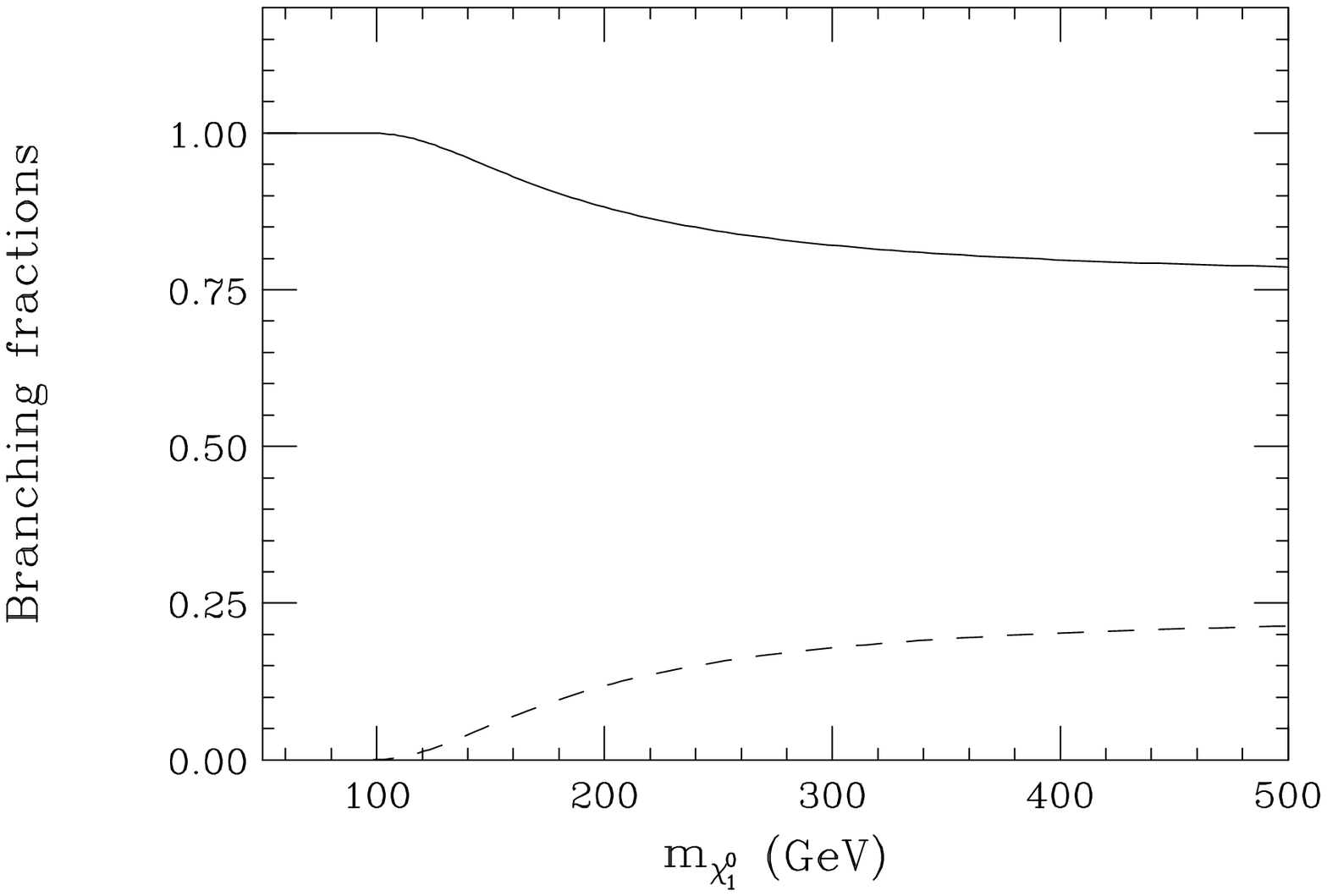}
{The branching ratios for $\na \to \gamma + G$ (solid line) 
and $\na \to Z^0 + G$ (dashed line) as a function of 
$m_{\na}$ for $\Lambda=M$.} 
For example, as discussed in section \ref{addhiggs},
$\Delta_-$ large and negative leads to a mostly 
Higgsino $\na$, which decays predominantly by 
$\na \to h^0 + G$.
The branching ratios in the minimal model 
for $\na \to \gamma +G$ and 
$\na \to Z^0 + G$ are shown in Fig. 
\ref{sfig1x} as a function of $m_{\na}$ for $\Lambda=M$.

In the minimal model, with $\na$ decaying within the detector,
the signatures are the same as those given above, but 
with an additional pair of  $\gamma \gamma$,
$\gamma Z^0$, or $Z^0 Z^0$.
At an $e^+ e^-$ collider $e^+ e^- \to \na \na \to 
\gamma \gamma + \Emiss$ becomes the discovery mode 
\cite{sbtalk, signatures, stump}.
At a hadron collider the reduction in $\EmissT$ 
from the secondary decay is 
more than compensated by the additional 
very distinctive visible energy.
The presence of hard photons 
significantly reduces the background compared
with standard supersymmetric signals 
\cite{sbtalk, signatures, tevatron, akkm, SUSY96}.
In addition, decay of $\na \to \gamma + G$
over a macroscopic distance leads to displaced
photon tracks, and of
$\na \to Z^0 + G$ to displaced charged particle tracks. 
Measurement of the displaced vertex distribution 
gives a measure of the supersymmetry breaking scale.

\jfig{sfig3x}{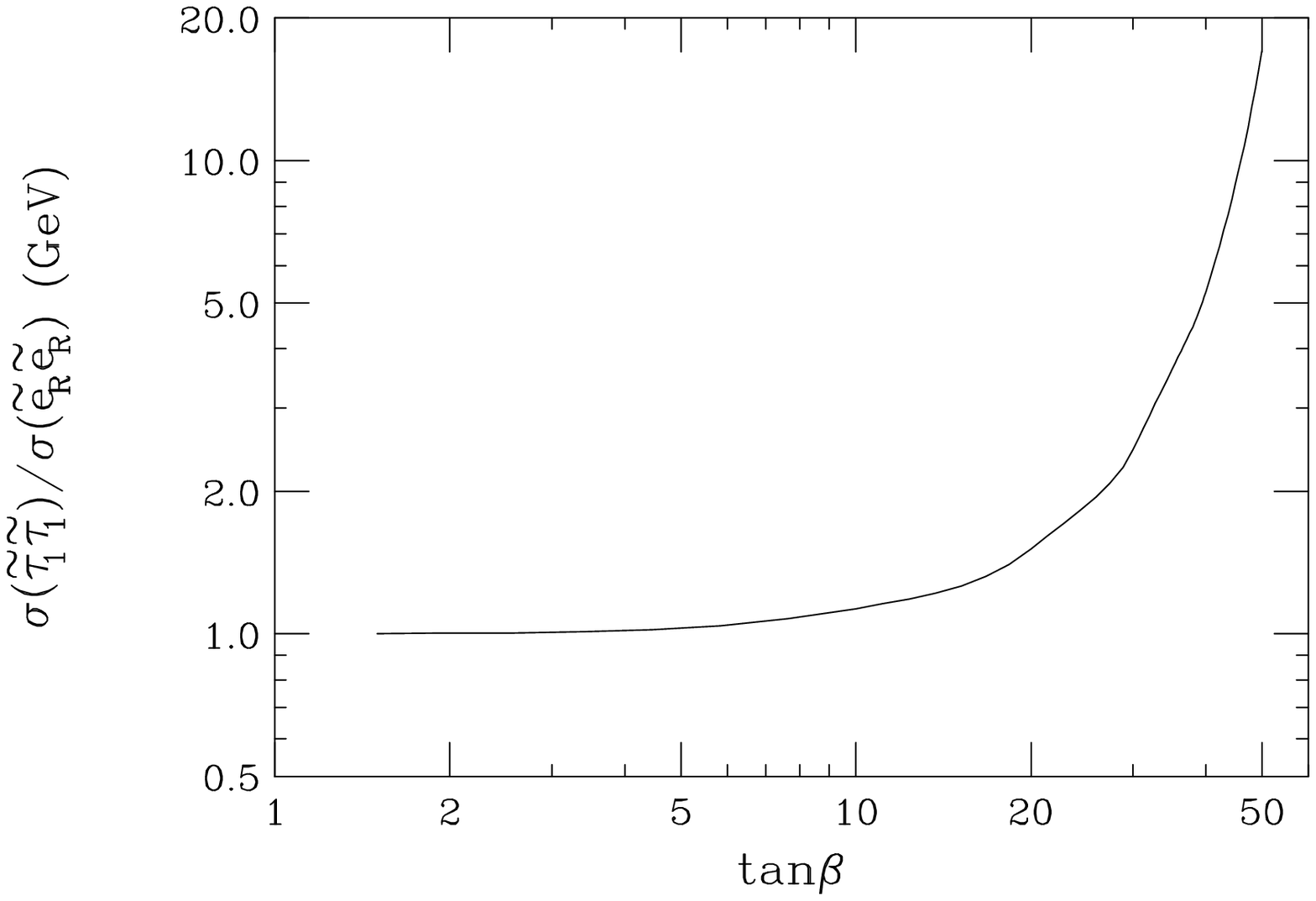}
{The ratio $\sigma( p \bar{p} \to \stau_1^+ \stau_1^-) / 
\sigma( p \bar{p} \to \eR^+ \eR^-)$ as a function of 
$\tan \beta$ for $\mbino(M)=115$ GeV and $\Lambda=M$.
The center of mass energy is 2 TeV.}
In the minimal model, for large $\tan \beta$ the 
$\stau_1$ can become significantly lighter than 
$\eR$ and $\tilde{\mu}_R$.
This enhances the $\stau_1^+ \stau_1^-$ production cross 
section at a hadron collider. 
The ratio  $\sigma( p \bar{p} \to \stau_1^+ \stau_1^-) / 
\sigma( p \bar{p} \to \eR^+ \eR^-)$ for $\sqrt{s}=2$ TeV 
is shown in Fig. \ref{sfig3x} as a function of 
$\tan \beta$ for $\mbino(M)=115$ GeV and $\Lambda=M$.
Measurement of this ratio gives a measure of the
$\tilde{\tau}_1$ mass. 
Within the minimal model this allows an indirect probe
of $\tan \beta$.

\subsubsection{Heavy Charged Particle Signatures}

In the minimal model 
the $\stau_1$ becomes lighter than $\na$
for $\tan \beta$ large enough, 
as discussed in section \ref{electroweaksection}.
The $\stau_1$ is then the lightest standard model superpartner. 
This is not a cosmological problem since the $\stau_1$ 
can decay to the Goldstino component of the gravitino,
$\stau_1 \to \tau + G$,  
on a cosmologically short time scale. 
However, if the supersymmetry breaking
scale is larger than a few 1000 TeV, 
at a collider this decay 
takes place well outside the detector.
The signature for supersymmetry in this case is 
heavy charged particles passing through the detector, rather than 
missing energy.
At an $e^+e^-$ collider the most relevant mode is then 
$e^+e^- \to \stau_1^+ \stau_1^-$. 
At a hadron collider the signatures are very different
since $\na$ decays by $\na \to \stau_1^{\pm} \tau^{\mp}$. 
Over much of the parameter space
the dominant chargino production then gives rise to 
the signatures
$p \bar{p} \to W^{\pm} Z^0  \tau^+ \tau^- \stau_1^+ \stau_1^-$, 
$W^{\pm} h^0  \tau^+ \tau^- \stau_1^+ \stau_1^-$, and 
$W^+ W^- \tau^+ \tau^-  \stau_1^+ \stau_1^-$.
The additional cascade decays 
$\nb \to \stau_1^{\pm} \tau^{\mp}$ and 
$\chi_1^{\pm} \to \stau_1^{\pm} \nu_{\tau}$ 
are also available through the $\stau_L$ component of 
$\stau_1$. 
Chargino production can therefore also give the signatures 
$p \bar{p} \to W^{\pm} \tau^{\pm} \tau^{\pm}
   \stau_1^{\mp} \stau_1^{\mp}$,
$Z^0 \tau^{\pm} \stau_1^{\pm} \stau_1^{\mp} + \EmissT$, 
$h^0 \tau^{\pm} \stau_1^{\pm} \stau_1^{\mp} + \EmissT$,
$\tau^{\pm} \stau_1^{\mp} \stau_1^{\mp} + \EmissT$,
$W^{\pm} \tau^{\pm} \stau_1^{\mp} \stau_1^{\pm} + \EmissT$,
and 
$\stau_1^+ \stau_1^- + \EmissT$. 
Finally, direct pair production gives the 
signature 
$p \bar{p} \to \stau_1^+ \stau_1^-$, 
while $\tilde{l}_{R}^+ \tilde{l}_{R}^-$ 
production gives 
$p \bar{p} \to l^+ l^{-} \tau^+ \tau^- \stau_1^+ \stau_1^-$
for $l = e, \mu$.

If the supersymmetry breaking scale is well below a few 
1000 TeV the $\stau_1$ decays within the detector
to a Goldstino by $\stau_1 \to \tau + G$.
The signature of heavy charged particle production is 
then lost, but missing energy results since the Goldstinos
escape the detector.
In the signatures given above then all 
the $\stau_1^{\pm}$ are replaced by $\tau^{\pm} + \EmissT$. 

The signature of heavy charged particles can also result
with multiple generations in the messenger sector. 
As discussed in section \ref{multiple}, messenger 
sectors with larger matter representations result
in gauginos which are heavier relative to the scalars
than in the minimal model. 
For $N \geq 3$, and over much of the parameter
space of the $N=2$ model, a right handed slepton 
is the lightest standard model superpartner. 
Because of the larger Yukawa coupling, the $\stau_1$ 
is always lighter than $\tilde{e}_R$ and $\tilde{\mu}_R$. 
However, for small to moderate $\tan \beta$ 
$m_{\tilde{\mu}_R} - m_{\stau_1} < m_{\tau} + m_{\mu}$,
and the decay 
$\tilde{\mu}_R^{\pm} \to \stau_1^+ \tau^- \mu^{\pm}$ 
through the $B$-ino component of off-shell $\chi_1^{0*}$
is kinematically blocked, and 
likewise for $\tilde{e}_R$ \cite{SUSY96}.
In addition, the second order electroweak decay 
$\tilde{\mu}_R^{+} \to \stau_1^{+} {\nu}_{\tau} \bar{\nu}_{\mu}$
is highly suppressed and not relevant for decay within the  
detector. 
In this case all three sleptons $\tilde{e}_R$, 
$\tilde{\mu}_R$, and $\stau_1$, are effectively
stable on the scale of the detector
for a supersymmetry breaking scale larger than a few 1000 TeV. 
At an $e^+e^-$ collider the most relevant signature
becomes $e^+e^- \to \lR^+ \lR^-$ with the sleptons
leaving a greater than minimum ionizing track in the detector. 
At a hadron collider $\chi_1^{\pm} \chi_2^0$ and
$\chi_1^+ \chi_1^-$ production gives the signatures 
$p \bar{p} \to W^{\pm} Z^0 l^+ l^{\prime -} \tilde{l}_R^- 
  \tilde{l}_R^{\prime +}$, 
$W^{\pm} h^0 l^+ l^{\prime -} \tilde{l}_R^- 
  \tilde{l}_R^{\prime +}$, and  
$W^+ W^- l^+ l^{\prime -} \tilde{l}_R^- \tilde{l}_R^{\prime +}$, 
while
direct slepton pair production gives 
$p \bar{p} \to  \tilde{l}_R^+ \tilde{l}_R^-$. 
If $\tan \beta$ is large then 
$m_{\tilde{\mu}_R} - m_{\stau_1} > m_{\tau} + m_{\mu}$,
so that the decay 
$\tilde{\mu}_R^{+} \to \stau_1^{\pm} \tau^{\mp} \mu^{+}$
can take place within the detector,
and likewise for $\eR$.
All the cascades then end with $\stau_1^{\pm}$.
The additional $\tau^{\pm} l^+$, $\tau^{\pm} l^-$ which result
from $\tilde{l}_R^{\pm}$ decay are very soft unless the
splitting $m_{\lR} - m_{\stau_1}$ is sizeable. 

If the supersymmetry breaking scale is below a few 1000 TeV, 
the sleptons can decay to the Goldstino by 
$\tilde{l}_R \to l + G$ within the detector. 
A missing energy signature then results from the escaping
Goldstinos, and all the $\tilde{l}_R^{\pm}$ in the above signatures
are replaced by $l^{\pm} + \EmissT$. 
If the decay $\lR \to l + G$ takes place over a macroscopic 
distance the spectacular signature of a greater than 
minimizing ionizing track with a kink to a minimum
ionizing track results \cite{signatures,SUSY96}. 
Again, measurement of the decay length distribution would
give a measure of the supersymmetry breaking scale. 
All these interesting heavy charged particle signatures should
not be overlooked in the search for supersymmetry at 
future colliders.

\section{Conclusions}

Gauge-mediated supersymmetry breaking has many consequences
for the superpartner mass spectrum, and phenomenological 
signatures. 
In a large class of gauge-mediated models 
(including all the single spurion models given in this paper)
the general features include:
\begin{itemize}
\item The natural absence of flavor changing neutral currents. 
\item A large hierarchy among scalars with different gauge
charges, {$m_{\tilde q_R}/m_{\tilde{l}_R}\lsim 6.3$}, and 
{$m_{\tilde l_L}/m_{\tilde{l}_R}\lsim 2.1$}, with the inequalities
saturated for a messenger scale of order the supersymmetry
breaking scale. 
\item Mass splittings between scalars with different gauge 
 quantum numbers are related by various sum rules. 
\item ``Gaugino unification'' mass relations.
\item{Precise degeneracy among the first two generation scalars, 
    and sum rules
    for the third generation that test the flavor symmetry of 
     masses at the messenger scale.}
\item Radiative electroweak symmetry breaking induced by 
heavy stops, even for a low messenger scale. 
\item Small $A$-terms.
\item The lightest standard model superpartner is either
$\na$ or $\lR^{\pm}$.
\item 
The possibility of 
the lightest standard model superpartner
decaying within the detector to its partner plus the Goldstino. 
\end{itemize}
The mass relations and sum rules hold in a very large class of
gauge-mediated models and represent fairly generic features. 
The possibility that the lightest standard model superpartners is 
a charged slepton leads to the dramatic signature of 
heavy charged particles leaving a greater than minimum ionizing
track in the detector. 
This signature should not be overlooked in searches for supersymmetry
at future colliders. 
The possibility that the lightest standard model
superpartner decays within the detector, either
$\na \to (\gamma, Z^0, h^0) + G$ or $\lR \to l + G$, 
leads to very distinctive signatures, and provides
the possibility of indirectly measuring the supersymmetry
breaking scale.

The minimal model of gauge-mediated supersymmetry breaking
is highly constrained, and gives the additional general 
features:
\begin{itemize}
\item Gauginos are lighter than the associated scalars, 
$m_3 < m_{\tilde{q}}$, 
$m_2 < m_{\lL}$, and 
$m_1 < m_{\lR}$. 
\item The Higgsinos are heavier than the electroweak gauginos, 
$3 m_1 \lsim |\mu| \lsim 6 m_1$. 
\item Absence of a light stop.
\item The mass of the lightest Higgs boson receives large 
radiative corrections from the heavy stops, 
{$80\gev\lsim m_{h^0}\lsim 140\gev$.}
\item Unless $\tan \beta$ is very large,
the lightest standard model superpartner is the 
mostly $B$-ino $\na$, which decays predominantly by 
$\na \to \gamma +G$. 
\item At a hadron collider the largest supersymmetric production 
cross section is for $\chi_1^{\pm} \chi_2^0$ and 
$\chi_1^+ \chi_1^-$. 
\item{Discernible deviation in $Br(b\to s\gamma)$ from the standard model
         with data from future $B$-factories.}
\end{itemize}

If superpartners are detected at a high energy collider, one of the most
important tasks will be to match the low energy spectrum with a more
fundamental theory.  
Patterns and relations among the superpartner masses
can in general give information about the messenger sector
responsible for transmitting supersymmetry breaking. 
As discussed in this paper, gauge-mediated supersymmetry
breaking leads to many distinctive patterns in the superpartner
spectrum. 
Any spectroscopy can of course 
be trivially mocked by postulates of
non-universal boundary conditions at any messenger scale. 
However, gauge-mediation in its minimal form represents 
a {\it simple} anzatz which is highly predictive. 
In addition, if decay of the lightest standard model superpartner
takes place within the detector, implying a low supersymmetry breaking
scale, the usual gauge interactions
are likely to play some role in the messenger sector.

The overall scale for the superpartner masses is of course a free
parameter.
However, the Higgs sector mass parameters set the scale
for electroweak symmetry breaking. 
Since all the superpartner masses are related to a single
overall scale with gauge-mediated supersymmetry breaking, 
it is reasonable that the states transforming under
$SU(2)_L$ have mass of order the electroweak scale. 
From the low energy point of view,
masses much larger than this scale would appear 
to imply that electroweak symmetry breaking is tuned, and 
that the electroweak scale is unnaturally small. 
Quantitative measures of tuning are of course subjective. 
However, 
when the overall scale is large compared to $\mZ$, 
tuning among the Higgs sector parameters arises in 
the minimization condition (\ref{mincona}) as a 
near cancelation between $(\tan^2 \beta-1)|\mu|^2$ and 
$\mHu^2 - \tan^2 \beta \mHd^2$, resulting in 
$\mZ^2 \ll |\mu|^2$. 
In this regime the near cancelation enforces constraints
among some of the Higgs sector parameters in order to obtain
proper electroweak symmetry breaking. 
As the overall superpartner scale is increased these
tuned constraints are reflected by ratios
in the physical spectrum which become independent of the 
electroweak scale. 
This tuning is visually apparent in Fig. \ref{sfig14n}
as the linear dependence of $\mA$ on $m_{\na}$
at large overall scales. 
The ``natural'' regime in which the Higgs sector parameters
are all the same order as the electroweak and superpartner scale
can be seen in Fig. \ref{sfig14n}
as the non-linear dependence of $\mA$ on $m_{\na}$. 
In Fig. \ref{sfig16n} this ``natural''
non-linear regime with light superpartners
is in the far lower
left corner, and hardly discernible in the linearly scaled plot. 
Although no more subjective than any measure of tuning, 
this bodes well for the prospects 
indirectly detecting the effects of superpartners and Higgs
bosons in precision measurements, and 
for directly producing superpartners at future colliders.

\medskip
\noindent
{\it Acknowledgements:} We would like to thank M. Carena, M. Dine, 
G. Giudice, H. Haber, S. Martin, 
M. Peskin, D. Pierce, A. Pomarol, 
and C. Wagner for constructive comments.
We would also like to thank 
the Aspen Center for Physics and CERN, where this work was partially 
completed. 



\appendix

\section{Soft Masses for a General Messenger Sector}

\label{appgeneral}

The messenger sector of the minimal model may be generalized
to fields $Q$ and $\overline{Q}$ forming any vector
representation of the standard model gauge group.
With a coupling to a single background spurion, 
$W = \lambda S Q \overline{Q}$,
the general expression 
for gaugino masses at the messenger scale is 
\beq
m_{\lambda_i} =  \sum_m C_{2,i} \Lambda ~ { \alpha_i \over 4 \pi}
\eq
where ${\rm Tr} (T^a T^b) = C_{2} \delta^{ab}$,
and $\sum_m$ is over all messenger representations.  
For a fundamental of $SU(N_c)$ $C_2 = 1/2$,
for a two index antisymmetric tensor $C_2 = N_c/2-1$,
and for an adjoint $C_2 = N_c$. 
The general expression for scalar masses at the messenger 
scale is 
\beq
m^2 = 2 \Lambda^2~ \sum_{m,i} C_{2,i} C_{3,i} 
   \left( \alpha_i \over 4 \pi \right)^2
\eq
where $\sum_a (T^a T^a) = C_{3}{\bf 1}$ is over the visible sector
field.
For a fundamental of $SU(N_c)$ $C_3 = (N_c^2 -1) / (2N_c)$,
for a two index antisymmetric tensor $C_3 = (N_c+1)(N_c-2)/N_c$, 
and for an adjoint $C_3 = N_c$. 

The successful prediction of 
$SU(3)_C \times SU(2)_L \times U(1)_Y$ gauge coupling
unification \cite{dimgeo}
is unaffected at lowest order by the messenger sector if all the 
$C_{2,i}$ $i=1,2,3$ are identical.  
In this case the effect of the messengers 
on the renormalization group evolution of gauge couplings
can be absorbed in a 
shift of the gauge coupling at the unification scale. 	
Messengers which can be embedded in GUT representations are a 
subset of all such possible messenger sectors. 
All messenger sectors of this type with a single spurion 
also yield the ``gaugino unification'' mass relations
at lowest order,
$m_3 : m_2 : m_1 =  \alpha_3 : \alpha_2 : \alpha_1$.
In addition, with a single spurion the ratios of gaugino to scalar
masses at the messenger scale depend on $C_2$. 
For $\alpha_3 \gg \alpha_2 > \alpha_1$ the approximate relations
$m^2 \simeq 2 C_2 C_{3,i} m_{\lambda_i}^2$ for $i=1,2,3$, 
result at the messenger scale. 
With a single spurion $\Rslash = \sqrt{C_2}$.

\section{A Non-minimal Messenger Sector}

\label{appnonmin}


The minimal model of gauge-mediated supersymmetry breaking
represents a very constrained and predictive theory for 
the soft supersymmetry breaking parameters. 
However, 
as discussed in section (\ref{subvariations}), 
non-minimal messenger sectors can 
modify the relations among the soft terms. 
Among the possible deviations away from the minimal model
are {\it i}) messenger masses which arise from a sector
not associated with supersymmetry breaking, thereby introducing
dependence on the messenger Yukawa couplings, 
{\it ii}) an approximate $U(1)_R$ symmetry, leading to gauginos much
lighter than the scalars, 
{\it iii}) messenger sectors which can be embedded in a 
GUT theory, but do not lead to the ``standard'' GUT relations
among gaugino masses. 
Here we present a single model which illustrates all of these 
features. 
The model is a generalization of the one given in Ref. \cite{tevatron}
with two generations of ${\bf 5} + \overline{\bf 5}$ of $SU(5)$
and two fields, $X$ and $S$, with a superpotential 
coupling
\beq
W = X ( \lambda_2 \ell_1 \bar{\ell}_1 + \lambda_3 q_1 \bar{q}_1 + \xi^2)
 + \lambda' S (\ell_1 \bar{\ell_2} 
               + q_1 \bar{q}_2  
        + ~1 \leftrightarrow 2~)
\eq
The field $S$ is taken to transform as a singlet under 
$SU(5)$, while $X$ and the spurion $\xi^2$ transform as the 
$SU(3)_C \times SU(2)_L \times U(1)_Y$ 
singlet component of ${\bf 24}$'s of $SU(5)$. 
The $({\bf 8}, {\bf 1}) + ({\bf 3},{\bf 2}) + (\bar{\bf 3},{\bf 2}) 
  + ( {\bf 1}, {\bf 3})$ components of $X$ can 
gain a mass at the GUT scale. 
With these representations $SU(5)$ invariance implies  
that $X$ couples to the messenger fields proportional to 
$U(1)_Y$, so that $3 \lambda_3 = - 2 \lambda_2$
at the GUT scale.  
The messenger Yukawa couplings are therefore $SU(5)$  
invariant, but not $SU(5)$ singlets. 
For $\xi \neq 0 $ supersymmetry is broken by the O'Raifeartaigh
mechanism. 
For $\lambda^{\prime} S > \xi$, the ground state is at 
$q_i = \bar{q}_i = \ell_i = \bar{\ell}_i = 0$, 
with $X$ and $S$ undetermined at tree level. 
For $S \neq 0$ there is a $U(1)_R$ symmetry which is broken
only for $X \neq 0$. 

The model exhibits the features mentioned above with  
$(\lambda_i X), \xi \ll \lambda' S$.
In this limit the messengers obtain a mass mainly from 
the $S$ expectation value, and receive soft supersymmetry breaking
masses only from the auxiliary component of $X$.
This misalignment of the scalar and fermion mass matrices
introduces dependence on the relative magnitude
of the Yukawa couplings $\lambda_2$ and $\lambda_3$. 
In this limit the $X$ superfield may be treated as an insertion
in the graphs which give rise to visible sector soft terms. 
The resulting gaugino masses are in the ratios
\beq
m_3 : m_2 : m_1 = \lambda_3^2 \alpha_3: 
 \lambda_2^2 \alpha_2 : 
  {1 \over 5}(2 \lambda_3^2 + 3 \lambda_2^2) \alpha_1
\eq
Likewise the scalar masses at the messenger scale
are in the approximate ratios
\beq
m_{\tilde{q}}^2 : m_{\lL}^2 : m_{\lR}^2 \simeq 
{4 \over 3} \lambda_3^2 \alpha_3^2 : 
{3 \over 4} \lambda_2^2 \alpha_2^2 : 
{3 \over 25} ( 2 \lambda_3^2 + 3 \lambda_2^2) \alpha_1^2
\eq
In addition to the sensitivity to the relative magnitude
of the Yukawa couplings, there is an approximate $U(1)_R$ 
symmetry in this limit.  
Since the gaugino masses require an insertion of both the  
scalar and auxiliary components of $X$, while the scalars 
require only auxiliary components, 
the gauginos are lighter than the scalars 
by ${\cal O}(\lambda_i X / \lambda' S)$.  
Finally, even though the model may be embedded in a GUT theory, 
the gauginos do not satisfy the ``standard'' gaugino unification
relation.
This results from the misalignment of the messenger
scalar and fermion mass matrices, and the fact that the couplings
transform under $SU(5)$.

This model may also be generalized to a messenger sector
with two  generations
of ${\bf 10} + \overline{\bf 10}$ of $SU(5)$,
with superpotential couplings
\beq
W = X ( \lambda_{3,2} Q_1 \bar{Q}_1  + \lambda_{3} u_1 \bar{u}_1
   + \lambda_1 e_1 \bar{e}_1 + \xi^2 ) + 
 \lambda' S (  Q_1 \bar{Q}_2 +
               u_1 \bar{u}_2 +
               e_1 \bar{e}_2 +
      ~ 1 \leftrightarrow 2~)
\eq
With these representations $SU(5)$ invariance again implies 
that $X$ couples proportional to $U(1)_Y$, so that 
$2 \lambda_1 = 12 \lambda_{3,2} = - 3 \lambda_3$ 
at the GUT scale. 
In the limit $(\lambda_i X), \xi \ll \lambda' S$,
the gaugino masses are in the ratio
\beq
m_3 : m_2 : m_1 = ( 2 \lambda_{3,2}^2 + \lambda_3^2) \alpha_3 : 
  3 \lambda_{3,2}^2 \alpha_2 : 
 { 1 \over 5} (\lambda_{3,2}^2 + 8 \lambda_3^2 + 6 \lambda_1^2) \alpha_1
\eq
This leads to the interesting hierarchy $m_2 \ll m_1 < m_3$,
which does not even approximately satisfy ``gaugino unification.''
The large $B$-ino mass arises from the large messenger positron
Yukawa coupling. 
The scalar masses are in the approximate ratios
\beq
m_{\tilde{q}}^2 : m_{\lL}^2 : m_{\lR}^2 \simeq 
{4 \over 3} (2 \lambda_{3,2}^2 + \lambda_3 ) \alpha_3^2  : 
{9 \over 4} \lambda_2^2 \alpha_2^2 : 
{3 \over 25} ( \lambda_{3,2}^2 + 8 \lambda_3^2 + 6 \lambda_1^2 ) 
    \alpha_1
\eq
Again, in this model 
the gauginos are lighter than the scalars by 
 ${\cal O}(\lambda_i X / \lambda' S)$.



\section{Decay to the Goldstino}

\label{appgoldstino}

The spontaneous breaking of global supersymmetry leads to the existence of 
a massless Goldstino fermion, the Goldstino. 
The lowest order couplings 
for emission or absorption of a single 
on-shell Goldstino 
are fixed by the supersymmetric 
Goldberger-Treiman low energy theorem to be proportional to the
divergence of the supercurrent 
\cite{fayet}
\beq
{\cal L } = - {1 \over F} \partial_{\mu} G^{\alpha}  j^{\mu}_{~\alpha}
~+~h.c.
\label{goldcoupling}
\eeq
where 
\beq
 j^{\mu}_{~\alpha} = \sigma^{\nu}_{\ai \aid} \bar{\sigma}^{\mu \aid \bi} 
  \psi_{\bi} D_{\nu} \phi + {1 \over 2 \sqrt{2}} \sigma^{\nu}_{\ai \aid} 
  \bar{\sigma}^{\rho \aid \bi} \sigma^{\mu}_{\bi \bid}  
  {\lambda^{a \bid}}^* F^a_{\nu \rho} 
\eeq
is the supercurrent, and the components of the 
chiral superfields and vector superfield strengths are given by 
\beq
\Phi = \phi + \sqrt{2} \theta^{\ai} \psi_{\ai} + \theta^2 F
\eeq
\beq
W_{\alpha}^a = \lambda_{\ai}^a + \left( \delta_{\ai}^{~\bi} D^a 
   - {i \over 2} \sigma^{\mu}_{\ai \aid} \bar{\sigma}^{\nu \aid \bi} 
F^a_{\mu \nu}  \right)  \theta_{\beta}  + 
 i \theta^2 \sigma^{\mu}_{\ai \aid} \partial_{\mu} {\lambda^{a \aid}}^*
\eeq
The physical Goldstino and supersymmetry breaking scales are given by  
\beq
G^{\ai} = {1 \over F}~
\sum_{\psi_i} \psi_i^{\ai} \langle F_i \rangle +
  {1 \over \sqrt{2}} \sum_{\lambda_i^a} \lambda_i^{a \ai} \langle D_i^a \rangle
\eq
\beq
F \equiv 
   \sum_{i} \langle F_i \rangle +
  {1 \over \sqrt{2}} \sum_{i,a} \langle D_i^a \rangle
\eq
where the sums are over all chiral and vector multiplets,
with auxiliary components $F_i$ and $D_i^a$ respectively, in both
the supersymmetry breaking and visible sectors. 

Only terms which break supersymmetry
in the low energy effective theory
contribute to $\partial_{\mu} j^{\mu}_{~\alpha}$, and therefore contribute
to Goldstino couplings. 
The derivative form of the coupling (\ref{goldcoupling})
may be obtained by applying a 
space-time dependent supersymmetry transformation to cancel the 
non-derivative Goldstino couplings arising from the effective
operators which couple the visible and supersymmetry breaking sectors. 
Either basis can in principle be used to compute Goldstino couplings.
The derivative basis is more often convenient since supersymmetry breaking
in the low energy theory 
then appears only on on-shell external states through equations of motion. 
This basis is especially useful for the lightest 
neutralino in the standard model,
$\na$,  which is a mixture of states that  
receive mass from both supersymmetric and supersymmetry breaking terms
in the low energy theory. 
 
In local supersymmetry the Goldstino becomes the longitudinal component
of the gravitino, giving a gravitino mass of 
\beq
m_G = {F \over \sqrt{3} M_p}
\eq
where $M_p = m_p / \sqrt{8 \pi}$ is the reduced Planck mass.
For a Messenger scale well below the Planck scale, the gravitino
is much lighter than the electroweak scale, and is naturally 
the lightest supersymmetric particle (LSP). 
The lightest standard model superpartner is the the next to lightest
supersymmetric particle (NLSP). 
In this case the gravitino mass 
and couplings of the spin ${3 \over 2}$ transverse components 
are completely irrelevant
for accelerator experiments. 
The global description in terms of the 
the spin ${1 \over 2}$ Goldstino component is therefore sufficient.


The Goldstino acts on supermultiplets like the supercharge,
and therefore transforms a superpartner into its partner. 
Since the Goldstino couplings are suppressed compared with electro-weak and 
strong interactions, 
decay to the Goldstino is only relevant for the 
lightest standard model superpartner (NLSP).
Assuming $R$-parity
conservation, the NLSP is quasi-stable and can only decay through coupling
to the Goldstino. 
For $\sqrt{F}$ below a few 1000 TeV, such a decay can take place
within a detector. 
With gauge-mediated supersymmetry breaking the natural candidate
for the NLSP 
is either a slepton or electro-weak neutralino. 
For a slepton NLSP the decay rate for $\tilde{l} \to l + G$ is 
\beq
\Gamma( \tilde{l} \to l + G) = { m_{\tilde{l}}^5 \over 16 \pi F^2}
\eq

It is also possible that the NLSP is the lightest electro-weak neutralino,
$\na$. 
It can decay through the gaugino components by $\na \to \gamma + G$ and 
$\na \to Z^0 + G$ if kinematically accessible, and through the 
Higgsino components by 
$\na \to Z^0 + G$, 
$\na \to h^0 + G$, $\na \to H^0 + G$, and $\na \to A^0 + G$ if 
kinematically accessible. 
The decay rates for gauge boson final states are   
\beq
\Gamma( \na \to \gamma + G) = | \cos \theta_W N_{1 \tilde{B}} + 
                                \sin \theta_W N_{1 \tilde{W}} |^2 
    ~{ m_{\na}^5 \over 16 \pi F^2}
\eq
$$
\Gamma( \na \to Z^0 + G) =   \left( |
                                \sin \theta_W N_{1 \tilde{B}} - 
                                \cos \theta_W N_{1 \tilde{W}} |^2 
              + {1 \over 2}  |  \cos \beta N_{1 {d}} - 
                                \sin \beta N_{1 {u}}  |^2  \right)
$$
\beq
    ~~~~~\times~ { m_{\na}^5 \over 16 \pi F^2}
    \left( 1 - { m_{Z^0}^2 \over m_{\na}^2} \right)^4 
\eq
The amplitudes for the 
decay $\na \to Z^0+G$ from gaugino and Higgsino components do not
interfere since the gaugino admixtures couple only to transverse $Z^0$ components, while
the Higgsino admixtures couple only to longitudinal $Z^0$ components. 
Even though $\na$ is a fermion, the decay to a gauge boson is isotropic in the
rest frame. 
This follows since $\na$ is Majorana, and can therefore
decay to both Goldstino helicities. 
The rate to the two Goldstino helicities sums to an isotropic distribution. 

The decay rates for Higgs boson final states are 
\beq
\Gamma( \na \to h^0 + G) =    |
                                \sin \alpha N_{1 {d}} - 
                                \cos \alpha N_{1 {u}} |^2 
    ~{ m_{\na}^5 \over 32 \pi F^2}
    \left( 1 - { m_{h^0}^2 \over m_{\na}^2} \right)^4 
\eq
\beq
\Gamma( \na \to H^0 + G) =    |
                                \cos \alpha N_{1 {d}} + 
                                \sin \alpha N_{1 {u}} |^2 
    ~{ m_{\na}^5 \over 32 \pi F^2}
    \left( 1 - { m_{H^0}^2 \over m_{\na}^2 } \right)^4 
\eq
\beq
\Gamma( \na \to A^0 + G) =    |
                                \sin \beta N_{1 {d}} +
                                \cos \beta N_{1 {u}} |^2 
    ~{ m_{\na}^5 \over 32 \pi F^2}
    \left( 1 - { m_{A^0}^2 \over m_{\na}^2 } \right)^4 
\eq
where the physical Higgs bosons, and Goldstone boson are related to 
$H_u^0$ and $H_d^0$ by \cite{hhguide}
$$
H_d^0 = {1 \over \sqrt{2}} \left( v_d +
  H^0 \cos \alpha - h^0 \sin \alpha + i A^0 \sin \beta
   - i G^0 \cos \beta \right)
$$
\beq
H_u^0 = {1 \over \sqrt{2}} \left( v_u +
  H^0 \sin \alpha + h^0 \cos \alpha + i A^0 \cos \beta
  + i G^0 \sin \beta \right)
\label{higgscomp}
\eq
These expressions for the decay rates agree with those obtained
in Ref. \cite{akkm}.
Because of the Goldstino derivative coupling to the gauge field strengths and
derivative of the scalar components, the decay to massive final states
suffers a $\beta^4$ suppression near threshold, where 
$\beta=\sqrt{1-4 (m_f^2/m_i^2)/(1+m_f^2/m_i^2)^2}$ is the 
massive final state velocity in the decaying rest frame.

With minimal boundary conditions, electro-weak symmetry breaking implies
that both $|\mu|$ and $m_{A^0}$ are somewhat larger than $m_{\na}$ and 
$m_{Z^0}$.
It is therefore interesting to consider the decay rates in this limit. 
For $\mu^2 - m_1^2 \gg \mZ^2$,
$\na$ is mostly $B$-ino.
In this limit the neutralino mass matrix may be diagonalized perturbatively
to find the small Higgsino admixtures in $\na$. 
To ${\cal O}(\mZ \mu / (\mu^2 - m_1^2))$
the $\na$ eigenvectors become 
\bea
N_{1 \tilde{B}} & \simeq & 1 \nonumber \\
N_{1 \tilde{W}} & \simeq & 0 \nonumber \\
N_{1 {d}} & \simeq & ~\sin \theta_W \sin \beta 
    ~{ \mZ ( \mu + m_1 \cot \beta) \over |\mu|^2 - m_1^2  } 
  \nonumber \\
N_{1 {u}} & \simeq & -\sin \theta_W \cos \beta 
    ~{ \mZ ( \mu + m_1 \tan \beta) \over |\mu|^2 - m_1^2  } 
\eea
The decay rates to gauge boson final states are then dominated by
the $B$-ino component
\beq
\Gamma( \na \to \gamma + G) \simeq
    { \cos^2 \theta_W m_{\na}^5 \over 16 \pi F^2}
\eq
\beq
\Gamma( \na \to Z^0 + G) \simeq
            { \sin^2 \theta_W m_{\na}^5 \over 16 \pi F^2}
    \left( 1 - { m_{Z^0}^2 \over m_{\na}^2 } \right)^4 
\eq
For decay to Higgs bosons, with $m_{A^0} \gg m_{\na}$, only 
$\na \to h^0 + G$ is open. 
For $m_{A^0}^2 \gg m_{Z^0}^2$ the Higgs boson
decoupling limit is reached in which 
the $h^0$ couplings become standard model like. 
In this limit the $h^0 - H^0$ mixing angle
is related to $\tan \beta$ by 
$\sin \alpha \simeq - \cos \beta$, and
$\cos \alpha \simeq \sin \beta$ \cite{hhguide}.
The $h^0$ components of $H^0_u$ and $H^0_d$ then align with the
expectation values, as can be seen from (\ref{higgscomp}). 
In the $B$-ino limit, using the approximate $\na$ eigenvalues
given above, the
$\na h^0 G$ coupling is then proportional 
to 
\beq
\sin \alpha N_{1d} - \cos \alpha N_{1u} \simeq
{ \mZ \sin \theta_W \over |\mu|^2 - m_1^2} \left( 
 { \mu \mZ^2 \sin 4 \beta \over 2 \mA^2} - m_1 \cos 2 \beta \right)
\eq
where 
$\cos(\alpha - \beta) \simeq \sin 4 \beta \mZ^2 / (2 \mA^2)$
\cite{hhguide}.
For $\mZ^2 / \mA^2 \ll m_1 / \mu$ 
the decay rate in the 
large $|\mu|$ and $\mA$ 
limits becomes 
\beq
\Gamma( \na \to h^0 + G) \simeq 
      { \sin^2 \theta_W  \cos^2 2 \beta  m_{\na}^5 \over 32 \pi F^2}
    \left( { \mZ m_{\na} \over |\mu|^2 - m_{\na}^2} \right)^2 
    \left( 1 - { m_{h^0}^2 \over m_{\na}^2 } \right)^4 
\eq
This is down by ${\cal O}(\mZ^2 m_{\na}^2 / \mu^4)$
 compared with
the gauge boson final states. 

The branching ratio for $\na \to h^0 + G$ is therefore always
quite
small in the minimal model.  
For small $m_{\na}$ the rate suffers the $\beta^4$ 
threshold suppression,
and for large $m_{\na}$ is suppressed by the rapid decoupling
of $h^0$. 
	For $m_{\na}^2 \gg \mZ^2$ 
the branching ratios to gauge bosons therefore 
dominate the two body decays
and approach
$$
{\rm Br}(\na \to \gamma + G) \simeq \cos^2 \theta_W \left( 
   1 + 4 \sin^2 \theta_W \mZ^2 / m_{\na}^2 \right) 
$$
\beq
{\rm Br}(\na \to Z^0 + G) \simeq \sin^2 \theta_W \left( 
   1 - 4 \cos^2 \theta_W \mZ^2 / m_{\na}^2 \right) 
\eq

\end{document}